\documentclass[
reprint,
 amsmath,
 amssymb,
 aps,
 prx,
 superscriptaddress, 
]{revtex4-2}

\usepackage[utf8]{inputenc}
\usepackage{graphicx}
\usepackage{enumitem}
\usepackage{overpic}
\usepackage{comment}
\usepackage{here}
\usepackage{ulem}
\usepackage{amsmath}
\usepackage{amssymb}
\usepackage{amsfonts}
\usepackage{physics}
\usepackage{mathbbol}
\usepackage{xfrac}
\usepackage{mathtools}

\usepackage{xcolor}
\definecolor{unibas01}{RGB}{165,215,210}  % mint
\definecolor{unibas05}{RGB}{210,5,55}     % red
\definecolor{unibas04}{RGB}{140,145,150}  % gray
\definecolor{unibas77}{RGB}{213,187,84}   % yellow
\colorlet{unibas-mint}{unibas01}
\colorlet{unibas-red}{unibas05}
\colorlet{unibas-gray}{unibas04}
\colorlet{unibas-yellow}{unibas77}

\usepackage{KeldyshInOutDiags} 

\newcommand{\comma}{~,}
\newcommand{\fullstop}{~.}
\newcommand{\newtext}[1]{{\color{black}#1}}
\newcommand{\ahat}{\hat{a}}
\newcommand{\dahat}{\hat{a}^\dagger}
\DeclareMathOperator{\bin}{\hat{\textit{b}}_{\text{in}}}
\DeclareMathOperator{\bout}{\hat{\textit{b}}_{\text{out}}}
\DeclareMathOperator{\dbin}{\hat{\textit{b}}_{\text{in}}^\dagger}
\DeclareMathOperator{\dbout}{\hat{\textit{b}}_{\text{out}}^\dagger}
\DeclareMathOperator{\rrangle}{\rangle\!\rangle}
\DeclareMathOperator{\llangle}{\langle\!\langle}

\usepackage[
  breaklinks=true,
  colorlinks=false,
  pdfusetitle=true,
  bookmarks=true,
  bookmarksopen=true,
  linktocpage=true,
  pdfstartview=FitH,
  pdfpagemode=UseOutlines
]{hyperref}

\begin{document}
\title{Path Integral Approach to Input--Output Theory}
\author{Aaron Daniel}
\affiliation{
Department of Physics and Swiss Nanoscience Institute, University of Basel, Klingelbergstrasse 82, CH-4056 Basel, Switzerland}
\author{Matteo Brunelli}
\affiliation{
Department of Physics and Swiss Nanoscience Institute, University of Basel, Klingelbergstrasse 82, CH-4056 Basel, Switzerland}
\affiliation{
JEIP, UAR 3573 CNRS, Collège de France, PSL Research University, 11 Place Marcelin Berthelot, 75321 Paris Cedex 05, France
}
\author{Aashish A. Clerk}
\affiliation{Pritzker School of Molecular Engineering, University of Chicago, Chicago, Illinois 60637, USA}
\author{Patrick P. Potts}
\affiliation{
Department of Physics and Swiss Nanoscience Institute, University of Basel, Klingelbergstrasse 82, CH-4056 Basel, Switzerland}
\date{\today}

\begin{abstract}
Input--output theory is a well--known tool in quantum optics and ubiquitous in the description of quantum systems probed by light. Owing to the generality of the setup it describes, the theory finds application in a wide variety of experiments in circuit and cavity QED.  
We present an approach to input--output theory using the Schwinger-Keldysh path integral formalism that gives us direct access to the full output field statistics such as the first and second order coherence functions. By making the rich toolbox of non--equilibrium quantum field theory accessible, our formalism greatly simplifies the treatment of nonlinear systems and provides a uniform way of obtaining perturbative results. We showcase this particular strength by computing the output field statistics of a Kerr nonlinear oscillator at finite temperatures through the use of diagrams and diagram summation techniques. We find a reduction in reflection that is not due to photon leakage but rather associated to the squeezing of the output light.
\end{abstract}

\maketitle

\section{Introduction}
\label{sec:introduction}
\begin{figure}[t]
    \centering
    \includegraphics[width=0.95\linewidth]{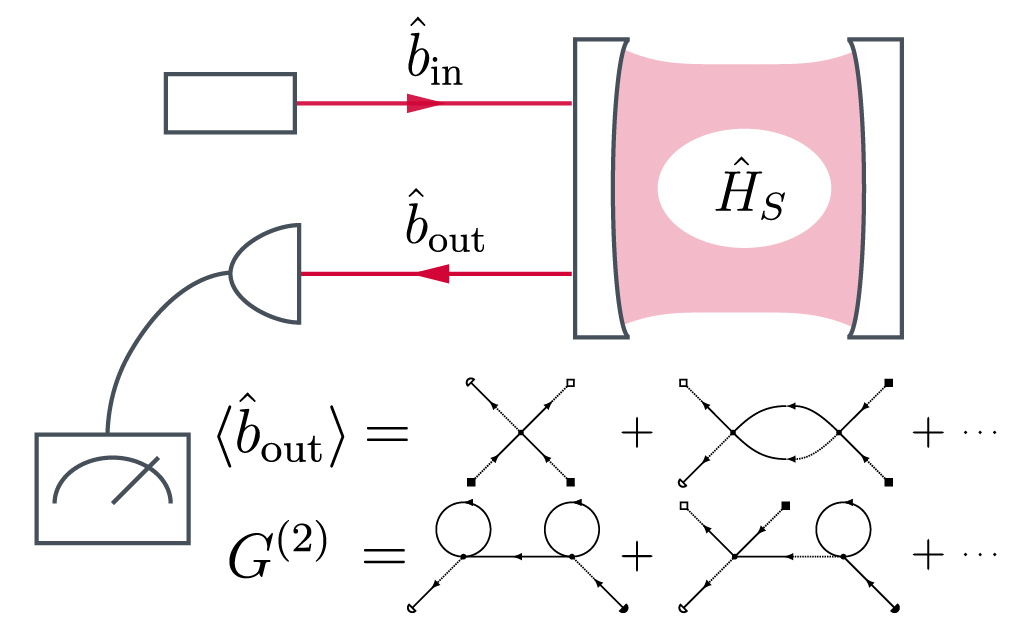}
    \caption{Sketch of the general setup described by input--output  theory. A system, placed inside a cavity and described by $\hat{H}_S$, is probed by light.
The incoming light is described by the input operator $\hat{b}_\text{in}$, the outgoing light which is measured by the output operator $\hat{b}_\text{out}$. Statistical quantities of the output light, such as the average output field $\langle \bout \rangle$ or the second order coherence function $G^{(2)}$, can be determined perturbatively with diagrams in our approach.}
    \label{fig:cavity_sketch}
\end{figure}

Numerous recent developments in quantum physics rely on light-matter coupling. On the one hand, light can be used for control and readout~\cite{RevModPhys.82.1155} and on the other hand, the coupling to light can fundamentally alter the properties of quantum systems~\cite{Huebener2021}. For an enhanced coupling strength, as well as targeted control over the emitted light, quantum systems can be embedded in cavities, see Fig.~\ref{fig:cavity_sketch}. Such setups are at the heart of the timely fields of circuit \cite{RevModPhys.93.025005} and cavity QED \cite{RevModPhys.87.1379}, optomechanics \cite{RevModPhys.86.1391}, as well as the emerging field of cavity materials engineering~\cite{Lu:25,Huebener2021,Mivehvar2021}, where the light-matter coupling is leveraged to alter the properties of many-body systems. 
Theoretically, these setups require the use of driven-dissipative theories such as master equations~\cite{breuer_book}, nonequilibrium field theories~\cite{Kamenev2011}, methods based on matrix-product states~\cite{Strathearn2018}, or a range of other techniques (see for instance Ref.~\cite{Mivehvar2021}). While these approaches are well established to describe the dynamics within the cavity, we are often interested in the properties of the light coming out of the cavity~\cite{walls_book,PhysRevA.30.1386,PhysRevA.31.3761,Yurke2004}. This light is what is typically measured and it can be further used as the input for another quantum system. Importantly, it may feature strikingly different behavior from the light within the cavity, for instance Wigner negativities for classical intra--cavity fields \cite{PhysRevResearch.3.023041} or unlimited squeezing \cite{PhysRevA.29.408,walls_book}.

Input-output theory allows for computing the light that leaves the cavity using quantum Langevin equations.  While constituting a powerful formalism, these equations are difficult to work with for nonlinear systems, where approximation tools are needed. 
In contrast to the dynamics within the cavity, there is a lack of theoretical methods for computing the light that leaves a nonlinear cavity. This constitutes a severe limitation in exploring the rich response of nonlinear systems embedded in cavities.
For a notable exception, see Refs.~\cite{PhysRevA.102.023717,PhysRevLett.123.123604}, where a master equation approach is developed to describe the light leaving a cavity that is driven by pulses and Ref.~\cite{PhysRevA.108.053703} for an approach to equal--time Green functions based on scattering theory. 

Here we fill this gap by combining input–output theory with the Schwinger-Keldysh path integral formalism. 
Using a dissipative action that only involves system degrees of freedom, and source fields for the output light, our formalism allows for calculating the statistical properties of the outgoing light, providing direct access to observables including the reflection, as well as the first and second order coherence functions. 
Through our approach, a rich toolbox of perturbative techniques becomes accessible, facilitating the treatment of nonlinear systems. In particular, we make use of diagrams that greatly simplify calculations and show promise for intuitive explanations of the interactions between the system and output modes. With our approach we capture a reduction in the reflection coefficient of a coherently driven Kerr cavity that is not due to photon loss and that cannot be captured by standard mean field theory. For the same system, we further demonstrate partial summation techniques that emerge from the diagrammatic treatment of the system. Our approach thereby offers a systematic method for determining the output field statistics of nonlinear quantum systems and establishes a connection between input-output theory, predominantly used in quantum optics and optomechanics, and the broader field of non--equilibrium field theory.

The manuscript is structured as follows. In Sec.~\ref{sec:HO_Keldysh} we introduce the general framework of our path integral approach to input--output theory and showcase it with the example of an empty cavity. We then consider nonlinear systems, e.g. the Kerr oscillator, in our framework in Sec.~\ref{sec:non_linear_system} and in Sec.~\ref{sec:diagrams} we detail our diagrammatic techniques. In Sec.~\ref{sec:kerr_output_statistics} we evaluate the perturbative results for the output field statistics of the Kerr oscillator and in Sec.~\ref{sec:derivation} we present a detailed derivation of the general formalism. 

\section{Introducing the Path Integral Approach to Input--Output Theory}
\label{sec:HO_Keldysh}
\subsection{Setup and Models}
The general setting of a quantum system placed inside a cavity that interacts with light introduced in Fig.~\ref{fig:cavity_sketch} is ubiquitous in modern physics and multiple approaches to modeling it exist. In this section we briefly mention existing approaches to modeling the intra--cavity system and then introduce our approach which focuses on the output field directly.  
One of the most prominent tools to model the dynamics of dissipative quantum systems are Lindblad master equations. In this framework we model the setup presented in Fig.~\ref{fig:cavity_sketch} as,   
\begin{equation}
    \dv{}{t} \hat{\rho} = -i [\hat{H}_S + \hat{H}_\mathrm{d}(t),\hat{\rho}] +  \kappa n_B \mathcal{D}[\dahat]\hat{\rho}  
    + \kappa (n_B+1) \mathcal{D}[\ahat]\hat{\rho} \comma \label{eq:qme_system}
\end{equation}
where $\mathcal{D}[\hat{O}] \hat{\rho} = \hat{O} \hat{\rho} \hat{O}^\dagger - \frac{1}{2}\{\hat{O}^\dagger \hat{O} \textcolor{blue}{,\hat{\rho}}\}$.  Here we introduced the system Hamiltonian,
\begin{align}
     \hat{H}_S &= \omega_S \dahat \ahat + \hat{H}_\text{int} \comma \label{eq:system_hamiltonian_LME} 
\end{align}
where $\hat{H}_{\rm int}$ denotes any non--linear term in the Hamiltonian, as well as any additional degrees of freedom that couple to the cavity mode. The input in this framework is captured through the drive term, 
\begin{align}
     \hat{H}_\mathrm{d}(t) &= - i \sqrt{\kappa} \qty(f(t) \dahat - \overline{f}(t) \ahat) \comma
\end{align}
with the input operator appearing implicitly through its average as $f(t) = \langle \bin(t) \rangle$. The second and third term on the right-hand side of Eq.~\eqref{eq:qme_system} describe the incoherent exchange of photons with the Markovian bath provided by the input and output modes, where the thermal occupation is given by
\begin{equation}
    \label{eq:nbin}
    \langle\!\langle\dbin(t)\bin(t')\rangle\!\rangle = \delta(t-t') n_B,
\end{equation}
where $\langle\!\langle \hat{A}\hat{B}\rangle\!\rangle = \langle \hat{A}\hat{B}\rangle-\langle \hat{A}\rangle\langle\hat{B}\rangle$.
In this description we do not have access to the output field $\bout$, we only model the intra--cavity system.  

In the framework of conventional input--output theory \cite{Gardiner}, see App.~\ref{sec:conventional_input_output_theory}, the dynamics of the very same setup are captured by the Langevin equation of the system mode,  
\begin{equation}
    \dv{t} \hat{a} = i \qty[ \hat{H}_S, \hat{a}] - \frac{\kappa}{2} \hat{a} - \sqrt{\kappa} \, \bin(t), \label{eq:eom_in}
\end{equation}
and the output field statistics become accessible through the input--output relation, 
\begin{align}
    \bout(t) = \bin(t) + \sqrt{\kappa}\, \hat{a}(t) \fullstop \label{eq:in_out_rel}
\end{align}
For linear systems, Eq.~\eqref{eq:eom_in} can be solved and the output statistics can be fully determined by the known statistics of the input field. Alternatively, Eq.~\eqref{eq:in_out_rel} can be used to cast the statistics of the output field into correlation functions of the intra-cavity mode, which may then be solved using established techniques. For higher-order correlation functions, this approach quickly becomes unwieldy due to the cross--terms between the input--field and the cavity mode \cite{PhysRevA.31.3761}. As it is known that the statistics of the output field may be strikingly different from the intra-cavity field~\cite{PhysRevResearch.3.023041,PhysRevA.29.408,walls_book}, a theoretical formalism that provides direct access to the former is highly desirable.

A further tool to model the intra--cavity system described by Eq.~\eqref{eq:qme_system} is given through the Schwinger--Keldysh formalism, see App.~\ref{sec:Keldysh_Formalism}. In this formalism, we capture the dynamics of the system through a path integral with the central object being the action~\cite{Sieberer_2016,THOMPSON2023169385,Kamenev2011}
\begin{align}
    S[\phi] = S_0[\phi] + S_{\rm d}[\phi] +  S_\text{int}[\phi] \fullstop \label{eq:system_action_0_d_int}
\end{align}
with
\begin{align}
    &S_0[\phi]   
    = \int dt \bigg[\Bar{\phi}^{\text{q}}(t) \left(i \partial_t - \omega_S+i\frac{\kappa}{2}\right) \phi^{\text{cl}}(t)+ \label{eq:system_action_0} \\
   & \Bar{\phi}^{\text{cl}}(t) \left(i \partial_t -\omega_S-i\frac{\kappa}{2}\right) \phi^{\text{q}}(t) + i \kappa  F \Bar{\phi}^{\text{q}}(t) \phi^{\text{q}}(t)\bigg]\nonumber \comma  \\
   & S_{\rm d}[\phi] =  i \sqrt{2 \kappa}\int dt \qty[f(t) \Bar{\phi}^{\text{q}}(t) - \Bar{f}(t) \phi^{\text{q}}(t)] \fullstop \label{eq:system_action_d}
\end{align}
Here $S_0$ describes the linear cavity coupled to a Markovian bath, $S_{\rm d}$ describes the coherent drive, and $S_\text{int}$  captures the interaction Hamiltonian $\hat{H}_\text{int}$. The classical and quantum field $\phi^\text{cl}$ and $\phi^\text{q}$ describe the dynamics of the intra-cavity field, the bar denotes complex conjugation, and we introduced the distribution function, 
\begin{equation}
    F = 2n_B + 1 \fullstop \label{eq:distribution_function}
\end{equation}
The classical field can be linked to the diagonal elements of the density matrix, while the quantum field captures coherences, see App.~\ref{sec:Keldysh_Formalism} for details.
In this formalism, source fields may be introduced in order to compute correlation functions of the intra-cavity field. The power of the Schwinger-Keldysh formalism lies in the wealth of approximative and diagrammatic techniques for evaluating correlation functions.

\subsection{Path-integral approach to input--output theory}
In the path integral approach to input--output theory, we access the correlation functions of the output field $\bout$ directly and we therefore rederive the path integral describing our system from the unitary description of the system together with the bath. Our goal is to access time- and normal ordered correlation functions, since they are intimately connected to photodetection~\cite{glauber:1963}. To that end we introduce source fields $\chi,\chi'$ for the $\bout(t)$ and $\dbout(t)$ results in a Moment Generating Functional (MGF), see  Sec.~\ref{sec:derivation} for a sketch of the derivation,
\begin{align}
\Lambda_{\text{out}}[\chi,\chi'] = \int \mathcal{D}[\phi] e^{i S_\text{out}[\phi, \chi,\chi']} \comma \label{eq:moment_generating_functional_1}
\end{align}
with the action 
\begin{align}
\label{eq:full_general_action}
   & S_{\text{out}}[\phi,\chi,\chi']
    = S[\phi] + \int dt i \chi(t) \chi'(t) n_B \\\nonumber
     & - \int dt \chi(t) \qty[f(t) + \sqrt{\frac{\kappa}{2}} F \phi^\text{q}(t) + \sqrt{\frac{\kappa}{2}}\phi^{\text{cl}}(t)] \\ \nonumber
     & - \int dt \chi'(t) \qty[\Bar{f}(t) - \sqrt{\frac{\kappa}{2}} F \Bar{\phi}^{\text{q}}(t) + \sqrt{\frac{\kappa}{2}} \Bar{\phi}^{\text{cl}}(t)]  
      \fullstop 
\end{align}
 Equations \eqref{eq:moment_generating_functional_1} and \eqref{eq:full_general_action} constitute the main technical result of our paper as any normal and time-ordered correlation function of the output field may be obtained by taking functional derivatives of the MGF
\begin{equation}
\begin{aligned}
     &\langle \dbout(t^-_1)  \dots \dbout(t^-_N) \bout(t^+_M) \dots \bout(t^+_1) \rangle \\&
    = \left. \qty(\prod_{l=1}^N  \frac{i\delta }{\delta \chi'(t^-_l)}) \qty(\prod_{p=1}^M  \frac{i\delta }{\delta \chi(t_p^+)})\Lambda_{\text{out}}\qty[\chi,\chi'] \right|_{\chi = \chi'=0}, \label{eq:general_correlation_function}
\end{aligned}
\end{equation}
where the time arguments obey the following ordering, 
 \begin{align}
     t^-_1 \leq &\dots \leq t^-_N\comma & t^+_1 \leq &\dots \leq t^+_M \fullstop \label{eq:time_ordering2}
 \end{align}
Similarly to actions including source fields for computing intra-cavity observables, Eq.~\eqref{eq:full_general_action} reduces to the action describing the system in Eq.~\eqref{eq:system_action_0_d_int} when the source fields are set to zero. These source fields are however markedly different than how source fields for intra-cavity observables enter. This is particularly evident by the term that is quadratic in the source fields, which describes the white noise in the output field that is also present in the input field [c.f.~Eq.~\eqref{eq:nbin}]. For $n_B=0$, the terms multiplying the source fields in Eq.~\eqref{eq:full_general_action} may be understood from the input-output relation in Eq.~\eqref{eq:in_out_rel}. For normally ordered correlation functions \newtext{and in the limit of zero temperature}, the input operators may be replaced by their averages $\bin(t)\rightarrow f(t)$. In addition, the source field $\chi$ couples to the field $\phi^+=(\phi^{\rm cl}+\phi^{\rm q})/\sqrt{2}$ while the field $\chi'$ couples to the field $\bar{\phi}^-=(\bar{\phi}^{\rm cl}-\bar{\phi}^{\rm q})/\sqrt{2}$. This particular coupling ensures normal ordering of the correlation functions, \newtext{as discussed in detail in Sec.~\ref{sec:derivation} below (see below Eq.~\eqref{eq:lambdainout}).}
At finite temperatures, $n_B\neq 0$, the output noise acquires a direct contribution from the input noise, the term quadratic in the source fields. Additionally, the quantum fields coupled to the source fields acquire a factor $F$, capturing correlations between the input noise and the noise leaking out of the cavity.

In the following it will be more convenient to make use of the cumulant generating functional (CGF) which is given as the logarithm of the MGF, 
\begin{align}
\label{eq:cumgen}
    \mathcal{S}_{\text{out}}[\chi,\chi'] = \ln(\Lambda_\text{out}[\chi,\chi']) \comma
\end{align}
and yields the cumulants of the output field, 
\begin{equation}
\begin{aligned}
\label{eq:cumulants}
 &\langle\!\langle \dbout(t^-_1)  \dots \dbout(t^-_N) \bout(t^+_M) \dots \bout(t^+_1) \rangle\!\rangle \\&
    =   \left. \qty(\prod_l  \frac{i\delta }{\delta \chi'(t^-_l)}) \qty(\prod_p  \frac{i\delta }{\delta \chi(t^+_p)})\mathcal{S}_{\text{out}}[\chi,\chi']  \right|_{\chi = \chi'=0},
   \end{aligned}
\end{equation}
here denoted by double brackets $\langle\! \langle \dots \rangle\! \rangle$.
%If one sets the source fields $\chi,\chi'$ to zero, the action again corresponds to $S_0$, a driven dissipative system that is modeled by the Lindblad master equation in Eq.~\eqref{eq:qme_system}.  

For the specific case of a linear cavity, i.e. $\hat{H}_{\rm int} = 0$, the path integral in the MGF  \ref{eq:moment_generating_functional_1} can be solved to find
\begin{equation}
\begin{aligned}
%\label{eq:HO_MGF}
    %\Lambda_\text{out}^\text{HO}[\chi,\chi'] 
    %&= 
    %e^{\mathcal{S}_\text{out}^\text{HO}[\chi,\chi',f]} \comma  \\ 
   & \mathcal{S}_\text{out}^0[\chi,\chi',f] = \ln \Lambda_\text{out}^0[\chi,\chi']\\
    &= \begin{aligned}[t] &-i \int dt \chi(t) \left[f(t)-i\kappa \int dt'G^R(t-t') f(t')\right] \\& -i \int dt \chi'(t) \left[\bar{f}(t)+i\kappa \int dt'G^A(t-t') f(t')\right] \\
    &- \int dt \, \chi(t) \chi'(t) n_B \comma \end{aligned}
\label{eq:HO_MGF}
\end{aligned}
\end{equation}
where we introduced the retarded and advanced Green functions capturing the dynamics of the system, $G^{R/A}$, see App.~\ref{sec:harmonic_oscillator} and Eq.~\eqref{eq:DHO_greens_functions_time} below. 
 As described in Eq.~\eqref{eq:cumulants}, we can find the cumulants of the output field by taking functional derivatives. For the linear cavity, only the first two cumulants are nonvanishing as the output light is in a Gaussian state. The average reads 
\begin{equation}
    \langle \hat{b}_{\text{out}}(\tau) \rangle_0  
    = f(\tau)- i \kappa \int dt\, G^R(\tau-t)f(t) 
    \label{eq:bout_HO}
    \comma
\end{equation}
Which is given by the sum of the reflected input light and the light leaking out of the cavity, as determined by the input-output relation in Eq.~\eqref{eq:in_out_rel}. The light that leaks out of the cavity also arises from the input, but is affected by the cavity through the retarded Green function.
We can interpret this term as an input--photon entering at time $t$ and then being propagated through the cavity to exit at time $\tau$.

The second cumulant reads
\begin{align}
\label{eq:nbout}
    \langle\! \langle \dbout(\tau_1) \bout(\tau_2) \rangle\! \rangle_0  
    &= n_B \delta(\tau_2-\tau_1) \comma 
\end{align}
and is thus identical to the noise in the input field, see Eq.~\eqref{eq:nbin}.
This is due to the process being Gaussian and is entirely expected but we note the simplicity with which this result arises. We further note that we did not need to specify the input signal and that the results obtained so far are valid for all input signals $f(t)$ describing an arbitrary time--dependent coherent drive. \newtext{Note that the fluctuations of the output light are equal to the fluctuations of the input light even though the corresponding cumulant of the intra--cavity field $\llangle \dahat(\tau_1) \ahat(\tau_2) \rrangle$ is non--trivial. We mention this to further illustrate the fact that the output field can show markedly different behavior than the intra--cavity system.}  The whole formalism laid out \newtext{in this section} can equally well be applied in frequency space, see App.~\ref{sec:KIO_in_frequency_space}.

\subsection{Glauber-Sudarshan P Functional}
The state of a bosonic mode can be described by the Glauber-Sudarshan $P$ function \cite{glauber_1963,sudarshan_1963}, which provides a phase space representation of quantum states as statistical mixtures of coherent states weighted by the $P$ function. The moments of the $P$ function are normal ordered expectation values of creation and annihilation operators. This description can be generalized to a continuum of modes, resulting in the $P$ functional~\cite{PhysRevLett.100.013605,PhysRevA.95.063805}. 
This generalization is needed to describe the output field, which consists of a continuum of modes. The moments of the $P$ functional are then exactly the expectation values obtained from the MGF. Consequently, the $P$ functional is obtained by taking the Fourier transform of the MGF, 
\begin{equation}
    P_{\text{out}}[\alpha] = \int \mathcal{D}[\chi,\chi'] e^{i \int dt \qty[\chi'(t) \alpha(t) + \Bar{\alpha}(t) \chi(t)]} \Lambda_{\text{out}}\qty[\chi,\chi']\fullstop
\end{equation}
The Glauber-Sudarshan $P$ function of a single mode, say $\hat{b}_{\text{out}}(\tau)$, is obtained by the path integral
\begin{equation}
    \label{eq:Pfunction}
    P(\alpha_\tau) = \int \mathcal{D}[\alpha] \delta(\alpha_\tau-\alpha(\tau))P_{\text{out}}[\alpha],
\end{equation}
where all fields $\alpha(t)$ are integrated out except the one at time $\tau$.
For the linear cavity, the Fourier transform can straightforwardly be evaluated to find, 
\begin{align}
\label{eq:pHO}
     P_{\text{out}}^0[\alpha]&= \frac{1}{n_B^N} e^{ \int dt \frac{|\alpha(t) - \langle \bout(t) \rangle_0|^2}{n_B}},
\end{align}
where $N$ denotes the number of output modes, which tends to infinity in the continuous-time representation used here.
The multi-time $P$-functional here describes a time--dependently displaced thermal state where the displacement is given by the average of the output field $\langle \bout(t) \rangle$, in complete agreement with the cumulants given in Eqs.~\eqref{eq:bout_HO} and \eqref{eq:nbout}. In discrete time, Eq.~\eqref{eq:pHO} is a product of $N$ independent $P$ functions, each describing a displaced thermal state. 
\vspace{2mm}
\newtext{\section{Derivation}}
\label{sec:derivation}
In this section we derive the MGF in Eq.~\eqref{eq:moment_generating_functional_1} with the general action in Eq.~\eqref{eq:full_general_action} which forms the central quantity in the path integral approach to input--output theory. 
We start with the general setting which consists of a system $S$ coupled to a bath of harmonic oscillators $B$
\begin{align}
    \hat{H} = \newtext{H_S(\dahat,\ahat)} + \hat{H}_B + \hat{V}, \label{eq:general_setting_total_hamiltonian}
\end{align}
\newtext{with $H_S$ being a normal--ordered function of the bosonic creation and annihilation operators $\dahat$ and $\ahat$, and }where the bosonic bath is described by the Hamiltonian,
\begin{equation}
    \hat{H}_B = \sum_k \omega_k \hat{b}_k^\dagger \hat{b}_k,~~\qty[\hat{b}_k, \hat{b}_{k'}^\dagger] = \delta_{k,k'} \fullstop
    \label{eq:general_setting_bath}
\end{equation}
The coupling is modeled by the rotating-wave interaction term,
\begin{equation}
    \hat{V} = \sum_k \qty( g_k \hat{a}^\dagger \hat{b}_k + g_k^* \hat{b}_k^{\dagger} \hat{a}) \fullstop 
    \label{eq:general_setting_interaction}
\end{equation}
\newtext{Furthermore we assume that the time--evolution starts in a factorized initial state between system and bath,
\begin{align}
    \hat{\rho}_0 = \hat{\rho}_S\otimes\hat{\rho}_B \fullstop
    \label{eq:factorized_initial_state}
\end{align}}
In order to describe this setting in the framework of Schwinger--Keldysh non--equilibrium field theory, we define the partition function of system and bath as,  
\begin{equation}
\label{eq:lambda}
\Lambda = \Tr{\hat{\rho}(t)}=\int \mathcal{D}[\phi,\varphi] e^{i S[\phi,\varphi]}, 
\end{equation}
where the action takes the form, 
\begin{align}
    S[\phi,\varphi] = S_S[\phi]+S_B[\varphi]+S_V[\phi,\varphi].
\end{align}
With the constituents 
\begin{widetext}
\begin{align}
    S_S[\phi] &= [t]\sum_{j=1}^{2N-1} \delta t_j \qty[i \Bar{\phi}_j \frac{\phi_j-\phi_{j-1}}{\delta t_j} - H_S(\Bar{\phi}_j, \phi_{j-1})]  
    + i \Bar{\phi}_0 \phi_0 -i \ln \rho_S(\Bar{\phi}_0,\phi_{2N-1})\comma \label{eq:action_system_all_modes} \\ 
    S_B[\varphi] &= \sum_{k} \sum_{j=1}^{2N-1} \delta t_j \qty[i \Bar{\varphi}_{k,j}\frac{\varphi_{k,j}-\varphi_{k,j-1}}{\delta t_j}-\omega_k \Bar{\varphi}_{k,j} \varphi_{k,j-1}] +i \sum_k \Bar{\varphi}_{k,0} \varphi_{k,0}  -i \ln \rho_B(\Bar{\varphi}_0, \varphi_{2N-1}) \comma 
    \end{align}
\end{widetext}
    \begin{equation}
    S_V[\phi,\varphi] = -\sum_k \sum_{j=1}^{2N-1} \delta t_j \qty[g_k \Bar{\phi}_j \varphi_{k,j-1} + g_k^* \Bar{\varphi}_{k,j}\phi_{j-1}].
\end{equation}
Here the fields $\phi_j$ denote the system mode and $\varphi_{k,j}$ the bath modes. Their indices denote time along the Keldysh contour, wich goes from the initial time $t_0$ to the final time $t_{N-1} = t_N$ and back to $t_{2N-1}=t_0$, see Fig.~\ref{fig:contour}. The time step thus obeys $\delta t_j = {\rm sign}(N-j)\delta t$, with ${\rm sign}(0)=0$.
We further introduced 
\newtext{$\rho_r(a,b) = \bra{a} \hat{\rho}_{r} \ket{b}\comma ~ r = S,B$ }with $\ket{a}$ denoting coherent states, c.f.~Eq.~\eqref{eq:coherent_state}. For a more detailed introduction to the Schwinger-Keldysh formalism, see App.~\ref{sec:Keldysh_Formalism}. 
\begin{figure}
\centering
\begin{tikzpicture}[decoration={markings,
%mark=at position 0.5cm with {\arrow[line width=1pt]{>}},
mark=at position 3cm with {\arrow[line width=1pt]{Latex}},
%mark=at position 7.85cm with {\arrow[line width=1pt]{>}},
%mark=at position 5.4cm with {\arrow[line width=1pt]{>}},
mark=at position 8cm with {\arrow[line width=1pt]{Latex}}
}
]
\pgfsetxvec{\pgfpoint{0.5cm}{0}}
\pgfsetyvec{\pgfpoint{0}{0.5cm}}
% The axes
\draw[help lines,<-] (-9,1) -- (4,1) coordinate (xaxis);

% The path
\path[draw,line width=0.8pt,postaction=decorate] (3,1.5) node[above] {$t_0$} -- (-7,1.5) node[above] {$t_{N-1}$} arc (90:270:0.5) -- (-7,0.5) node[below] {$t_N$} -- (3,0.5) node[below] {$t_{2N-1}$};
  
% The labels
\node at (-9.5,1) {$t$};
\node at (-8.5,1.5) {$+\infty$};
\node at (4,1.5) {$-\infty$};
\node at (3,1.5) [circle,fill,inner sep=1.5pt]{};
\node at (-7,1.5) [circle,fill,inner sep=1.5pt]{};
\node at (-7,0.5) [circle,fill,inner sep=1.5pt]{};
\node at (3,0.5) [circle,fill,inner sep=1.5pt]{};
\end{tikzpicture}
\caption{The closed time contour of the path integral in the Schwinger-Keldysh formalism. \newtext{The top part of the contour from $t_0$ to $t_{N-1}$ is called the forward branch and the fields on it are labeled by a plus sign superscript $\phi^+$. The lower part of the contour from $t_N$ to $t_{2N-1}$ is called the backwards branch and fields on it are denoted by a minus superscript $\phi^-$.}}
    \label{fig:contour}
\end{figure}

We now make the usual set of assumptions that are made in the derivation of conventional input--output theory~\cite{Gardiner}, see App.~\ref{sec:conventional_input_output_theory}.  In order to assure Markovianity in our setup, we assume the Fourier transform of our system operator $\hat{a}$ to be peaked at a specific frequency $\Omega$ with a width $\gamma$. We further introduce the bath spectral density 
\begin{align}
    \rho(\omega) = \sum_k |g_k|^2 \delta(\omega-\omega_k) \comma
\end{align}
which we assume to be well approximated around the frequency $\Omega$ by a constant value, 
\begin{align}
    \rho(\omega) \approx 2 \pi \kappa \comma \omega \in (\Omega - \gamma, \Omega + \gamma) \fullstop
\end{align}
Now we introduce the input-- and output fields which are determined by the states of the bath at the initial $t_0/t_{2N-1}$ and final $t_{N-1}/t_N$ times, 
\begin{align}
    &\varphi_{\text{in}}^{+/-}(t) = \frac{1}{\sqrt{\kappa}} \sum_k i g_k e^{-i \omega_k (t-t_0)} \varphi_{k}(t_{0/2N-1}) \comma \\
    &\varphi_{\text{out}}^{+/-}(t) = \frac{1}{\sqrt{\kappa}} \sum_k ig_k e^{i \omega_k(t-t_{N-1})} \varphi_{k}(t_{N-1/N}) \fullstop
\end{align}
The two possible values for the initial and final times come from the fact that the Keldysh contour goes from the distant past to the distant future and back. The superscript $\pm$ denotes the forward ($+$) and backward ($-$) branch of the Keldysh contour.
The input field describes the relevant linear combination of bath fields that interact with the system at time $t$ \textit{before} the interaction, while the output field describes the same mode \textit{after} the interaction~\cite{RevModPhys.82.1155}.

Since only the bath fields at the initial and final time contribute to the input and output fields, we integrate out all bath fields at intermediate times in Eq.~\eqref{eq:lambda} resulting in 
\begin{align}
    \Lambda = \int \mathcal{D}[\phi, \varphi_\text{in}, \varphi_\text{out}] e^{S^{\text{io}}\qty[\phi, \varphi_{\text{in}},\varphi_{\text{out}}]}\comma 
\end{align}
with the input--output action, 
\begin{multline}
    S^{\text{io}}\qty[\phi, \varphi_{\text{in}},\varphi_{\text{out}}] = S^{\text{io}}_S[\phi] + S^{\text{io}}_B[\varphi_{\text{in}},\varphi_{\text{out}}] \\ + S^{\text{io}}_V[\phi,\varphi_{\text{in}},\varphi_{\text{out}}], \label{eq:keldysh_input_output_action_with_input_output_fields}
\end{multline}
with the following terms, here already given in the continuum limit, 
\begin{widetext}
\begin{align}
    S^{\text{io}}_S[\phi] &= \int_{t_0}^{t_N} dt [\Bar{\phi}^+(t)\qty(i \partial_t + i \frac{\kappa}{2}) \phi^+(t) 
    - \Bar{\phi}^-(t)\qty(i \partial_t- i \frac{\kappa}{2}) \phi^-(t)  
    -H_S(\Bar{\phi}^+(t),\phi^+(t))+H_S(\Bar{\phi}^-(t), \phi^-(t)) ], 
     \nonumber\\
    S^{\text{io}}_B\qty[ \varphi_{\text{in}}, \varphi_{\text{out}}] &=i \int_{t_0}^{t_N} dt \qty(\Bar{\varphi}_{\text{in}}^+(t)~\Bar{\varphi}_{\text{out}}^+(t)~\Bar{\varphi}_{\text{out}}^-(t) ~\Bar{\varphi}_{\text{in}}^-(t))\mqty(1 & 0 & 0 & 0 \\ 
                                        -1 & 1 & 0 & 0 \\ 
                                         0 & -1 & 1 & 0 \\ 
                                         0 & 0 & -1 & 1 )  \mqty(\varphi_{\text{in}}^+(t) \\ 
                                                            \varphi_{\text{out}}^+(t)\\
                                                            \varphi_{\text{out}}^-(t)\\
                                                            \varphi_{\text{in}}^-(t)) \label{eq:io_action_bath}
   - i \ln \rho_B(\varphi_\text{in}),\nonumber\\
    S^{\text{io}}_V[\varphi_{\text{in}},\varphi_{\text{out}}] &= i \sqrt{\kappa} \int_{t_0}^{t_N}dt \qty[
    \Bar{\phi}^+(t) \varphi_{\text{in}}^+(t)-\Bar{\varphi}_{\text{out}}^+(t) \phi^+(t) - \Bar{\phi}^-(t) \varphi_{\text{out}}^-(t) + \Bar{\varphi}_{\text{in}}^-(t) \phi^-(t)] .
\end{align}
\newtext{In this step we dropped the boundary term of the system $\hat{\rho}_S$ since we assume the initial time $t_0$ to be in the distant past.}

As for the input and output fields, the plus and minus superscripts on the system fields denote the forward and backward branch of the Keldysh contour. In particular, 
\newtext{\begin{equation}
    \phi_j \rightarrow \begin{cases}
    \phi^+(t_0+j\delta t) \hspace{2cm} \text{for }\hspace{.2cm} j < N,\\
    \phi^{-}(t_0+[2N-1-j]\delta t)\hspace{.76cm} \text{for }\hspace{.2cm} j \geq N.
    \end{cases}
\end{equation}}

As the initial bath state, we consider a displaced thermal state, 
\begin{align}
\label{eq:rhob}
    \hat{\rho}_B = \bigotimes_k \qty(1-e^{-\beta \omega_k})\hat{D}_k(\alpha_k) e^{-\beta \omega_k \hat{b}_k^\dagger \hat{b}_k} \hat{D}_k^{\dagger}(\alpha_k) \comma
\end{align}
with the displacement operator $\hat{D}_k(\alpha) = e^{\alpha \hat{b}_k^\dagger - \Bar{\alpha} \hat{b}_k}$, see \cite{gerry_knight_2004}, and the inverse temperature of the thermal state $\beta$. By assuming that $e^{-\beta\omega_k}\simeq e^{-\beta\Omega}$ for all bath frequencies $\omega_k$ that influence the system, \newtext{the bath state $ \rho_B(\Bar{\varphi}_0, \varphi_{2N-1})$ appearing in the action can be expressed purely in terms of the input fields as}
\begin{equation}
    \rho_B(\varphi_{\rm in}) = \qty(1-e^{-\beta \Omega})^N 
    \exp\left\{\int_{t_0}^{t_N} dt e^{-\beta \Omega} \Bar{\varphi}_{\text{in}}^+(t)\varphi_{\text{in}}^-(t)+(1-e^{-\beta \Omega})\qty[f(t) \Bar{\varphi}_{\text{in}}^+(t)+\Bar{f}(t) \varphi_{\text{in}}^-(t)-|f(t)|^2]\right\} \fullstop
\end{equation}
\end{widetext}
Note that $N$ goes to infinity in the continuum limit and the prefactor $\qty(1-e^{-\beta \Omega})^N$ is simply a shorthand for the normalization constant. The average of the input field is determined by the displacement of all bath modes
\begin{align}
    f(t_j) = \sum_{k=0}^{N-1} \frac{\alpha_k}{\sqrt{N\delta t}} e^{-i\omega_k (t_j -t_0)}\fullstop \label{eq:f_definition}
\end{align}
For $f=0$, we obtain a thermal state, whereas for zero temperature, we get a coherent state. 
At this stage we introduce the Keldysh rotation \cite{Keldysh1964ud} in order to make use of the symmetries inherent in the approach. To this end, we introduce, for system and bath fields, the classical and quantum fields which are defined as
\begin{align}
    \psi^{\text{cl}}(t) &= \frac{1}{\sqrt{2}}\qty(\psi^+(t) + \psi^-(t)) \comma \\ \psi^{\text{q}}(t) &= \frac{1}{\sqrt{2}}\qty(\psi^+(t) - \psi^-(t)) \fullstop \label{eq:keldysh_rot}
\end{align}
If we make a stationary phase approximation of the action in Eq.~\eqref{eq:keldysh_input_output_action_with_input_output_fields}, we arrive at relations between the input and output fields that are analogous to the operator relation in Eq.~\eqref{eq:in_out_rel},
for the classical and quantum fields, 
\begin{align}
    \varphi_{\text{out}}^{\text{cl}}(t) &= \varphi_{\text{in}}^{\text{cl}}(t) + \sqrt{\kappa} \phi^{\text{cl}}(t), \\ \varphi_{\text{out}}^{\text{q}}(t) &= \varphi_{\text{in}}^{\text{q}}(t) + \sqrt{\kappa} \phi^{\text{q}}(t)=0 \fullstop
\end{align}

We now introduce source fields $\chi$ and $\chi'$ for the output field and its hermitian conjugate, 
\begin{multline}
\label{eq:lambdainout}
    \Lambda_{\text{out}} [\chi,\chi']   
    = \int \mathcal{D}[\phi,\varphi_{\text{in}},\varphi_{\text{out}}] e^{i S^{\text{io}[\phi,\varphi_{\text{in}},\varphi_{\text{out}}]}} \\ \times e^{-i \int_{t_0}^{t_N} dt \qty[\chi(t) \varphi_{\text{out}}^+(t) + \chi'(t) \Bar{\varphi}_{\text{out}}^-(t)]}  \fullstop
\end{multline}
Note that the unbarred output field is located on the forward branch of the Keldysh contour while the barred field is located on the backward branch. This guarantees that any correlation function derived from this generating functional will be automatically normal-- and time--ordered \newtext{according to} Eq.~\eqref{eq:time_ordering2}. \newtext{Normal ordering is ensured because the barred fields, related to creation operators, are evaluated on the backward-in-time contour ($-$) and thus after the unbarred fields, related to annihilation operators, which are evaluated on the forward-in-time contour ($+$). To obtain the MGF introduced in Eq.~\eqref{eq:moment_generating_functional_1} with the action given in Eq.~\eqref{eq:full_general_action}, we integrate out the input- and output fields in Eq.~\eqref{eq:lambdainout}.}
Note that this generating functional is still general and holds for any system action $S_S[\phi]$, as long as the bath is Markovian and can be described by a displaced thermal state. This derivation can be extended to include arbitrary Gaussian bath states that also involve squeezing, as well as multiple baths. 
\section{Non--Linear Systems}
\label{sec:non_linear_system}
Having introduced the basic framework of our novel approach to input--output theory and after treating a non--interacting system with the linear cavity, we now turn towards solving systems that contain nonlinear terms. 
In Eq.~\eqref{eq:system_action_0_d_int} we had already 
split up the action into a Gaussian part 
which corresponds to the action of a linear cavity, and the non-Gaussian, interacting part $S_\text{int}$. For completeness, we provide $S_\text{int}[\phi]$ for a general single-mode interaction Hamiltonian in App.~\ref{sec:Kerr_oscillator_details}.

As it turns out, the separation between the linear cavity and the interaction 
can also be made on the level of the generating functionals.
We implement this by shifting the system fields in the path integral by their average under the linear cavity action 
\begin{equation}
    \langle \phi^{\text{cl}/\text{q}}\rangle_0\qty[\chi,\chi'] =\frac{\int \mathcal{D}[\phi] \phi^{\text{cl}/\text{q}} e^{i S_{\text{out}}^0\qty[\phi,\chi,\chi']}}{\Lambda_{\text{out}}^0[\chi,\chi']} \comma \label{eq:shift_average_def}
\end{equation}
where $S_{\text{out}}^0$ denotes the action in Eq.~\eqref{eq:full_general_action} with $S_{\rm int}$ set to zero. We find
\begin{align}
    \langle\phi^\mathrm{q}(t) \rangle_0[\chi'] &= \sqrt{\frac{\kappa}{2}}\int dt' G^A(t-t') \chi'(t')\comma \label{eq:phiqavg}\\ 
    \langle \Bar{\phi}^\mathrm{q}(t) \rangle_0 [\chi] &= \sqrt{\frac{\kappa}{2}}\int dt'  G^R(t'-t) \chi(t')  \comma \label{eq:phibarqavg}\\
   \langle \phi^\mathrm{cl}(t) \rangle_0 [\chi'] &=  -F \langle \phi^\text{q}(t)\rangle_0[\chi'] \label{eq:phiclavgwithsourcefield} \\\nonumber&-i \sqrt{2\kappa} \int dt' G^R(t-t') f(t')  \comma \\ \label{eq:phibarclavgwithsourcefield}
    \langle\Bar{\phi}^\mathrm{cl}(t) \rangle_0 [\chi] &=   F \langle \Bar{\phi}^\text{q}(t)\rangle_0 [\chi] \\\nonumber&+ i \sqrt{2\kappa} \int dt' G^A(t'-t) \Bar{f}(t').
\end{align}
In the absence of source fields, these expressions recover $\langle\phi^{\rm q}(t)\rangle_0 = 0$ and $\langle\phi^{\rm cl}(t)\rangle_0 = \sqrt{2}\langle\hat{a}\rangle_0$, as expected from the conventional Schwinger-Keldysh formalism. In the presence of source fields, the quantum field obtains a finite average proportional to the source fields and the classical field obtains a similar contribution that involves the distribution function $F$.
With the help of these averages, we define new integration variables as, 
\begin{align}
\label{eq:shift}
    \delta \phi^{\text{cl}} &= \phi^\text{cl} - \langle \phi^\text{cl} \rangle_0[\chi'] \comma& \delta \phi^\text{q} &= \phi^\text{q} - \langle \phi^\text{q} \rangle_0[\chi'] \comma \\ \nonumber
    \delta \overline{\phi^{\text{cl}}} &= \overline{\phi^\text{cl}} - \langle \overline{\phi^\text{cl}} \rangle_0[\chi] \comma& \delta \overline{\phi^\text{q}} &= \overline{\phi^\text{q}} - \langle \overline{\phi^\text{q}} \rangle_0[\chi] \fullstop
\end{align}
Note that here we shift the fields and their complex conjugates independently see App.~\ref{sec:shift} for a justification of this step. 
With this transformation we can now separate the MGF as,
\begin{align}
        \Lambda_\mathrm{out}[\chi,\chi] = \Lambda_\mathrm{out}^0[\chi,\chi'] \Lambda_\mathrm{out}^\mathrm{int}[\chi,\chi'] \comma 
\label{eq:moment_generating_functional_factorization}
\end{align}
where the MGF of the linear cavity is known from the previous section, see Eq.~\eqref{eq:HO_MGF} and the interaction functional takes the form, 
\begin{align}
\label{eq:intfunc}
  \Lambda_\mathrm{out}^\mathrm{int}[\chi,\chi'] &= \int \mathcal{D}[\delta \phi] e^{i S_0[\delta \phi] + i S_\mathrm{int}[\delta \phi, \chi,\chi']} \fullstop 
\end{align}
Here $S_0$ is given in Eq.~\eqref{eq:system_action_0} and $S_\mathrm{int}[\delta\phi,\chi,\chi']$ is obtained from $S_{\rm int}[\phi]$ by replacing $\phi^{\rm cl}$ and $\phi^{\rm q}$ according to Eq.~\eqref{eq:shift}.

In effect, we thereby shift our treatment towards the deviations around the exactly solvable, non--interacting linear cavity. 
For a given interaction Hamiltonian $\hat{H}_\text{int}$ we can now access the statistics of the output field perturbatively in the interaction strength. For this, we employ a cumulant expansion \cite{doi:10.1143/JPSJ.17.1100}. The interaction functional in Eq.~\eqref{eq:intfunc} can be written as an average over the action $S_0$
\begin{equation}
\label{eq:avgint}
    \Lambda_{\mathrm{out}}^\mathrm{int}[\chi,\chi'] = \left\langle e^{iS_\text{int}[\delta\phi,\chi,\chi']} \right\rangle_0.
\end{equation}
We note that whenever averages over $\delta\phi$ are taken, we use the subscript $0$ to denote averages over the linear cavity action without a drive (or source fields), such that $\langle\delta\phi^{\rm cl/q}\rangle_0=0$. Equation \eqref{eq:avgint} can be understood as the moment generating function of the random variable $S_\text{int}[\delta\phi,\chi,\chi']$. Similarly to Eq.~\eqref{eq:cumgen}, the moment generating function can be written as the exponent of the cumulant generating function. We may thus write
\begin{equation}
        \Lambda_\mathrm{out}^\mathrm{int}[\chi,\chi'] = 
        e^{ i  \langle S_\mathrm{int}[\delta \phi, \chi,\chi']  \rangle_0 -\frac{1}{2}\langle \! \langle S_\mathrm{int}^2[\delta \phi, \chi,\chi'] \rangle \!\rangle_0  + \dots} \fullstop \label{eq:expansion_Kerr_interaction_MGF}
\end{equation}
See App.~\ref{sec:cumulant_exp} for a detailed explanation of this step. 
The cumulants in the last expression can be obtained through the moments
\begin{equation}
    \langle S^n_\mathrm{int}[\delta \phi, \chi,\chi'] \rangle_0
     = \int \mathcal{D}[\phi] S^n_\mathrm{int}[\delta \phi, \chi,\chi'] e^{i S_0\qty[\delta \phi]} \fullstop \label{eq:S_int_avgs}
\end{equation}
Since the MGF factorizes, the CGF will split into a sum 
\begin{align}
    \mathcal{S}_{\text{out}}[\chi,\chi'] = \mathcal{S}_{\text{out}}^0[\chi,\chi'] + \mathcal{S}_\text{int}[\chi,\chi'] \fullstop
\end{align}
To compute for example the first order approximate expression \newtext{in the interaction parameter} for the average output field $\langle \bout(\tau) \rangle$ we only need to take into account the first term of the cumulant expansion in Eq.~\eqref{eq:expansion_Kerr_interaction_MGF} and evaluate Eq.~\eqref{eq:S_int_avgs} for $n = 1$, 
\begin{align}
    &\langle \bout(\tau) \rangle = 
    i \left. \frac{\delta \mathcal{S}_\mathrm{out}[\chi,\chi']}{\delta \chi(\tau)}\right|_{\chi = \chi'=0} \nonumber \\ 
    &\simeq \langle \bout(\tau) \rangle_0
   +\left.\frac{i \delta}{\delta \chi(\tau)}    \langle i S_{\mathrm{int}}[\delta \phi,\chi,\chi'] \rangle_0 \right|_{\chi=\chi'=0}  \fullstop
    \label{eq:average_output_field_begin}
\end{align}
The averages of the form of Eq.~\eqref{eq:S_int_avgs} are taken with respect to a Gaussian action and may therefore be evaluated using Wick's theorem. They are thus fully determined by Eqs.~\eqref{eq:phiqavg}-\eqref{eq:phiclavgwithsourcefield}, as well as the Green functions
\begin{align}
    G^R(t-t') &= -i\langle\delta\phi^{\rm cl}(t)\delta\bar{\phi}^{\rm q}(t')\rangle_0,\nonumber\\
    \label{eq:Greens} G^A(t-t') &= -i\langle\delta\phi^{\rm q}(t)\delta\bar{\phi}^{\rm cl}(t')\rangle_0=\left[G^R(t'-t)\right]^*,\\
    G^K(t-t') &= -i\langle\delta\phi^{\rm cl}(t)\delta\bar{\phi}^{\rm cl}(t')\rangle_0\nonumber\\\nonumber&=F[G^R(t-t')-G^A(t-t')],
\end{align}
where the last equality reflects the fluctuation-dissipation theorem~\cite{Kamenev2011}, which holds for the Gaussian action without the drive.
For the linear cavity, the Green functions are, 
\begin{align}
   G^A(t-t') &= i \theta(t'-t) e^{-i \omega_S (t-t')} e^{\sfrac{\kappa}{2}(t-t')} \comma \nonumber\\
   G^R(t-t') &= -i \theta(t-t') e^{-i \omega_S (t-t')} e^{-\sfrac{\kappa}{2}(t-t')} \comma \label{eq:DHO_greens_functions_time}\\ 
   G^K(t-t') &= - i F e^{-i \omega_S (t-t')}e^{-\sfrac{\kappa}{2}|t-t'|} \fullstop \nonumber
\end{align}
As discussed in detail in the next section, a diagrammatic approach can be employed to compute the correlation functions of the output fields perturbatively.

\begin{figure}
    \centering
    \vspace{2mm}
    \begin{tikzpicture}
    \pgfsetxvec{\pgfpoint{0.5cm}{0}}
  \pgfsetyvec{\pgfpoint{0}{0.5cm}}
        \node at (-0.5,7) {a)};
        \node at (-0.5,4) {b)};
        \coordinate (shift) at (0,6.5);
        \coordinate (P1) at ($(0,0) +(shift) $); 
        \coordinate (P2) at ($(4,0) +(shift) $); 
        \coordinate (P3) at ($(5,0) +(shift) $); 
        \coordinate (P4) at ($(9,0) +(shift) $); 
        \coordinate (P5) at ($(10,0) +(shift) $); 
        \coordinate (P6) at ($(14,0) +(shift) $); 
        \Gret[0.5]{P1}{P2};
        \node at ($(2,0.75)+(shift) $) {$i G^R(t,t')$};
        \node at ($(0.5,-0.75)+(shift) $) {$\delta \phi^\mathrm{cl}(t)$}; 
        \node at ($(3.5,-0.75)+(shift) $) {$\delta \Bar{\phi}^\mathrm{q}(t')$};
        \Gadv[0.5]{P3}{P4};
        \node at ($(7,0.75)+(shift) $) {$i G^A(t,t')$};
        \node at ($(5.5,-0.75)+(shift) $) {$\delta \phi^\mathrm{q}(t)$}; 
        \node at ($(8.5,-0.75)+(shift) $) {$\delta \Bar{\phi}^\mathrm{cl}(t')$};
        \Gkel[0.5]{P5}{P6};
        \node at ($(12,0.75)+(shift) $) {$i G^K(t,t')$};
        \node at ($(10.5,-0.75)+(shift) $) {$\delta \phi^\mathrm{cl}(t)$}; 
        \node at ($(13.5,-0.75)+(shift) $) {$\delta \Bar{\phi}^\mathrm{cl}(t')$};
    %\end{tikzpicture}
    
    %\begin{tikzpicture}
    \pgfsetxvec{\pgfpoint{0.75cm}{0}}
    \pgfsetyvec{\pgfpoint{0}{0.75cm}}
    
        \coordinate (V4) at (2,2); 
        \coordinate (V2) at (6,2); 
        \coordinate (V3) at (10,2);
        \coordinate (V1) at (2,0.5); 
        \coordinate (V5) at (6,0.5);
        \coordinate (V6) at (10,0.5);

        \vertexcoordinates{V1}
        \node at (V1) [circle,fill,color=white]{};
        \vertexcoordinates{V2}
        \node at (V2) [circle,fill,color=white]{};
        \vertexcoordinates{V3}
        \node at (V3) [circle,fill,color=white]{};
        \vertexcoordinates{V4}
        \node at (V4) [circle,fill,color=white]{};
        \vertexcoordinates{V5}
        \node at (V5) [circle,fill,color=white]{};
        \vertexcoordinates{V6}
        \node at (V6) [circle,fill,color=white]{};

        \node at ($(V1L) + (1,0.5)$) {$\langle \Bar{\phi}^\text{cl}(t) \rangle_0$};
        \createinput[0.75]{V1L}{V1}
        \node at ($(V2L) + (1,0.5)$) {$\langle \phi^\text{q}\rangle^{'}_0(t-\tau)$};
        \createoutput[0.75]{V2L}{V2}
        \node at ($(V3L) + (1,0.5)$) {$-F \langle \phi^\text{q} \rangle^{'}_0(t-\tau)$};
        \createoutputK[0.75]{V3L}{V3}
        \node at ($(V4L) + (1,0.5)$) {$\langle \phi^\text{cl}(t) \rangle_0$};
        \destroyinput[0.75]{V4L}{V4}
        \node at ($(V5L) + (1,0.5)$) {$\langle \Bar{\phi}^\text{q}\rangle^{'}_0(t-\tau) $};
        \destroyoutput[0.75]{V5L}{V5}
        \node at ($(V6L) + (1,0.5)$) {$F \langle \Bar{\phi}^\text{q} \rangle^{'}_0(t-\tau)$};
        \destroyoutputK[0.75]{V6L}{V6}
    \end{tikzpicture}
    \caption{Diagram building blocks. a) Conventional way of depicting the Green functions that appear in the Keldysh formalism \cite{Altland2010}. The full (dashed) lines stand for classical (quantum) fields and the arrows point from creation to annihilation.
    b) Diagrammatic depiction of the averages. 
    Squares denote sources of photons (filled) and anti-photons (empty) provided by the input field. Half-circles denote sinks of photons (empty) and anti-photons (filled) reflecting the detection in the output field. Every filled (empty) square thus implies a factor of $\langle \bin \rangle$ ($\langle\dbin \rangle$) while the number of filled (empty) half-circles is determined by the number of output operators $\dbout$ ($\bout$) in the correlation function that is being considered.
    }
    \label{fig:operator_diags}
\end{figure}

\section{Diagrammatic Approach to the Kerr Oscillator}
\label{sec:diagrams}
As an example of a non--linear system we consider the Kerr oscillator with the interaction Hamiltonian, 
\begin{align}
    \hat{H}_\text{int} =  U \hat{a}^\dagger \hat{a}^\dagger \hat{a} \hat{a} \comma \label{eq:Kerr_Hamiltonian}
\end{align}
and the corresponding interaction action \cite{PhysRevResearch.5.033107, PhysRevA.103.033711}, 
\begin{equation}
     S_\mathrm{int}[\phi]  = -U \int dt \qty(\Bar{\phi}^\mathrm{cl} \phi^\mathrm{cl} + \Bar{\phi}^\mathrm{q} \phi^\mathrm{q}) \qty(\Bar{\phi}^\mathrm{q} \phi^\mathrm{cl} + \Bar{\phi}^\mathrm{cl} \phi^\mathrm{q})\label{eq:Kerr_interaction_action_pure} \comma
\end{equation}
where we suppressed the time arguments of the fields for ease of notation.
To obtain $S_{\rm int}[\delta\phi,\chi,\chi']$, We shift the system fields as described in Eq.~\eqref{eq:shift}. In order to compute correlation functions of the output field, we first introduce a pictorial representation for each of the terms appearing due to the interaction action and then explain how to form valid diagrams, i.e. correction terms for the respective correlation function, out of these terms. 

\subsection{Pictorial Representation}
The correlation functions of the output field may be expressed through the Green functions in Eqs.~\eqref{eq:Greens} and \eqref{eq:DHO_greens_functions_time}, and the functional derivatives of the averages 
in Eqs.~\eqref{eq:phiqavg}-\eqref{eq:phiclavgwithsourcefield}. After performing the functional differentiation one can represent all terms through the averages of the classical cavity field 
\begin{align}
   \langle \phi^\mathrm{cl}(t) \rangle_0  &=  -i \sqrt{2\kappa} \int dt' G^R(t-t') f(t')  \comma \\ 
    \langle\Bar{\phi}^\mathrm{cl}(t) \rangle_0 &=  i \sqrt{2\kappa} \int dt' G^A(t'-t) \Bar{f}(t') \comma 
\end{align}
and the functional derivatives of the quantum field averages appearing either with our without a prefactor given by the distribution function $F$. For the functional derivatives we introduce the notation, 
\
\begin{align}
    \langle \phi^{\text{q}} \rangle^{'}_0(t-\tau) &= \left. \frac{i\delta }{\delta \chi'(\tau)} \langle \phi^{\text{q}}(t) \rangle [\chi'] \right|_{\chi'=0}\comma \\ 
    \langle \Bar{\phi}^{\text{q}} \rangle^{'}_0(t-\tau) &= \left. \frac{i\delta }{\delta \chi(\tau)} \langle \Bar{\phi}^{\text{q}}(t) \rangle [\chi] \right|_{\chi=0}\nonumber\fullstop
\end{align}
To be able to represent all terms pictorially, we introduce representations for the three Green functions and the six possible external legs in Fig.~\ref{fig:operator_diags}. These representations serve as the building blocks for our diagrams.

As an example, we consider the average output field to first order in $U$, c.f.~Eq.~\eqref{eq:average_output_field_begin}, 
\begin{align}
   \langle \bout(\tau) \rangle\ &\simeq~ \langle  \bout(\tau) \rangle_0  \label{eq:bout_first_order_approx}
    \\\nonumber&  \overbrace{-2i U \int dt  \, 2 n_B \langle \Bar{\phi}^\text{q} \rangle^{'}_0(t-\tau)  \langle \phi^\text{cl}(t) \rangle_0}^{\text{Diagram a) in Fig.~\ref{fig:diagrams_bout_first_order}}}   \\\nonumber&  \underbrace{-iU \int dt \langle \Bar{\phi}^\text{q}\rangle^{'}_0(t-\tau)\langle \phi^\text{cl}(t) \rangle_0^2 \langle \Bar{\phi}^\text{cl}(t) \rangle_0}_{\text{Diagram b) in Fig.~\ref{fig:diagrams_bout_first_order}}}  \fullstop 
\end{align}
This equation may be understood as an effective coherent drive provided by the Kerr nonlinearity. Indeed, we may obtain Eq.~\eqref{eq:bout_first_order_approx} through the replacement $\hat{H}_{\rm int}\rightarrow 2U\hat{a}\langle \hat{a}^\dagger\hat{a}^\dagger a\rangle_0+H.c.$ 

The perturbative terms can be represented
through the diagrams in Fig.~\ref{fig:diagrams_bout_first_order}. To obtain these diagrams, we depict the integral over $t$, together with the multiplication by $-i U$, as a vertex that connects to four arrows. Each arrow belongs to either one of the Green functions or one of the terms depicted in Fig.~\ref{fig:operator_diags}. The only Green function that contributes to Eq.~\eqref{eq:bout_first_order_approx} is the equal-time Keldysh Green function, which equals to $2n_B$ in our convention, see App.~\ref{sec:loops} for details. We also call this a thermal loop. Lastly, the diagrams come with associated multiplicities which are explained in Sec.~\ref{sec:evaluating_the_diagrams}.

Similarly to Feynman diagrams, we can interpret the diagrams in Fig.~\ref{fig:diagrams_bout_first_order} in terms of photons and anti-photons. While photons move along the arrows, anti-photons move against the arrows. The filled squares denote photons that enter the cavity from the input drive. They are associated to the operator $\bin$, because a photon that enters the cavity is removed from the drive. The empty squares denote anti-photons from the input drive, associated to the operator $\dbin$. Empty half-circles denote photons that are detected in the output field. They are associated to the operator $\bout$ because a detection removes a photon from the output field. Causality is ensured by the squares being associated to times before the vertex and half-circles being associated to times after the vertex. The cavity field can therefore only be affected by the input field at previous times and the detection can only be affected by the cavity field at times previous to the detection. We do not include a time axis in our diagrams. Instead, causality ensures that time grows as we move from squares to vertices to half-circles.

\subsection{Constructing the Diagrams}
\label{sec:diagram_rules}
We may determine arbitrary cumulants of the output field [c.f.~Eq.~\eqref{eq:cumulants}] perturbatively by finding the corresponding diagrams. 
The mathematical terms making up the $n$-th order contribution to any cumulant arise from functional differentiation of $\langle\!\langle (iS_{\rm int})^n\rangle\!\rangle/n!$, see Eq.~\eqref{eq:expansion_Kerr_interaction_MGF}. The diagrams that represent these terms include exactly $n$ vertices \newtext{and need to be fully connected}. Furthermore, due to the way the source fields enter the CGF, the diagrams require one empty (filled) half-circle for each $\bout$ ($\dbout)$ that appears in the cumulant. In order for the diagrams to correspond to actual terms that appear in the interaction action, they need to obey a set of rules. Here we provide these rules without derivation. For a justification of each rule, see App.~\ref{sec:origin_diag_rules}.
\begin{figure}[t]
\centering
\begin{tikzpicture}
\pgfsetxvec{\pgfpoint{0.75cm}{0}}
\pgfsetyvec{\pgfpoint{0}{0.75cm}}
%\draw[step=1,gray,very thin] (-1,-1) grid (10,5);
  \coordinate (V1) at (2, 2); 
  \coordinate (V2) at (7, 2);
  \vertexcoordinates{V1}
  \vertexcoordinates{V2}
  %%% Diagram a) 
  \node at (-1,4) {a)};
  \destroyinput[0.75]{V1BL}{V1}
  \destroyoutput[0.75]{V1TL}{V1}
  \clloop[0.75]{V1}{0}
  \node at (0.9,2) {$-iU \int dt$};
  \node at (1.95,3.5) {$\langle \Bar{\phi}^\text{q} \rangle^{'}_0(t-\tau)$};
  \node at (1.5,0.5) {$\langle \phi^\text{cl}(t) \rangle_0$};
  \node at (3,2) {$2 n_B$};
  %%% Diagram b) 
  \node at (4,4) {b)};
  \destroyinput[0.75]{V2BR}{V2}
  \destroyoutput[0.75]{V2TL}{V2}
  \destroyinput[0.75]{V2BL}{V2}
  \createinput[0.75]{V2TR}{V2}
  \node at (6.95,3.5) {$\langle \Bar{\phi}^\text{q} \rangle^{'}_0(t-\tau)$};
  \node at (6.5,0.5) {$\langle \phi^\text{cl}(t) \rangle_0$};
  \node at (8.5,1.5) {$\langle \phi^\text{cl}(t) \rangle_0$};
  \node at (8.5,2.5) {$\langle \Bar{\phi}^\text{cl}(t) \rangle_0$};
  \node at (5.9,2) {$-i U \int dt$};
\end{tikzpicture}
  \caption{First order diagrams for $\langle \bout \rangle$.  
  a) An input photon (filled square) travels into the cavity, interacts with a thermal excitation and then exits the cavity to be detected at time $\tau$. The multiplicity is equal to two, since all legs connecting to the vertex are distinct. b) Two photons and one anti-photon from the input interact, resulting in one photon exiting the cavity towards detection at time $\tau$. The multiplicity is one since two legs are identical.}
    \label{fig:diagrams_bout_first_order}
\end{figure}
\begin{enumerate}
    \item Every vertex connects to two ingoing and to two outgoing lines. 
    
        \begin{center}
        \begin{tikzpicture}
        \pgfsetxvec{\pgfpoint{0.4cm}{0}}
        \pgfsetyvec{\pgfpoint{0}{0.4cm}}
        %%%
        \coordinate (V1) at (2,2); 
        \coordinate (V2) at (10,2); 
        %%%
        \vertexcoordinates{V1}; 
        \vertexcoordinates{V2}; 
        %%% Correct diagram
        \destroyinput[0.4]{V1BR}{V1}
        \destroyoutput[0.4]{V1TL}{V1}
        \destroyinput[0.4]{V1BL}{V1}
        \createinput[0.4]{V1TR}{V1}
        %%% Wrong diagram
        \destroyinput[0.4]{V2BR}{V2}
        \destroyoutput[0.4]{V2TL}{V2}
        \destroyinput[0.4]{V2BL}{V2}
        \destroyinput[0.4]{V2TR}{V2}
        %%% Checkmark and cross
        \node at (2,-0.5) {\textcolor{unibas-mint}{$\checkmark$}};
        \node at (10,-0.5) {\textcolor{unibas-red}{$\cross$}};
        \end{tikzpicture}
        \end{center}
   The left diagram is valid but the right diagram not since it contains three ingoing lines at the same vertex.
    
    \item Every vertex connects to either one or three dashed lines.
    \begin{center}
        \begin{tikzpicture}
        \pgfsetxvec{\pgfpoint{0.4cm}{0}}
        \pgfsetyvec{\pgfpoint{0}{0.4cm}}
        %%%
        \coordinate (V1) at (2,2); 
        \coordinate (V2) at (8,2); 
        \coordinate (V3) at (14,2);
        %%%
        \vertexcoordinates{V1}; 
        \vertexcoordinates{V2}; 
        \vertexcoordinates{V3}; 
        %%% 1st Correct diagram
        \destroyinput[0.4]{V1BR}{V1}
        \destroyoutput[0.4]{V1TL}{V1}
        \destroyinput[0.4]{V1BL}{V1}
        \createinput[0.4]{V1TR}{V1}
        %%% 2d correct diagram
        \createoutput[0.4]{V2BR}{V2}
        \destroyoutput[0.4]{V2TL}{V2}
        \destroyoutput[0.4]{V2BL}{V2}
        \destroyinput[0.4]{V2TR}{V2}
        %%% Wrong diagram3
        \destroyinput[0.4]{V3BR}{V3}
        \destroyoutput[0.4]{V3TL}{V3}
        \createoutput[0.4]{V3BL}{V3}
        \destroyinput[0.4]{V3TR}{V3}
        %%% Checkmark and cross
        \node at (2,-0.5) {\textcolor{unibas-mint}{$\checkmark$}};
        \node at (8,-0.5) {\textcolor{unibas-mint}{$\checkmark$}};
        \node at (14,-0.5) {\textcolor{unibas-red}{$\cross$}};
        \end{tikzpicture}

    \end{center}
    The left and center diagram show the two possible ways that lines can attach to a vertex - either one full line or three full lines. The rightmost diagram is not valid since the vertex connects to two full lines and two \newtext{dashed} lines.
    \item The only loops that exist are those of the Keldysh Green function. 
    \begin{center}
        \begin{tikzpicture}
        \pgfsetxvec{\pgfpoint{0.4cm}{0}}
        \pgfsetyvec{\pgfpoint{0}{0.4cm}}
        \coordinate (V1) at (2,2); 
        \coordinate (V2) at (8,2);
        \coordinate (V3) at (14,2);

        \vertexcoordinates{V1}; 
        \vertexcoordinates{V2}; 
        \vertexcoordinates{V3}; 

        %%%%
        \destroyoutput[0.4]{V1TL}{V1}
        \destroyinput[0.4]{V1BL}{V1}
        \clloop[0.4]{V1}{0}
        %%%
        \destroyoutput[0.4]{V2TL}{V2}
        \destroyinput[0.4]{V2BL}{V2}
        \draw[thick,dash pattern=on 1.2\pgflinewidth off 1.2\pgflinewidth, decoration={markings, mark=at position 1 with {\arrow[scale=0.4]{Latex[length=10pt]}}}, postaction={decorate}] (V2) ++(0:1) circle [radius=1];
        %%%
        \destroyoutputK[0.4]{V3TL}{V3}
        \destroyinput[0.4]{V3BL}{V3}
        \draw[thick,dash pattern=on 1.2\pgflinewidth off 1.2\pgflinewidth, decoration={markings, mark=at position 1 with {\arrow[scale=0.4]{Latex[length=10pt]}}}, postaction={decorate}] (V3) arc(180:360:1) ;
        \draw[thick, decoration={markings, mark=at position 0 with {\arrow[scale=0.4]{Latex[length=10pt]}}}, postaction={decorate}] (16,2) arc(0:180:1) ;
        %%% Checks and crosses
        \node at (3,0) {\textcolor{unibas-mint}{$\checkmark$}};
        \node at (9,0) {\textcolor{unibas-red}{$\cross$}};
        \node at (15,0) {\textcolor{unibas-red}{$\cross$}};
        \end{tikzpicture}

    \end{center}

    Loops of \newtext{dashed} lines or mixed loops cannot exist. 
\end{enumerate}

\begin{figure*}[t]
\centering
\begin{tikzpicture}
\pgfsetxvec{\pgfpoint{0.445cm}{0}}
  \pgfsetyvec{\pgfpoint{0}{0.445cm}}
  
  % Define the start, middle, and end points
  % V : Vertex 
  % L,R : Left, Right 
  % T,B : Top, Bottom 
  \coordinate (SVVV1) at (2, 2); 
  \coordinate (SVVV2) at (6, 2);
  \vertexcoordinates{SVVV1}
  \vertexcoordinates{SVVV2}
  \coordinate (SVVVV1) at (12, 2); 
  \coordinate (SVVVV2) at (16, 2);
  \vertexcoordinates{SVVVV1}
  \vertexcoordinates{SVVVV2}
  \coordinate (V1) at (22, 2); 
  \coordinate (V2) at (26, 2);
  \vertexcoordinates{V1}
  \vertexcoordinates{V2}
  \coordinate (VV1) at (32, 2); 
  \coordinate (VV2) at (36, 2);
  \vertexcoordinates{VV1}
  \vertexcoordinates{VV2}

  \coordinate (VVV1) at (2, -3); 
  \coordinate (VVV2) at (6, -3);
  \vertexcoordinates{VVV1}
  \vertexcoordinates{VVV2}
  \coordinate (VVVV1) at (12, -3); 
  \coordinate (VVVV2) at (16, -3);
  \vertexcoordinates{VVVV1}S
  \vertexcoordinates{VVVV2}
  \coordinate (SV1) at (22, -3); 
  \coordinate (SV2) at (26, -3);
  \vertexcoordinates{SV1}
  \vertexcoordinates{SV2}
  \coordinate (SVV1) at (32, -3); 
  \coordinate (SVV2) at (36, -3);
  \vertexcoordinates{SVV1}
  \vertexcoordinates{SVV2}

  %%% Diagram a) 
  \node at (-1,4) {a)};
  
  \clloop[0.445]{V1}{90}
  \createinput[0.445]{V2TL}{V2}
  \destroyinput[0.445]{V2TR}{V2}
  \destroyinput[0.445]{V2BR}{V2}
  \destroyoutput[0.445]{V1BL}{V1}[0.3]
  \Gret[0.445]{V1}{V2}
  
  %%% Diagram b) 
  \node at (9,4) {b)};
  \destroyinput[0.445]{VV1BR}{VV1}
  \destroyinput[0.445]{VV1TL}{VV1}
  \createinput[0.445]{VV2BR}{VV2}
  \destroyoutput[0.445]{VV1BL}{VV1}
  \clloop[0.445]{VV2}{90}
  \Gadv[0.445]{VV2}{VV1}

  %%% Diagram c) 
  \node at (19,4) {c)};
  \createinput[0.445]{VVV1BR}{VVV1}
  \destroyinput[0.445]{VVV1TL}{VVV1}
  \destroyinput[0.445]{VVV2BR}{VVV2}
  \destroyoutput[0.445]{VVV1BL}{VVV1}
  \clloop[0.445]{VVV2}{90}
  \Gret[0.445]{VVV1}{VVV2}

  %%% Diagram f) 
  \node at (29,4) {d)};
  %\destroyinput[0.445]{VVVV1TL}{VVVV1}
  \destroyinput[0.445]{SVV2R}{SVV2}
  \destroyoutput[0.445]{SVV1L}{SVV1}
  %\createinput[0.445]{VVVV2TR}{VVVV2}
  \Gkelup[0.445]{SVV1}{SVV2}
  \Gkel[0.445]{SVV1}{SVV2}
  \Gadvdown[0.445]{SVV2}{SVV1}

%%%% Second line 
  %%% Diagram d) 
  \node at (-1,-1) {e)};
  \destroyinput[0.445]{SV2TR}{SV2}
  \createinput[0.445]{SV2BR}{SV2}
  \destroyinput[0.445]{SV1TL}{SV1}
  \destroyoutput[0.445]{SV1BL}{SV1}
  \Gkelup[0.445]{SV2}{SV1}
  \Gretdown[0.445]{SV1}{SV2}
  
  %%% Diagram e) 
  \node at (9,-1) {f)};
  \createinput[0.445]{VVVV1TL}{VVVV1}
  \destroyinput[0.445]{VVVV2BR}{VVVV2}
  \destroyoutput[0.445]{VVVV1BL}{VVVV1}
  \destroyinput[0.445]{VVVV2TR}{VVVV2}
  \Gkelup[0.445]{VVVV1}{VVVV2}
  \Gretdown[0.445]{VVVV1}{VVVV2}

  %%% Diagram f) 
  \node at (19,-1) {g)};
  \destroyinput[0.445]{SVVV1BR}{SVVV1}
  \createinput[0.445]{SVVV2TL}{SVVV2}
  \destroyinput[0.445]{SVVV1TL}{SVVV1}
  \destroyinput[0.445]{SVVV2TR}{SVVV2}
  \createinput[0.445]{SVVV2BR}{SVVV2}
  \destroyoutput[0.445]{SVVV1BL}{SVVV1}
  \Gadv[0.445]{SVVV2}{SVVV1}

  %%% Diagram h) 
  \node at (29,-1) {h)};
  \createinput[0.445]{SVVVV1BR}{SVVVV1}
  \createinput[0.445]{SVVVV2TL}{SVVVV2}
  \destroyinput[0.445]{SVVVV1TL}{SVVVV1}
  \destroyinput[0.445]{SVVVV2TR}{SVVVV2}
  \destroyinput[0.445]{SVVVV2BR}{SVVVV2}
  \destroyoutput[0.445]{SVVVV1BL}{SVVVV1}
  \Gret[0.445]{SVVVV1}{SVVVV2}
  
\end{tikzpicture}
  \caption{Second order diagrams for $\langle\bout\rangle$. a)-e) Diagrams that can be obtained by concatenating first order diagrams and can thus be obtained from a finite-temperature mean-field approach. f)-h) Diagrams containing doubly-- and triple--connected vertices.}
\label{fig:bout_2nd_order_diags}
\end{figure*}

These three rules are sufficient to find all contributing diagrams for a specific cumulant. However, taking into account the application of Wicks theorem and the causality structure of the Green functions yields additional constraints that ensure certain expressions to invariably go to zero. These constraints can also be formulated on a diagrammatic level and result in the following additional rules.
\begin{enumerate}
\setcounter{enumi}{3}
    \item 
    Whenever two vertices are connected by multiple Green functions, the connecting Green functions have to be arranged in such a way that all \newtext{dashed} lines connect to the same vertex. 
    
    %\begin{figure}
        \begin{center}
        \begin{tikzpicture}
        \pgfsetxvec{\pgfpoint{0.4cm}{0}}
        \pgfsetyvec{\pgfpoint{0}{0.4cm}}
        %%%
        \coordinate (V1) at (0,2);
        \coordinate (V2) at (4,2); 
        \coordinate (VV1) at (10,2);
        \coordinate (VV2) at (14,2);
        %%%
        \vertexcoordinates{V1}; 
        \vertexcoordinates{V2}; 
        \vertexcoordinates{VV1}; 
        \vertexcoordinates{VV2};
        %%% Correct diagram
        \destroyoutput[0.4]{V1TL}{V1}
        \destroyinput[0.4]{V1BL}{V1}
        \Gkelup[0.4]{V2}{V1}
        \Gretdown[0.4]{V1}{V2}
        \createinput[0.4]{V2BR}{V2}
        \destroyinput[0.4]{V2TR}{V2}
        %%% Wrong diagram
        \destroyinput[0.4]{VV1BL}{VV1}
        \destroyoutputK[0.4]{VV1TL}{VV1}
        \Gretup[0.4]{VV2}{VV1}
        \Gretdown[0.4]{VV1}{VV2}
        \destroyinput[0.4]{VV2BR}{VV2}
        \createinput[0.4]{VV2TR}{VV2}
        %%% Checkmark and cross
        \node at (2,-0.75) {\textcolor{unibas-mint}{$\checkmark$}};
        \node at (12,-0.75) {\textcolor{black}{$\cross$}};
        %%% This is a hack for alignment with the text inside the enumerate environment: 
        \node at (-3,0) {\textcolor{white}{.}};
        \end{tikzpicture}
        \end{center}
    
   The right diagram will invariably equal zero because the \newtext{dashed} lines of the connecting Green functions connect to opposite vertices.
   \item a) Any diagram that contains exactly two detector symbols at the same vertex, one filled and one unfilled, where exactly one connects to the vertex through a dashed line invariably vanishes. 
    
    %\begin{figure}
        \begin{center}
        \begin{tikzpicture}
        \pgfsetxvec{\pgfpoint{0.4cm}{0}}
        \pgfsetyvec{\pgfpoint{0}{0.4cm}}
        %%%
        \coordinate (V1) at (0,2);
        \coordinate (V2) at (4,2); 
        \coordinate (VV1) at (12,2);
        \coordinate (VV2) at (14,2);
        %%%
        \vertexcoordinates{V1}; 
        \vertexcoordinates{V2}; 
        \vertexcoordinates{VV1}; 
        %%% Correct diagram
        \destroyoutput[0.4]{V1TL}{V1}
        \createoutputK[0.4]{V1BL}{V1}
        \Gkelup[0.4]{V2}{V1}
        \Gretdown[0.4]{V1}{V2}
        \createinput[0.4]{V2BR}{V2}
        \destroyinput[0.4]{V2TR}{V2}
        %%% Wrong diagram
        \createoutput[0.4]{VV1BL}{VV1}
        \destroyoutputK[0.4]{VV1TL}{VV1}
        \destroyinput[0.4]{VV1BR}{VV1}
        \createinput[0.4]{VV1TR}{VV1}
        %%% Checkmark and cross
        \node at (2,-0.75) {\textcolor{black}{$\cross$}};
        \node at (12,-0.75) {\textcolor{black}{$\cross$}};
        %%% This is a hack for alignment with the text inside the enumerate environment: 
        \node at (-3,0) {\textcolor{white}{.}};
        \end{tikzpicture}
        \end{center}

    b) Any diagram that contains exactly three detector legs at one vertex, whereof two are identical, and one or two connect to the vertex through a dashed line, invariably vanishes.
    \begin{center}
        \begin{tikzpicture}
        \pgfsetxvec{\pgfpoint{0.4cm}{0}}
        \pgfsetyvec{\pgfpoint{0}{0.4cm}}
        %%%
        \coordinate (V1) at (0,2);
        \coordinate (V2) at (4,2); 
        \coordinate (VV1) at (9,2);
        \coordinate (VV2) at (13,2);
        %%%
        \vertexcoordinates{V1}; 
        \vertexcoordinates{V2}; 
        \vertexcoordinates{VV1}; 
        \vertexcoordinates{VV2};
        %%% Correct diagram
        \destroyoutput[0.4]{V1TL}{V1}
        \createoutputK[0.4]{V1BL}{V1}
        \createoutput[0.4]{V1BR}{V1}
        \Gret[0.4]{V2}{V1}
        \destroyoutput[0.4]{V2TR}{V2}
        \destroyinput[0.4]{V2BR}{V2}
        \createinput[0.4]{V2TL}{V2}
        %%% Wrong diagram
        \destroyoutputK[0.4]{VV1TL}{VV1}
        \createoutput[0.4]{VV1BL}{VV1}
        \createoutput[0.4]{VV1BR}{VV1}
        \Gret[0.4]{VV2}{VV1}
        \destroyoutput[0.4]{VV2TR}{VV2}
        \destroyinput[0.4]{VV2BR}{VV2}
        \createinput[0.4]{VV2TL}{VV2}
        %%% Checkmark and cross
        \node at (2,-0.75) {\textcolor{unibas-mint}{$\checkmark$}};
        \node at (12,-0.75) {\textcolor{black}{$\cross$}};
        %%% This is a hack for alignment with the text inside the enumerate environment: 
        \node at (-3,0) {\textcolor{white}{.}};
        \end{tikzpicture}
        \end{center}
    
    \newtext{\item Any diagram without sources, i.e. without square symbols, vanishes.
    \begin{center}
        \begin{tikzpicture}
        \pgfsetxvec{\pgfpoint{0.4cm}{0}}
        \pgfsetyvec{\pgfpoint{0}{0.4cm}}
        %%%
        \coordinate (V1) at (0,2);
        \coordinate (V2) at (6,2);
        \coordinate (V3) at (10,2);
        %%%
        \vertexcoordinates{V1}; 
        \vertexcoordinates{V2};  
        \vertexcoordinates{V3};
        %%% Correct diagram
        \destroyoutput[0.4]{V1TL}{V1}
        \createoutputK[0.4]{V1BL}{V1}
        \createoutputK[0.4]{V1BR}{V1}
        \destroyoutputK[0.4]{V1TR}{V1}
        %%% Wrong diagram
        \createoutput[0.4]{V2BL}{V2}
        \destroyoutputK[0.4]{V2TL}{V2}
        \Gkelup[0.4]{V2}{V3}
        \Gadvdown[0.4]{V3}{V2}
        \destroyoutput[0.4]{V3BR}{V3}
        \createoutput[0.4]{V3TR}{V3}
        %%% Checkmark and cross
        \node at (0,-0.75) {\textcolor{black}{$\cross$}};
        \node at (8,-0.75) {\textcolor{black}{$\cross$}};
        %%% This is a hack for alignment with the text inside the enumerate environment: 
        \node at (-3,0) {\textcolor{white}{.}};
        \end{tikzpicture}
        \end{center}}
   \item At zero temperature, two vertices cannot be connected by three lines. 
    
    %\begin{figure}
        \begin{center}
        \begin{tikzpicture}
        \pgfsetxvec{\pgfpoint{0.4cm}{0}}
        \pgfsetyvec{\pgfpoint{0}{0.4cm}}
        %%%
        \coordinate (V1) at (0,2);
        \coordinate (V2) at (4,2); 
        %%%
        \vertexcoordinates{V1}; 
        \vertexcoordinates{V2}; 
        %%% Correct diagram
        \destroyoutput[0.4]{V1L}{V1}
        \Gkelup[0.4]{V2}{V1}
        \Gretdown[0.4]{V1}{V2}
        \Gkel[0.4]{V1}{V2}
        \destroyinput[0.4]{V2R}{V2}
        %%% Wrong diagram
        \draw[->] (7,2) -- (10,2);
        \node at (11,2) {$\displaystyle 0$};
        \node at (8.5,3) {$\displaystyle n_B \to 0$};
        %%% This is a hack for alignment with the text inside the enumerate environment: 
        \node at (-3,0) {\textcolor{white}{.}};
        \end{tikzpicture}
        \end{center}
\end{enumerate}

The terms contributing, e.g., to the second order correction of the average output field can now be determined by finding all diagrams that (i) contain two vertices, (ii) one empty half-circle and are (iii) consistent with the diagrammatic rules outlined above. Some of the resulting diagrams are illustrated in Fig.~\ref{fig:bout_2nd_order_diags}. 

\subsection{Evaluating the diagrams}
\label{sec:evaluating_the_diagrams}
Once all diagrams that contribute to the desired cumulant are found, they need to be translated to the equivalent mathematical expressions in order to evaluate them. This is done by replacing all the constituents of the diagram with the corresponding expression according to Fig.~\ref{fig:operator_diags} and each vertex with a multiplication of $-iU$ and integration over its time argument, 
\begin{center}
        \begin{tikzpicture}
        \pgfsetxvec{\pgfpoint{0.4cm}{0}}
        \pgfsetyvec{\pgfpoint{0}{0.4cm}}
        %%%
        \coordinate (V1) at (0,2);
        %%%
        \vertexcoordinates{V1}; 
        %%% Wrong diagram
        \draw[->] (1,2) -- (3.5,2);
        \node at (6,2) {$\displaystyle -i U \int dt \fullstop$};
        \end{tikzpicture}
\end{center}
Additionally, each diagram gets an integer factor that we call its multiplicity which is comprised of three different numbers: 
\begin{enumerate} 
    \item 
    Each vertex contributes a multiplicity factor of two if and only if all four legs connecting to the vertex are different, i.e. each building block in Fig.~\ref{fig:operator_diags} appears at most once, and at most two legs connect to detector symbols. Hence we get for the full diagram, 
    $$2^{\# \text{vertices with distinct legs and $<3$ detectors}} \fullstop$$
    \begin{center}
        \begin{tikzpicture}
        \pgfsetxvec{\pgfpoint{0.4cm}{0}}
        \pgfsetyvec{\pgfpoint{0}{0.4cm}}
        %%%
        \coordinate (V1) at (0,2);
        \coordinate (V2) at (4,2); 
        %%%
        \vertexcoordinates{V1}; 
        \vertexcoordinates{V2}; 
        %%% Correct diagram
        \destroyinput[0.4]{V1TL}{V1}
        \createoutput[0.4]{V1BL}{V1}
        \Gkelup[0.4]{V2}{V1}
        \Gadvdown[0.4]{V2}{V1}
        \createinput[0.4]{V2BR}{V2}
        \destroyoutputK[0.4]{V2TR}{V2}
        %%% Checkmark and cross
        \draw[unibas-red, thick, dashed] (-2.1,-0.2) rectangle (2,4.1);
        \end{tikzpicture}
        \end{center}
    In this diagram the left vertex inside the red box has two equal legs, the two outgoing classical field lines, whereas the right vertex has four different legs. The vertices thus contribute to the multiplicity with a factor of two.
   
    \item Each diagram is multiplied by the factorial of the number of empty half--circles and the factorial of the number of filled half--circles,  
    \begin{equation*}
         (\#~ 
        {
        \tikz {\fill[black] (0,0) arc(90:-90:0.1) -- (0,0) --cycle;
        \draw[thick] (0,0) arc(90:-90:0.1) -- (0,0) --cycle ;}
        })! \times (\#~ 
        {
        \tikz {\fill[white] (0,0) arc(90:-90:0.1) -- (0,0) --cycle;
        \draw[thick] (0,0) arc(90:-90:0.1) -- (0,0) --cycle ;}
        })! \fullstop
    \end{equation*}
    \item Lastly we multiply by the number of ways connections between vertices can be realigned. This factor is found by cutting up each propagator that starts and ends at a vertex and counting all possible ways of reconnecting the arrows such that one recovers the original diagram.  
    \vspace{.15cm}
    \begin{center}
        \begin{tikzpicture}
        \pgfsetxvec{\pgfpoint{0.4cm}{0}}
        \pgfsetyvec{\pgfpoint{0}{0.4cm}}
        %%%
        \coordinate (V1) at (0,2);
        \coordinate (V2) at (4,2); 
        %%%
        \vertexcoordinates{V1}; 
        \vertexcoordinates{V2}; 
        %%% Correct diagram
        \destroyinput[0.4]{V1TL}{V1}
        \createoutput[0.4]{V1BL}{V1}
        \Gkelup[0.4]{V2}{V1}
        \Gadvdown[0.4]{V2}{V1}
        \createinput[0.4]{V2BR}{V2}
        \destroyoutputK[0.4]{V2TR}{V2}
        %%% Red switching arrow and cut line
        \draw[->,>=to,unibas-red] (1.2,2.75) to[out=-45, in=45] (1.2,1.25);
        \draw[->,>=to,unibas-red] (1.2,1.25) to[out=45, in=-45] (1.2,2.75);
        \draw[unibas-red, thick, dashed] (2,-0.25) to (2,4.25);
        \end{tikzpicture}
        \end{center}
        After cutting the propagator lines, there are two ways of reconnecting the dangling legs to recover the diagram. It therefore needs to be multiplied by a factor of two.
\end{enumerate}
A more detailed discussion of the multiplicities and their origins is given in App.~\ref{sec:mult_of_diags}. We illustrate the evaluation procedure with the following diagram, which contributes to the second order correction of the cumulant $\llangle \bout \bout \rrangle$, 

\begin{widetext}
\begin{center}
\begin{tikzpicture}
    %% This rescales the diagrams: 
    \pgfsetxvec{\pgfpoint{0.5cm}{0}}
    \pgfsetyvec{\pgfpoint{0}{0.5cm}}
    %% Grid for orientation 
    %\draw[step=0.25cm,gray,very thin] (-5,-3) grid (25,12);
    %%%%%%%%%%%%%%%%%%%%%%%%%%%%%%%%%%%%%%%%%%%%%%%%%%%%%%%%%%%
    %% Dressed Advanced Green function
    % Coordinates 
        \coordinate (V1) at (0,0); 
        \coordinate (V2) at (4,0); 
        \vertexcoordinates{V1} 
        \vertexcoordinates{V2}
    % Diagrams and math symbols 
    %% Left vertex
    \destroyoutputK[0.5]{V1BL}{V1}
    \destroyinput[0.5]{V1TL}{V1}
    %% Connections 
    \Gretdown[0.5]{V2}{V1}
    \Gkelup[0.5]{V1}{V2}
    %% Right vertex 
    \destroyoutput[0.5]{V2BR}{V2}
    \destroyinput[0.5]{V2TR}{V2}
\end{tikzpicture}
\end{center}
\vspace{-3mm}
\begin{align}
=  \underbrace{2^2 \times 2!}_{\text{multiplicity}} \times\underbrace{(-iU)^2 \iint dt_1 dt_2}_{\text{vertices}}   
\biggl(\underbrace{\langle \phi^\text{cl}(t_1)\rangle_0 \langle \overline{\phi}^\text{q}\rangle^{'}_0(t_1-\tau_1)}_{\text{left vertex legs}} 
\underbrace{G^K(t_1,t_2) G^R(t_2,t_1)}_{\text{vertex connections}}
\underbrace{\langle \overline{\phi}^\text{q}\rangle^{'}_0(t_1-\tau_1) \langle \phi^\text{cl}(t_2)\rangle_0 }_{\text{right vertex legs} }\biggr) \fullstop 
     \label{eq:example_diag_evaluation}
\end{align}
\end{widetext}

\begin{figure*}
    \centering
    \begin{overpic}[grid = false,tics = 5,width=\linewidth]{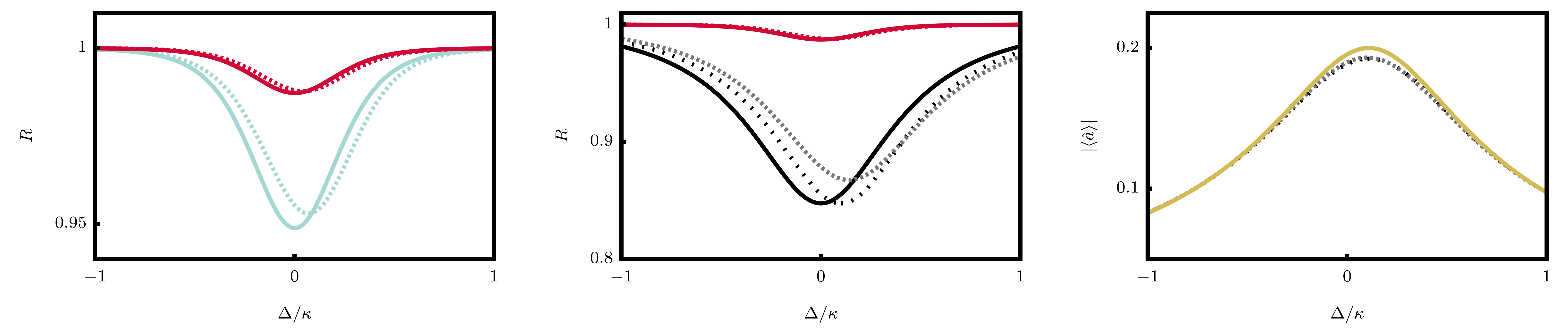}
    \put (2,19) {a)}
    \put (35,19) {b)}
    \put (68,19) {c)}
    \put (6.5,10) {\scriptsize{$\frac{\langle \hat{b}_\text{in} \rangle}{\sqrt{\kappa}} = 0.2$}}
    % Solid red line
    \put(6,8){
      \begin{tikzpicture}
        \draw[unibas-red, very thick] (0,0) -- (0.4,0);
      \end{tikzpicture}
    }
    \put(9,8){ \scriptsize{Approx.}}

    % Dashed red line
    \put(6,6){
      \begin{tikzpicture}
        \draw[unibas-red, very thick, yshift=-1cm, dash pattern=on 1pt off 1pt ] (0,0) -- (0.4,0);
      \end{tikzpicture}
    }
    \put(9,6){ \scriptsize{Exact}}

    \put (22.5,10) {\scriptsize{$\frac{\langle \hat{b}_\text{in} \rangle}{\sqrt{\kappa}} = 0.3$}}
    % Solid mint line
    \put(22,8){
      \begin{tikzpicture}
        \draw[unibas-mint, very thick] (0,0) -- (0.4,0);
      \end{tikzpicture}
    }
    \put(25,8){ \scriptsize{Approx.}}

    % Dashed mint line
    \put(22,6){
      \begin{tikzpicture}
        \draw[unibas-mint, very thick, yshift=-1cm, dash pattern=on 1pt off 1pt ] (0,0) -- (0.4,0);
      \end{tikzpicture}
    }
    \put(25,6){ \scriptsize{Exact }}
    %%%%%%%%%%%%%%%%%%%%%%%%%%%%%%%%%%%%%%%%%%%%%%%%%%%%%%%%%%%%%%%%%
    %% Middle Figure 
    % Solid mint line
    \put(40,10){
      \begin{tikzpicture}
        \draw[black, very thick] (0,0) -- (0.4,0);
      \end{tikzpicture}
    }
    \put(43,10){ \scriptsize{Approx.}}
    % Dashed mint line
    \put(40,8){
      \begin{tikzpicture}
        \draw[ unibas-gray,very thick, yshift=-1cm, dash pattern=on 1pt off 1pt ] (0,0) -- (0.4,0);
      \end{tikzpicture}
    }
    \put(43,8){ \scriptsize{Num.}}
    % Dashed mint line
    \put(40,6){
      \begin{tikzpicture}
        \draw[ very thick, yshift=-1cm, dash pattern=on 1pt off 2.5pt ] (0,0) -- (0.4,0);
      \end{tikzpicture}
    }
    \put(43,6){ \scriptsize{Loop sum.}}
    
    %%%%%%%%%%%%%%%%%%%%%%%%%%%%%%%%%%%%%%%%%%%%%%%%%%%%%%%%%%%%%%%%%%%%
    %% Right Figure 
    % Solid line
    \put(82,10){
      \begin{tikzpicture}
        \draw[unibas-yellow, very thick] (0,0) -- (0.4,0);
      \end{tikzpicture}
    }
    \put(85,10){ \scriptsize{MF}}
    % Dashed line
    \put(82,8){
      \begin{tikzpicture}
        \draw[ unibas-gray,very thick, yshift=-1cm, dash pattern=on 1pt off 1pt ] (0,0) -- (0.4,0);
      \end{tikzpicture}
    }
    \put(85,8){ \scriptsize{Num.}}
    % Dashed line
    \put(82,6){
      \begin{tikzpicture}
        \draw[ very thick, yshift=-1cm, dash pattern=on 1pt off 2.5pt ] (0,0) -- (0.4,0);
      \end{tikzpicture}
    }
    \put(85,6){ \scriptsize{Loop sum.}}
    
    %%%%%%%%%%%%%%%%%%%%%%%%%%%%%%%%%%%%%%%%%%%%%%%%%%%%%%%%%%%%%%%%%%%%%%%%%%%%%%%%
    \end{overpic}
    \vspace{-6mm}
    \caption{Reflection probability. a)
    The \newtext{dashed} lines correspond to the exact solution \cite{PhysRevA.94.033841} and the full lines show the second-order approximation obtained with our diagrammatic method. For the red curves, the drive strength is $\langle \bin \rangle /\sqrt{\kappa} = 0.2$  and for the mint curves $\langle \bin \rangle/\sqrt{\kappa} = 0.3$. b) Reflection probability of the Kerr oscillator for finite temperature, $n_B = 0.25$ (black). The red lines are the same as in a). We observe that the result for the reflection coefficient where we include all loop diagrams to infinite order (\newtext{dashed} line) performs better than the true second order approximation (full line) compared to the numerics (dashed line). c) absolute value of the intra--cavity field. Both partial summation methods, the mean field (MF) approach (yellow line) as well as the loop summation (\newtext{dashed} line) describe the intra--cavity dynamics well. For the reflection coefficient, the MF result however yields a constant value of one. For all plots $\sfrac{U}{\kappa}=0.1$. }
    \label{fig:reflection_coeff}
\end{figure*}

\section{Partial Summation Methods}
Our diagrammatic approach yields an expansion in the interaction $U$ in terms of diagrams for any given correlation function. In many cases, physical effects are dominated by specific classes of diagrams whose contribution cannot be reliably captured at finite order~\cite{Peskin:1995ev}. Partial summation methods can address this limitation by summing infinite subsets of diagrams that follow a specific pattern. In this section we explain how one can interpret a mean field theory approach to the average output field as a partial summation and we detail a partial summation technique for diagrams including thermal loops. 
\vspace{-.75cm}
\subsection{Finite-temperature mean-field theory}
The average output field $\langle \bout \rangle$, through the input--output relation from Eq.~\eqref{eq:in_out_rel}, is directly linked to the average intra--cavity mode $\langle \ahat \rangle$ and knowledge of one quantity directly implies knowledge of the other. The calculation of the average output field with our diagrammatic approach can therefore straightforwardly be connected to conventional approaches of determining the intra--cavity field average. 

We note that some of the diagrams contributing to the second order correction of the average output field may be obtained by concatenating the first-order diagrams from Fig.~\ref{fig:diagrams_bout_first_order}. This is done by taking a first-order diagram of $\langle\bout\rangle$, removing an input (anti-)photon, and replacing it by the outgoing leg of another first-order diagram of $\langle\bout\rangle$ ($\langle\dbout\rangle$). This pattern suggests that these diagrams can be captured by a partial summation technique. It turns out that the result obtained by this partial summation corresponds to a mean--field approach for the intra--cavity mode. In order to show this correspondence we model the intra--cavity mode with the Langevin equation in Eq.~\eqref{eq:eom_in}. We treat the Kerr resonator in a rotating frame, i.e. $\omega_s \to -\Delta$, and find, c.f.~Eq.~\eqref{eq:Kerr_Hamiltonian},
\begin{equation}
    \label{eq:eomkerr}
    \dv{t} \langle \hat{a}\rangle = -\left(\frac{\kappa}{2}+i\Delta\right)\langle\hat{a}\rangle -\sqrt{\kappa}f(t)-2iU\langle\hat{a}^\dagger\hat{a}\hat{a}\rangle.
\end{equation}
To approximate the last term, we write $\hat{a}=\langle\hat{a}\rangle+\delta\hat{a}$ and we assume that thermal fluctuations are dominating, such that expectation values of $\delta\hat{a}$ are determined by a thermal state with occupation $n_B$. This results in
\begin{equation}
    \langle\hat{a}^\dagger\hat{a}\hat{a}\rangle\simeq \langle\hat{a}^\dagger\rangle\langle\hat{a}\rangle^2+2n_B\langle\hat{a}\rangle.
\end{equation}
Substituting this approximation into Eq.~\eqref{eq:eomkerr} results in a non-linear equation for $\langle\hat{a}\rangle$, our finite-temperature mean-field equation that can be interpreted as introducing a self-consistent detuning depending on the cavity mode and temperature. This equation captures all diagrams that are obtained by concatenating the first-order diagrams shown in Fig.~\ref{fig:diagrams_bout_first_order}. To see this, we write
\begin{equation}
    \langle\hat{a}\rangle = \sum_{n=0}^\infty \langle\hat{a}\rangle_n,
\end{equation}
where $\langle\hat{a}\rangle_n$ is proportional to $U^n$. 
For $n>0$, the coefficients are given as follows, 
\begin{equation}
    \langle\hat{a}\rangle_n = \begin{aligned}[t] 2U\int dt'&G^R(t-t')\bigg[2n_B\langle\hat{a}\rangle_{n-1} \\ 
    &+\sum_{l,k,p}\delta_{l+k+p,n-1}\langle a\rangle_l\langle a\rangle_k\langle a\rangle^*_p\bigg]. \end{aligned}
    \label{eq:avgn}
\end{equation}
For $n=1$, Eq.~\eqref{eq:avgn} reproduces the diagrams in Fig.~\ref{fig:diagrams_bout_first_order}. For $n=2$, we simply replace one input term $\langle\phi^{\rm cl}\rangle_0$ by $\sqrt{2}\langle\hat{a}\rangle_1$. For our diagrams, this means removing one input leg and replacing it with a first-order diagram where the detector is removed. Similarly, an $n$-th order diagram can be broken down in lower-order diagrams. Repeating this process implies that all diagrams that appear in Eq.~\eqref{eq:avgn} are built by concatenating first-order diagrams. We note here that even though the mean-field solution describes the behavior of the intra--cavity mode $\hat{a}$ well, see Fig.~\ref{fig:reflection_coeff} c), it fails to capture the behavior of the output field. Indeed, the reflection coefficient  \newtext{(see Eq.~\eqref{eq:reflection_coeff_def} below for a definition)} obtained from the mean field solution is equal to one for all parameters considered, see Fig.~\ref{fig:reflection_coeff} a) and b) for the actual behavior.   
\subsection{Summation of loop diagrams}
\label{sec:loop_summation}
As it conveniently turns out, all diagrams containing loops can be automatically accounted for by implementing a shift in the detuning in the Green functions, see App.~\ref{sec:resummation},

\begin{equation}
\begin{tikzpicture}
    %% This rescales the diagrams: 
    \pgfsetxvec{\pgfpoint{0.25cm}{0}}
    \pgfsetyvec{\pgfpoint{0}{0.25cm}}
    %% Grid for orientation 
    %\draw[step=0.25cm,gray,very thin] (-8,-10) grid (25,20);
    %% Dressed Keldysh Green function 
    % Coordinates 
        \coordinate (VVV1) at (6,15); 
        \coordinate (VVV2) at (10,15); 
        \coordinate (VVV3) at (12,15); 
        \coordinate (VVV4) at (16,15);
        \coordinate (VVV5) at (18,15); 
        \coordinate (VVV6) at (22,15); 
    % Diagrams and math symbols 
        \node at (-0.5,15) {$\displaystyle i G^K_{U}(t-t') = \sum\limits_{l,p=0}^\infty$};
        \node at (5.5,15) {$\displaystyle \Biggl[$};
        \Gadv[0.4]{VVV2}{VVV1}
        \clloop[0.4]{VVV2}{100}
        \node at (11.5,15) {$\displaystyle \Biggr]^l$};
        \Gkel[0.4]{VVV4}{VVV3}
        \node at (16.5,15) {$\displaystyle \Biggl[$};
        \Gret[0.4]{VVV6}{VVV5}; 
        \clloop[0.4]{VVV5}{100}; 
        \node at (22.5,15) {$\displaystyle \Biggr]^p \comma$};
    % Equation 
        \node at (10.5,10) {$\displaystyle = F e^{i (\Delta + 4 n_B U) (t-t')}e^{-\sfrac{\kappa}{2}|t-t'|} \comma$};
     %%%%%%%%%%%%%%%%%%%%%%%%%%%%%%%%%%%%%%%%%%%%%%%%%%%%%%%%%%%
    %% Dressed retarded Green function 
    % Coordinates
    %    \coordinate (VV1) at (6,5); 
    %    \coordinate (VV2) at (10,5); 
    %    \coordinate (VV3) at (12,5); 
    %    \coordinate (VV4) at (16,5);
    % Diagrams and math symbols 
    %    \node at (-0.7,5) {$\displaystyle i G^R_U(t-t') =  \sum\limits_{n=0}^\infty $};
    %    \node at (10.5,5) {$\displaystyle \Biggl[$};
    %    \Gret[0.4]{VV2}{VV1}
    %    \clloop[0.4]{VV3}{100}
    %    \node at (17,5) {$\displaystyle \Biggr]^n \comma$};
    %    \Gret[0.4]{VV4}{VV3}
    % Equation 
    %    \node at (12.5,0) {$\displaystyle = \theta(t-t') e^{i (\Delta + 4 n_B U) (t-t')} e^{-\sfrac{\kappa}{2}(t-t')} \comma$};
    %%%%%%%%%%%%%%%%%%%%%%%%%%%%%%%%%%%%%%%%%%%%%%%%%%%%%%%%%%%
    %% Dressed Advanced Green function
    % Coordinates 
    %    \coordinate (V1) at (6,-5); 
    %    \coordinate (V2) at (10,-5); 
    %    \coordinate (V3) at (12,-5); 
    %    \coordinate (V4) at (16,-5);
    % Diagrams and math symbols 
    %    \node at (-0.7,-5) {$ \displaystyle i G^A_{U}(t-t') = \sum\limits_{n=0}^\infty $};
    %    \node at (5.5,-5) {$\displaystyle \Biggl[$};
    %    \Gadv[0.4]{V2}{V1}
    %    \clloop[0.4]{V2}{100}
    %    \node at (11.5,-5) {$\displaystyle \Biggr]^n 
    %    $};
    %    \Gadv[0.4]{V4}{V3}
    %    \node at (16.5,-5) {$\displaystyle \comma 
    %    $};
    % Equation 
    %    \node at (12.6,-10) {$\displaystyle = -\theta(t'-t) e^{i (\Delta + 4 n_B U) (t-t')} e^{\sfrac{\kappa}{2}(t-t')}\fullstop$};
    %%%%%%%%%%%%%%%%%%%%%%%%%%%%%%%%%%%%%%%%%%%%%%%%%%%%%%%%%%%
   
    \end{tikzpicture}
    \label{eq:resummed_Green_functions_diags_K}
\end{equation}

\begin{equation}
\begin{tikzpicture}
    %% This rescales the diagrams: 
    \pgfsetxvec{\pgfpoint{0.25cm}{0}}
    \pgfsetyvec{\pgfpoint{0}{0.25cm}}
    %% Grid for orientation 
    %\draw[step=0.25cm,gray,very thin] (-8,-10) grid (25,20);
    
     %%%%%%%%%%%%%%%%%%%%%%%%%%%%%%%%%%%%%%%%%%%%%%%%%%%%%%%%%%%
    %% Dressed retarded Green function 
    % Coordinates
        \coordinate (VV1) at (6,5); 
        \coordinate (VV2) at (10,5); 
        \coordinate (VV3) at (12,5); 
        \coordinate (VV4) at (16,5);
    % Diagrams and math symbols 
        \node at (-0.7,5) {$\displaystyle i G^R_U(t-t') =  \sum\limits_{n=0}^\infty $};
        \node at (10.5,5) {$\displaystyle \Biggl[$};
        \Gret[0.4]{VV2}{VV1}
        \clloop[0.4]{VV3}{100}
        \node at (17,5) {$\displaystyle \Biggr]^n \comma$};
        \Gret[0.4]{VV4}{VV3}
    % Equation 
        \node at (12.5,0) {$\displaystyle = \theta(t-t') e^{i (\Delta + 4 n_B U) (t-t')} e^{-\sfrac{\kappa}{2}(t-t')} \comma$};
    %%%%%%%%%%%%%%%%%%%%%%%%%%%%%%%%%%%%%%%%%%%%%%%%%%%%%%%%%%%
   
    \end{tikzpicture}
    \label{eq:resummed_Green_functions_diags_R}
\end{equation}

\begin{equation}
\begin{tikzpicture}
    %% This rescales the diagrams: 
    \pgfsetxvec{\pgfpoint{0.25cm}{0}}
    \pgfsetyvec{\pgfpoint{0}{0.25cm}}
    %% Grid for orientation 
    %\draw[step=0.25cm,gray,very thin] (-8,-10) grid (25,20);
    
    %%%%%%%%%%%%%%%%%%%%%%%%%%%%%%%%%%%%%%%%%%%%%%%%%%%%%%%%%%%
    %% Dressed Advanced Green function
    % Coordinates 
        \coordinate (V1) at (6,-5); 
        \coordinate (V2) at (10,-5); 
        \coordinate (V3) at (12,-5); 
        \coordinate (V4) at (16,-5);
    % Diagrams and math symbols 
        \node at (-0.7,-5) {$ \displaystyle i G^A_{U}(t-t') = \sum\limits_{n=0}^\infty $};
        \node at (5.5,-5) {$\displaystyle \Biggl[$};
        \Gadv[0.4]{V2}{V1}
        \clloop[0.4]{V2}{100}
        \node at (11.5,-5) {$\displaystyle \Biggr]^n 
        $};
        \Gadv[0.4]{V4}{V3}
        \node at (16.5,-5) {$\displaystyle \comma 
        $};
    % Equation 
        \node at (12.6,-10) {$\displaystyle = -\theta(t'-t) e^{i (\Delta + 4 n_B U) (t-t')} e^{\sfrac{\kappa}{2}(t-t')}\fullstop$};
    %%%%%%%%%%%%%%%%%%%%%%%%%%%%%%%%%%%%%%%%%%%%%%%%%%%%%%%%%%%
   
    \end{tikzpicture}
    \label{eq:resummed_Green_functions_diags_A}
\end{equation}
Using these \textit{dressed} Green functions in all diagrams that do not contain loops, as well as the 0-th order term, we recover all diagrams that include loops. For instance, all diagrams in Fig.~\ref{fig:bout_2nd_order_diags} that include loops are recovered by dressing the Green functions in Fig.~\ref{fig:diagrams_bout_first_order}\,b). The diagram in Fig.~\ref{fig:diagrams_bout_first_order}\,a) is obtained by dressing the Green function in the 0-th order term in Eq.~\eqref{eq:bout_HO}. A partial summation of all diagrams up to infinite order is obtained by summing over all diagrams with up to $n$ vertices that do not contain loops and dressing the Green functions. Interestingly, the result of this partial summation may outperform the exact $n$-th order approximation, see Fig.~\ref{fig:reflection_coeff}\,b) for an example. The exact $n$-th order approximation may be recovered by simply neglecting all terms of higher order than $n$ in the Kerr parameter $U$. Employing the dressed Green functions therefore removes the need to account for loop diagrams altogether, simplifying the task of finding all contributing diagrams.

\begin{figure}
    \centering
    \begin{tikzpicture}
       \pgfsetxvec{\pgfpoint{0.45cm}{0}}
\pgfsetyvec{\pgfpoint{0}{0.45cm}}
%% Plot
        \node[anchor=north west] at (-1,16.5) {\includegraphics[scale=0.9]{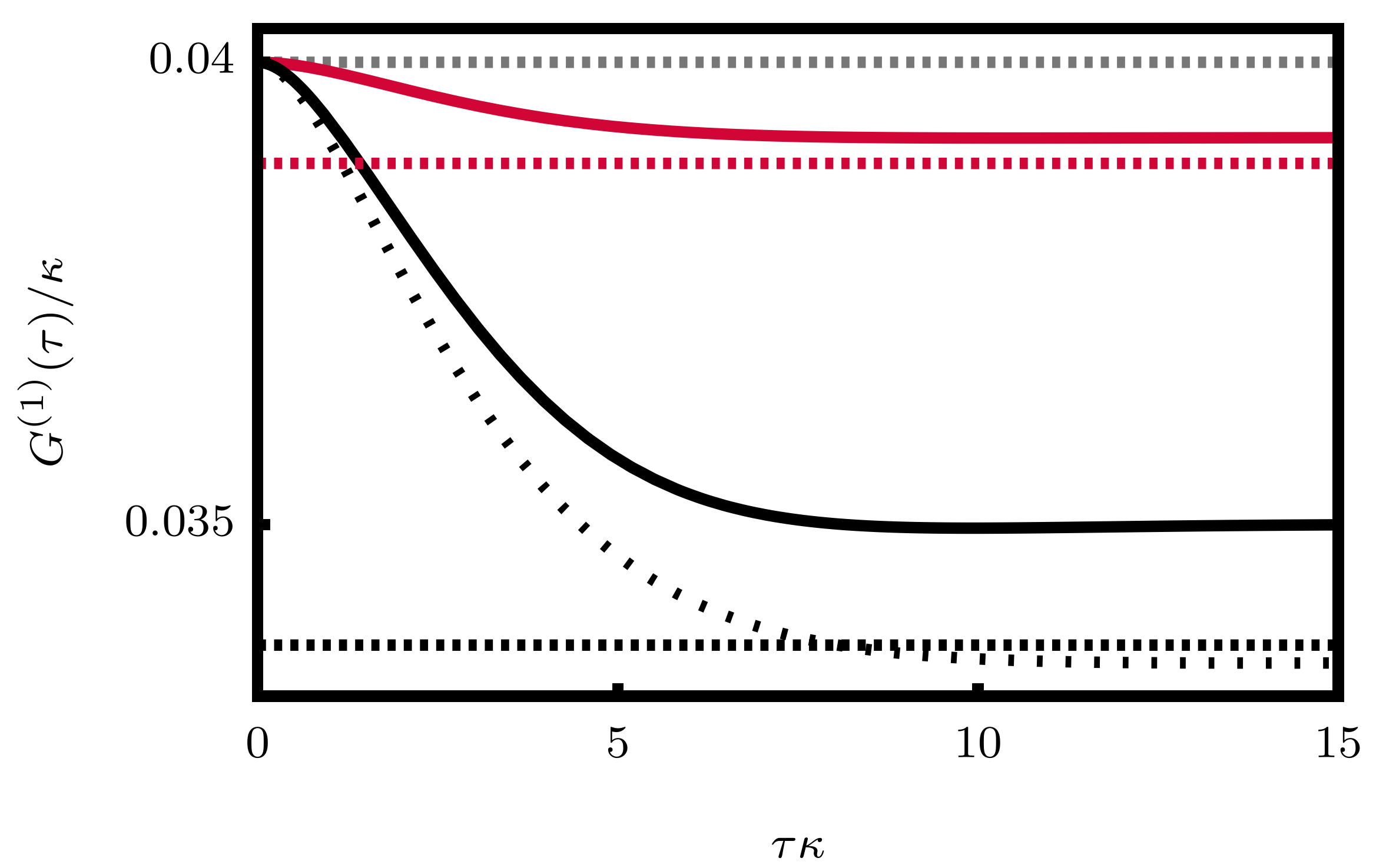}};
        \node at (-1,15.75) {a)};
%\draw[step=1,gray,very thin] (0,0) grid (38,10);
        %%%%%%%%%%%%%%%%%%%%%%%%%%%%%%%%%%%%%%%%%%%%%%
    \draw[unibas-red, very thick] (7,13) -- (8,13);
    \node at (12,13) {\scriptsize{Zero temp.}};
    % Dashed mint line
    \draw[black,very thick] (7,12) -- (8,12);
    \node at (12,12) { \scriptsize{Finite temp.}};
    % Dashed mint line
    \draw[ very thick,  dash pattern=on 1pt off 2.5pt ] (7,11) -- (8,11);
    \node at (12,11) { \scriptsize{Finite temp. loop sum.}};
    
    %%%%%%%%%%%%%%%%%%%%%%%%%%%%%%%%%%%%%%%%%%%
        \coordinate (V1) at (2,2); 
        \coordinate (V2) at (6,2); 
        \coordinate (V3) at (11,2); 
        \coordinate (V4) at (15,2); 
        \vertexcoordinates{V1}
        \vertexcoordinates{V2}
        \vertexcoordinates{V3}
        \vertexcoordinates{V4}

        %% First diagram 
        \node at (-1,4) {b)};
        \destroyinput[0.45]{V3TL}{V3}
        \destroyoutput[0.45]{V3BL}{V3}
        \Gretup[0.45]{V4}{V3}
        \Gadvdown[0.45]{V3}{V4}
        \createoutput[0.45]{V4BR}{V4}
        \createinput[0.45]{V4TR}{V4}
        
        %% Second diagram 
        \node at (11,4) {c)};
        \destroyinput[0.45]{V1TL}{V1}
        \destroyoutput[0.45]{V1BL}{V1}
        \Gadv[0.45]{V2}{V1}
        \destroyinput[0.45]{V1BR}{V1}
        \createoutputK[0.45]{V2BR}{V2}
        \createinput[0.45]{V2TR}{V2}
        \createinput[0.45]{V2TL}{V2}

\end{tikzpicture}
    \caption{ a) Perturbative $G^{(1)}$ function for the Kerr oscillator at zero temperature and at finite temperature, $n_B = 0.25$, for the second order approximation (full lines) and for the loop summation (\newtext{dashed} line) for $\sfrac{\Delta}{\kappa} = 0.3,~ \sfrac{\langle \bin \rangle}{\sqrt{\kappa}} = 0.2 $ and $\sfrac{U}{\kappa} = 0.1$. We plot the exact solution at $\tau = 0$ (gray dashed line) and the exact solutions for $\tau\rightarrow\infty$ at zero temperature (red dashed line) and finite temperature (black dashed line). b) \& c) Example diagrams that feature in the second order correction to the $G^{(1)}$ function.
    }
\label{fig:g1}
\end{figure}

\section{Output Field Statistics of the Kerr Oscillator}
\label{sec:kerr_output_statistics}

Following the procedure outlined in the previous section, we now turn to the output field of the coherently driven Kerr oscillator. For simplicity, we consider a drive that is time independent in a rotating frame, $f(t) = f$.  

Here we focus on four quantities: the reflection, the  $G^{(1)}$--function, the squeezing spectrum, and the $g^{(2)}$-- function. At zero temperature, we benchmark our results for equal-time correlators with the exact results of Ref.~\cite{PhysRevA.94.033841}. At finite temperatures, we provide numerical results to benchmark our results for equal-time correlators \cite{kramer2018quantumoptics}.

\subsection{Reflection}
The reflection amplitude is defined as
\begin{equation}
    \frac{\langle\bout\rangle}{\langle \bin\rangle} = \sqrt{R}e^{-i\theta}, \label{eq:reflection_coeff_def}
\end{equation}
with $R$ denoting the reflection probability. Having determined the average output field $\langle \bout \rangle$ to second order we can directly access the second order approximation of the reflection probability, see Fig.~\ref{fig:reflection_coeff}. In the limit of no Kerr nonlinearity, $U=0$, we find $R=1$, since the output light is just the reflected displaced thermal state provided by the input. For finite Kerr parameter $U$, the reflection probability decreases, since the output light is no longer a displaced thermal state. 
We note that all photons are still being reflected, $\langle \dbout\bout\rangle=\langle \dbin\bin\rangle$, hence no photons are being lost in this system. Instead, the Kerr nonlinearity reduces the average displacement of the output mode while squeezing it as shown below. The finite-temperature mean-field theory reproduces the intra--cavity field well in this regime, see Fig.~\ref{fig:reflection_coeff}\,c). The reflection probability obtained through this approach is however identically one for all parameters considered here. The mean--field approach fails to capture the effect of the Kerr--nonlinearity on the output field since the effect cannot be captured by a (self-consistent) shift in the detuning alone.  

For zero temperature, we can benchmark our results with the exact solution for the cavity mode from Ref.~\cite{PhysRevA.94.033841}. We observe that our approximate solution captures the reduced reflection well for small drive strength. However, it fails to capture the shift in the effective resonant frequency observed for larger drive strengths. At finite temperatures, we observe a significant reduction in the reflection probability, as shown in Fig.~\ref{fig:reflection_coeff}\,b), as well as a shift in the resonant frequency even for weak drives. While this shift is not captured by our second-order approximation, it is partly captured by the loop summation method at finite temperatures.

\subsection{First order coherence function}
We now turn to the first order coherence function of the output light, 
\begin{multline}
    G^{(1)}(\tau) = \langle \dbout(\tau) \bout(0) \rangle \\ 
    = \langle \! \langle \dbout(\tau) \bout(0) \rangle\!\rangle + \langle \dbout(\tau) \rangle \langle \bout(0) \rangle\ \fullstop
\end{multline}
Since all photons impinging on the cavity eventually have to be reflected, we find $G^{(1)}(0) = \langle \dbin(t)\bin(t)\rangle$.
In the long-time limit, $\langle \dbout(t) \bout(t') \rangle$ factorizes resulting in, 
\begin{align}
    G^{(1)}(\tau \to \infty) = |\langle \bout \rangle|^2 \comma  \label{eq:g1_longtimelimit}
\end{align}
which, at zero temperature, we may compare to the exact result from Ref.~\cite{PhysRevA.94.033841}. For finite temperature, we evaluate the right-hand side of Eq.~\eqref{eq:g1_longtimelimit} numerically.
Since the reflection is smaller than one, $R<1$, we find that $G^{(1)}(0)>G^{(1)}(\tau \to \infty)$. As a function of $\tau$, the $G^{(1)}$--function monotonously interpolates between these limits, see Fig.~\ref{fig:g1}. Example diagrams that contribute to the $G^{(1)}$--function are also shown in Fig.~\ref{fig:g1}. At finite temperature, we observe a marked reduction in the first-order coherence for long times, consistent with the lower reflection probability shown in Fig.~\ref{fig:reflection_coeff} b) .

\subsection{Squeezing spectrum}
We consider the squeezing spectrum as outlined in Ref.~\cite{PhysRevA.30.1386}. To this end we define the two Hermitian quadrature operators, 
\begin{equation}
\hat{X}_\pm(t) = \frac{1}{2}\qty(e^{-i \theta/2} \bout(t) \pm e^{i\theta/2} \dbout(t)) \fullstop 
\label{eq:quadrature_operators}
\end{equation}
The normal--ordered variances of these quadratures can be expressed through the cumulants of the output field as follows, 
\begin{align}
&\langle \! \langle \colon \hat{X}_{\pm}(t)\hat{X}_{\pm}(t') \colon \rangle \! \rangle =  \nonumber
    \\  &\,\frac{1}{2}\Re{e^{-i \theta} \langle \!\langle \bout(t) \bout(t') \rangle\!\rangle \pm \langle \!\langle \dbout(t) \bout(t') \rangle \!\rangle } \fullstop
\end{align}

%%%%%%%%%%%%%%%%%%%%%%%%%%%%%%%%%%%%%%%%%%%%%%%%%%%%%%%
\begin{figure}
    \centering
    \vspace{2mm}
    \begin{tikzpicture}
       \pgfsetxvec{\pgfpoint{0.45cm}{0}}
\pgfsetyvec{\pgfpoint{0}{0.45cm}}
%\draw[step=1,gray,very thin] (-1,-6) grid (9,5);
        \coordinate (V1) at (13,20.5); 
        \coordinate (V2) at (13,15.5); 
        \coordinate (V3) at (0,8); 
        \coordinate (V4) at (4,8); 
        \coordinate (V5) at (9,8); 
        \coordinate (V6) at (13,8); 
        
        \vertexcoordinates{V1}
        \vertexcoordinates{V2}
        \vertexcoordinates{V3}
        \vertexcoordinates{V4}
        \vertexcoordinates{V5}
        \vertexcoordinates{V6}
        
        %% First diagram 
        \node at (12.75,22.5) {b)};
        \destroyoutputK[0.45]{V1TL}{V1}
        \destroyoutput[0.45]{V1TR}{V1}
        \destroyinput[0.45]{V1BL}{V1}
        \destroyinput[0.45]{V1BR}{V1}
        %% Second diagram 
        \node at (12.75,17.5) {c)};
        \createoutputK[0.45]{V2TL}{V2}
        \createoutput[0.45]{V2TR}{V2}
        \createinput[0.45]{V2BL}{V2}
        \createinput[0.45]{V2BR}{V2}

        %% Third diagram 
        \node at (2,10) {d)};
        \destroyoutputK[0.45]{V3TL}{V3}
        \destroyoutput[0.45]{V3BL}{V3}
        \Gkelup[0.45]{V3}{V4}
        \Gretdown[0.45]{V3}{V4}
        \destroyinput[0.45]{V4BR}{V4}
        \destroyinput[0.45]{V4TR}{V4}

        %% Fourth diagram 
        \node at (9,10) {e)};
        \destroyinput[0.45]{V5TL}{V5}
        \destroyoutput[0.45]{V5BL}{V5}
        \destroyinput[0.45]{V5BR}{V5}
        \Gkel[0.45]{V6}{V5}
        \destroyinput[0.45]{V6TL}{V6}
        \createinput[0.45]{V6BR}{V6}
        \destroyoutput[0.45]{V6TR}{V6}
        
        %% Plot
        \node[anchor=north west] at (-2.5,23.5) {\includegraphics[scale=1]{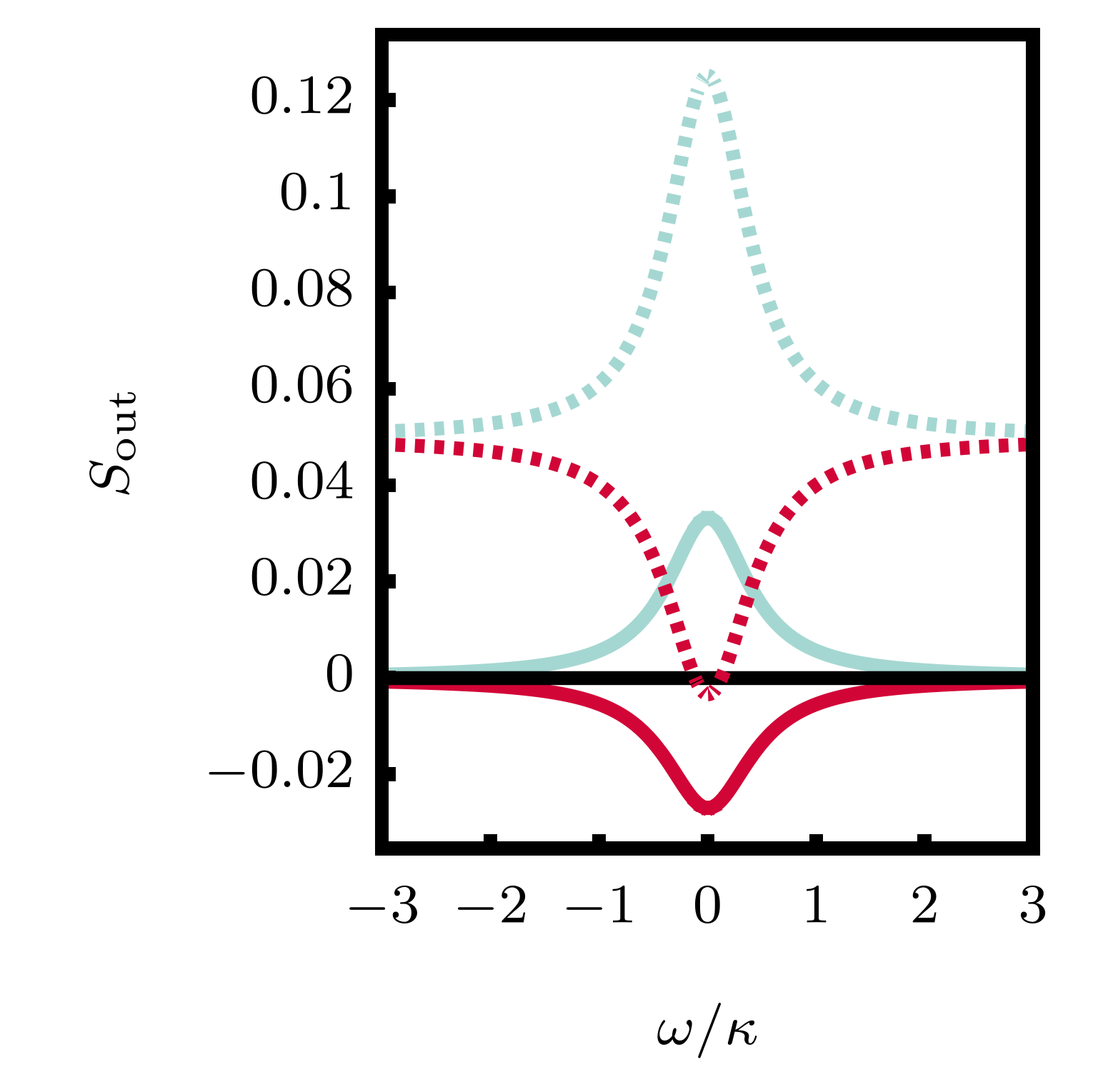}}; 
        \node at (-1.5,22.5) {a)};
\end{tikzpicture}
    \caption{a) perturbative squeezing spectrum of the Kerr oscillator for the $\hat{X}_+$ quadrature (red) and the $\hat{X}_-$ quadrature (mint) for $U/\kappa = 0.1,~\langle \bin \rangle/\sqrt{\kappa} = 0.2$, zero temperature (full lines) and $n_B = 0.25$ (dashed lines) and with $\theta = 1.49$ \newtext{and $\Delta/\kappa = 0.05$} where we find the maximal squeezing in the $\hat{X}_+ $ quadrature. b)-e) Example diagrams that contribute to the squeezing spectra. Diagram b) shows the only first order correction to the correlator $\langle \!\langle \bout \bout \rangle \!\rangle$ and diagram c) is the conjugated diagram for $\langle\!\langle \dbout \dbout \rangle \!\rangle$. Diagram d) shows a second order correction term to $\langle \!\langle \bout \bout \rangle \!\rangle$ quadratic in the input field while diagram e) is quartic in the input field. }
    \label{fig:squeezing_spectrum_diags}
\end{figure}

Since the cumulants are time-translationally invariant, we can determine the squeezing spectra of the quadratures as the Fourier transforms,
\begin{align}
    S_{\text{out},\pm}[\omega] = \int_{-\infty}^\infty \langle \! \langle \colon \hat{X}_{\pm}(\tau)\hat{X}_{\pm}(0) \colon \rangle \! \rangle e^{i \omega \tau } d\tau \fullstop
\end{align}
As illustrated in Fig.~\ref{fig:squeezing_spectrum_diags},
the output light of the Kerr oscillator is squeezed at zero temperature, for finite temperature squeezing is strongly reduced.

\subsection{Second order coherence function}
In order to quantify the bunching behavior of the output photons, we evaluate the normalized second order coherence function of the output light at zero temperature, 
\begin{equation}
    g^{(2)}(\tau)  = \frac{\langle \dbout(0) \dbout(\tau) \bout(\tau) \bout(0)\rangle}{ \langle \dbout(0) \bout(0) \rangle^2} \fullstop \label{eq:g2_def}
\end{equation}
The denominator is determined by $G^{(1)}(0) = \langle \dbin(0)\bin(0)\rangle$. The moment in the numerator can be expressed through cumulants of the output field as discussed in App.~\ref{sec:g2_through_cumulants} (see also Ref.~~\cite{Peccati2011}).  

\begin{figure}[t!]
    \centering
    \begin{tikzpicture}
       \pgfsetxvec{\pgfpoint{0.45cm}{0}}
\pgfsetyvec{\pgfpoint{0}{0.45cm}}
%\draw[step=1,gray,very thin] (0,0) grid (38,10);
        \coordinate (V1) at (2,2); 
        \coordinate (V2) at (6,2); 
        \coordinate (V3) at (11,2); 
        \coordinate (V4) at (15,2); 
        \vertexcoordinates{V1}
        \vertexcoordinates{V2}
        \vertexcoordinates{V3}
        \vertexcoordinates{V4}
        %% First diagram 
        \node at (-0.75,4) {c)};
        \destroyoutput[0.45]{V1BR}{V1}
        \createoutput[0.45]{V1BL}{V1}
        \createoutput[0.45]{V1TL}{V1}
        \Gkel[0.45]{V2}{V1}
        %\clloop[0.45]{V2}{-90}
        \createinput[0.45]{V2TL}{V2}
        \destroyinput[0.45]{V2BR}{V2}
        \destroyoutput[0.45]{V2TR}{V2}
        %% Second diagram 
        \node at (11,4) {d)};
        \createoutput[0.45]{V3BR}{V3}
        \destroyoutput[0.45]{V3BL}{V3}
        \destroyoutput[0.45]{V3TL}{V3}
        \Gkel[0.45]{V3}{V4}
        \createinput[0.45]{V4BR}{V4}
        \destroyinput[0.45]{V4TL}{V4}
        %\clloop[0.45]{V4}{-90}
        \createoutput[0.45]{V4TR}{V4}

        %% Plot
        \node[anchor=north west] at (-1.75,15) {\includegraphics[width=\linewidth]{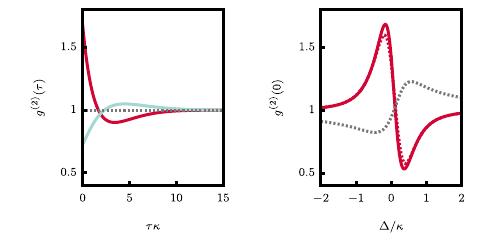}}; 
        \node at (-0.75,14) {a)}; 
        \node at (10,14) {b)}; 
\end{tikzpicture}
%\vspace{-2mm}
    \caption{a) Perturbative $g^{(2)}$ function of the Kerr oscillator as a function of time \newtext{at zero temperature and $\langle \bin \rangle / \sqrt{\kappa} = 0.2$} for $\Delta/\kappa = 0.2$ (mint) and $\Delta/\kappa = -0.2$ (red). b) Perturbative $g^{(2)}$ function at $\tau=0$ (full line) and the exact solution (dashed line) from Ref.~\cite{PhysRevA.94.033841}. The intra-cavity $g^{(2)}$ function (grey) behaves markedly different from the output field. c) \& d): Two diagrams that contribute to the second order correction of the cumulant $\llangle \dbout \dbout \bout \bout \rrangle$ that features in the $g^{(2)}$ function. Diagram b) can be found from diagram a) by reversing all arrows and filling (emptying) all filled (empty) symbols -- an operation that will always yield another valid diagram for a cumulant that contains the same number of creation and annihilation operators. }
    \label{fig:sout_plot}
\end{figure}

As illustrated in Fig.~\ref{fig:sout_plot}\,a), depending on the detuning of the coherent drive, the output light may be bunched or anti-bunched. Importantly, the $g^{(2)}$--function of the output light behaves markedly different from the intra-cavity $g^{(2)}$--function, see Fig.~\ref{fig:sout_plot}\,b).

%%%%%%%%%%%%%%%%%%%%%%%%%%%%%%%%%%%%%%%%%%%

% Here we put the derivation in the first submission: 
%\input{derivation}

\section{Conclusions and outlook}
We introduced a novel approach to input–output theory \cite{Gardiner}, based on the Schwinger--Keldysh path integral formalism \cite{Kamenev2011}. 
This approach allows for using the theoretical tools from path-integral approaches to directly compute the properties of the outgoing light, which can behave strikingly different from the light inside the cavity. 
We illustrate our formalism with a diagrammatic treatment of the Kerr nonlinear oscillator. Our approach allows for including finite temperatures in a straightforward way, by dressing the propagators of the theory. Furthermore, partial summation methods to infinite order in the Kerr nonlinearity are available, either based on the dressed propagators or on a mean-field approach.

Although the specific diagrammatic rules and pictorial representations used for the Kerr oscillator are system specific, the overall approach is general, and we expect the diagram rules to generalize straightforwardly to other models.

Our formalism can be extended to include squeezing, either through a squeezed bath or a parametric drive. A particularly interesting scenario is provided by the Kerr parametric oscillator, which exhibits non-classical behavior in the output field even though the cavity field is characterized by a non-negative Wigner function \cite{PhysRevResearch.3.023041}. Further extensions of our formalism include the generalization to few-level systems as well as spin and fermionic degrees of freedom embedded in the cavity. This may extend our formalism to the emerging field of cavity materials engineering ~\cite{Lu:25,Huebener2021,Mivehvar2021}.

\begin{acknowledgments}
We acknowledge helpful discussions with Peter Samuelsson and Brennan Hughes. 
This work was supported by the Swiss National Science Foundation (Eccellenza Professorial Fellowship PCEFP2\_194268). A.C. acknowledges support from the Simons Foundation through a Simons Investigator Award (Grant No.\,669487). A.D. acknowledges financial support by the QCQT PhD school. M.\,B.\ acknowledges funding from the European Research Council (ERC) under the European Union’s Horizon 2020 research and innovation program (Grant agreement No.\ 101002955 - CONQUER).
\end{acknowledgments}

\appendix 
\section{Path Integral Formalism}
\label{sec:Keldysh_Formalism}
In the following we give a short overview of the Schwinger-Keldysh formalism for non-equilibrium quantum field theory. We will stick closely to the exposition in Ref.~\cite{Kamenev2011}, recent developments and experimental applications of the formalism can be found in \cite{Sieberer_2016}. We follow the notation of Ref.~\cite{Kamenev2011} and therefore use a bar to denote complex conjugation. 
We start by introducing the coherent states of the cavity mode $\ahat$ and the bath modes $\hat{b}_k$,
\begin{align}
\label{eq:coherent_state}
    \ahat \ket{\phi} = \phi \ket{\phi},~~\hat{b}_k \ket{\varphi_k} = \varphi_k \ket{\varphi_k},
\end{align}
where the coherent states, 
\begin{align}
    \ket{\phi} &= \sum_{n=0}^\infty \frac{\phi^n}{\sqrt{n!}} \ket{n} = \sum_{n=0}^\infty \frac{\phi^n}{n!}\qty(\hat{a}^\dagger)^n \ket{0} = e^{\phi \hat{a}^\dagger}\ket{0}\label{eq:def_coherent_state},
\end{align}
are not normalized. This leads to the following identity for the inner product of two coherent states, 
\begin{align}
    \bra{\phi}\ket{\phi'} &= e^{\overline{\phi}\phi'} \label{eq:overlap_coherent_states}. 
\end{align}
Some heavily used identities for coherent states are mentioned in the following, 
\begin{align}
    \int d[\phi] e^{-|\phi|^2} \ket{\phi}\bra{\phi} &= \mathbb{1},\\ 
    d[\phi] :&= d(\Re{\phi})d(\Im{\phi})/\pi, \label{eq:complex_differential} \\
    \bra{\phi} H_S(\hat{a}^\dagger,\hat{a})\ket{\phi} &= H_S (\overline{\phi},\phi') \bra{\phi}\ket{\phi'}, \label{eq:hamilton_func_def}
\end{align}
where the last relation holds for any normally ordered operator $H_S(\hat{a}^\dagger,\hat{a})$.
To denote the tensor product of a state from the cavity and the states from the bath we introduce the following notation, 
\begin{align}
    \ket{\Psi} &= \ket{\phi} \bigotimes_k \ket{\varphi_k}\comma & d\qty[\Psi] &= d[\phi]\prod_k d[\varphi_k] \comma \\ 
    \Bar{\Psi}\Psi' &= \Bar{\phi}\phi'+\sum_k \Bar{\varphi}_k \varphi_k' \comma
\end{align}
with it we express the identity in the combined Hilbert space as follows, 
\begin{equation}
    \mathbb{1} = \int d\qty[\Psi] e^{-|\Psi|^2} \ket{\Psi}\bra{\Psi}.
    \label{eq:id_combined}
\end{equation}
As a consequence we compute the trace of an operator $\hat{O}$ as, 
\begin{equation}
    \Tr{\hat{O}} = \int d\qty[\Psi] e^{-|\Psi|^2} \bra{\Psi}\hat{O}\ket{\Psi}.
\end{equation}
As is customary in quantum field theory, we consider the partition function of our state, here given by the trace of the density matrix $\hat{\rho}$ at time $t$ \newtext{with arbitrary initial state $\hat{\rho}_0$},
\begin{equation}
    Z = \Tr{\hat{\rho}(t)} = \Tr{\hat{U}(t)\hat{\rho}_0\hat{U}^\dagger(t)} 
    = \Tr{\hat{\rho}_0} = 1,
\end{equation} 
where we used cyclic permutability of the trace, the time evolution operator \newtext{$\hat{U}(t) = e^{-i \hat{H} (t-t_0)} $ with $\hat{H}$ given in Eq.~\eqref{eq:general_setting_total_hamiltonian}} and the normalization condition. As a next step, we  write $\hat{U}(t)$ and its Hermitian conjugate as a product of time-evolution operators for the infinitesimal time $\hat{U}(\delta t)$. We then insert identities of the type from Eq. \eqref{eq:id_combined} in between these infinitesimal time evolutions resulting in,
\begin{widetext}
\begin{align}
    &Z = \int \mathcal{D}\qty[\Psi] e^{-\sum_{j=0}^{2N-1} |\Psi_j|^2} \bra{\Psi_{2N-1}} e^{i \delta t \hat{H}}\ket{\Psi_{2N-2}}\bra{\Psi_{2N-2}} e^{i \delta t \hat{H}}\ket{\Psi_{2N-3}}\cdots\\
    &\bra{\Psi_{N+1}}e^{i \delta t \hat{H}} \ket{\Psi_{N}} \bra{\Psi_N} \mathbb{1} \ket{\Psi_{N-1}}\bra{\Psi_{N-1}} e^{-i \delta t \hat{H}} \ket{\Psi_{N-2}}\cdots \bra{\Psi_1}e^{-i \delta t \hat{H}} \ket{\Psi_0}  \bra{\Psi_0} \hat{\rho}_0 \ket{\Psi_{2N-1}}.  \nonumber 
\end{align}

We now turn to evaluating the infinitesimal time evolution steps, 
\begin{align}
    \bra{\Psi_j}e^{-i \delta t_j \hat{H}} \ket{\Psi_{j-1}} &= \bra{\Psi_j} \qty(\mathbb{1} - i \delta t_j \hat{H} + \order{\delta t_j^2} ) \ket{\Psi_{j-1}}= e^{\Bar{\Psi}_j\Psi_{j-1}} - i \delta t_j H(\Bar{\Psi}_j,\Psi_{j-1})e^{\Bar{\Psi}_j\Psi_{j-1}}  + \order{\delta t_j^2} \nonumber \\
    &\approx e^{\Bar{\Psi}_j\Psi_{j-1}} e^{-i\delta t_j H(\Bar{\Psi}_j,\Psi_{j-1})} \comma  \label{eq:keldyhs_inf_step}
\end{align} 
\newtext{with $H(\Bar{\Psi}_j,\Psi_{j-1})$ defined analogously to Eq.~\eqref{eq:hamilton_func_def}.}
\end{widetext}
Here we introduced the time--step, 
\begin{align}
\delta t_j = \text{sign}(N-j) \delta t \fullstop
\end{align}
Notice that the last relation in Eq. \eqref{eq:keldyhs_inf_step} is exact in the limit $\delta t \to 0$. 
We further use the common notation for path integrals, 
\begin{align}
    \mathcal{D}\qty[\Psi] = \prod_{j=0}^{2N-1} d\qty[\Psi_j].
\end{align}
With this notation and the evaluated infinitesimal time step evolution, we can write the Keldysh partition function as a path integral, 
\begin{align}
       &Z = \int \mathcal{D}\qty[\Psi]e^{i S[\Psi]}, \label{eq:general_integral}\\
       &S[\Psi]  = \begin{aligned}[t]
       \sum_{j=1}^{2N-1} &\delta t_j \qty[i \Bar{\Psi}_j \frac{\Psi_j-\Psi_{j-1}}{\delta t_j}-H(\Bar{\Psi}_j,\Psi_{j-1})] \\ 
       &+ i \Bar{\Psi}_0 \Psi_0 -i \ln\rho_0(\Bar{\Psi}_0,\Psi_{2N-1}) \fullstop 
       \end{aligned}
       \label{eq:general_action}
\end{align}
The sum in the action can be interpreted as a (discrete) time integral along the closed-time contour sketched in Fig.~\ref{fig:contour}. Time first increases along the forward-in-time branch that arises from the time-evolution operator $\hat{U}(t)$. Time then decreases along the backward-in-time branch that comes from $\hat{U}^\dagger(t)$. \newtext{To distinguish between fields on these branches, we use plus superscripts to denote fields on the forward branch and minus superscripts for the backwards branch, 
\begin{align}
    \Psi_j^+ &= \Psi_j\comma & \Psi^-_j &= \Psi_{2N-1-j} & \text{for } j \in \qty{0,\dots,N} \fullstop
\end{align}
Using these new fields we can rewrite the action in Eq.~\eqref{eq:general_action},
\begin{align}
    S[\Psi] = \begin{aligned}[t] &\sum_{j=1}^{N-1} \delta t \qty[i \Bar{\Psi}_j^+\frac{\Psi_j^+-\Psi_{j-1}^+}{\delta t}-H(\Bar{\Psi}_j^+\Psi_{j-1}^+)] \\
    - &\sum_{j=1}^{N-1} \delta t \qty[i \Bar{\Psi}_j^- \frac{\Psi_j^- - \Psi_{j-1}^-}{\delta t} - H( \Bar{\Psi}_j^-\Psi_{j-1}^-)] \\
    + &i \Bar{\Psi}^+_0 \Psi^+_0 - i \ln \rho_0(\Bar{\Psi}^+_0,\Psi^-_{0}).\end{aligned}
    \label{eq:general_action_plusminusfields}
\end{align}
\subsection*{Example: Harmonic Oscillator}
We consider the action of a simple harmonic oscillator 
\begin{align}
    \hat{H}_S = \omega_S \dahat \ahat \comma
\end{align}
in the limit of zero coupling to the bath, i.e. $\kappa = 0$ and we evaluate the action according to \newtext{Eq.~\eqref{eq:general_action_plusminusfields}}
\begin{align}
    S[\phi] = \begin{aligned}[t] &\sum_{j=1}^{N-1} \delta t \qty[i \Bar{\phi}_j^+\frac{\phi_j^+-\phi_{j-1}^+}{\delta t}-\omega \Bar{\phi}_j^+\phi_{j-1}^+] 
    \\ 
    - &\sum_{j=1}^{N-1} \delta t \qty[i \Bar{\phi}_j^- \frac{\phi_j^- - \phi_{j-1}^-}{\delta t} - \omega \Bar{\phi}_{j-1}^-\phi_{j}^-] \\ 
    &+ i \Bar{\phi}^+_0 \phi^+_0 - i \ln \rho_S(\Bar{\phi}^+_{0},\phi^-_{N}) \fullstop
    \end{aligned}
\end{align}
As a notational shorthand, we can introduce the continuum limit and if we neglect the boundary terms, the action takes the following form
\begin{align}
    S[\phi] &= \begin{aligned}[t]\int_{t_0}^{t_N} dt &\left[\Bar{\phi}+(t) (i \partial_t -\omega ) \phi^+(t)\right. \\ &-\left. \Bar{\phi}^-(t) (i \partial_t - \omega) \phi^-(t)\right] \fullstop \end{aligned}
\end{align}
We now introduce the Keldysh rotation, i.e. the classical and quantum field, 
\begin{align}
    \phi^{\rm cl} &= \frac{1}{\sqrt{2}}\qty(\phi^+ + \phi^-)\comma & \phi^{\rm q} &= \frac{1}{\sqrt{2}}\qty(\phi^+ - \phi^-) \fullstop
\end{align}
Using these new fields, the action takes the following form
\begin{align}
    S[\phi] &= \begin{aligned}[t]\int_{t_0}^{t_N}dt &\left[\Bar{\phi}^{\text{q}}(t) (i \partial_t - \omega) \phi^{\text{cl}}(t) \right. \\ &\left.+ \Bar{\phi}^{\text{cl}}(t) (i \partial_t -\omega) \phi^{\text{q}}(t)\right] \fullstop \end{aligned}
\end{align}
}

\newtext{\subsection*{Dissipative Dynamics}
So far we have been modeling unitary dynamics of the system states $\phi$ governed by some Hamiltonian $\hat{H}_S$. In order to arrive at the dissipative dynamics we describe in the main text, we consider a joined state $\Psi$ that consists of a system state $\phi$ and the bath modes $\varphi_k$ and time--evolution governed by the Hamiltonian $\hat{H}$ in Eq.~\eqref{eq:general_setting_total_hamiltonian}. Since this Hamiltonian is bilinear in the bath creation and annihilation operators, we can perform the resulting Gaussian integral over the bath modes and arrive at a dissipative description of the system alone. With the initial state given by Eqs.~\eqref{eq:factorized_initial_state} and \eqref{eq:rhob}, we find  
\begin{align}
    Z = \int \mathcal{D}[\phi] e^{iS[\phi]} \comma
\end{align}
with the system action,
}

\begin{widetext}
\newtext{
\begin{align}
    S_0[\phi] &= \begin{aligned}[t] 
    \int dt &\bigg[\Bar{\phi}^{\text{q}}(t) \left(i \partial_t - \omega_S+i\frac{\kappa}{2}\right) \phi^{\text{cl}}(t) +    \Bar{\phi}^{\text{cl}}(t) \left(i \partial_t -\omega_S-i\frac{\kappa}{2}\right) \phi^{\text{q}}(t) + i \kappa  (2 n_B + 1) \Bar{\phi}^{\text{q}}(t) \phi^{\text{q}}(t)\bigg] 
    \end{aligned} \nonumber \\ 
 &= 
     \iint dt dt'  
    \bigg[\mqty(\Bar{\phi}^{\text{cl}}(t)\\ \Bar{\phi}^{\text{q}}(t))^T 
    \mqty(
    0 & \qty[G^{-1}]^A(t-t') \\ 
    \qty[G^{-1}]^R(t-t') & \qty[G^{-1}]^K(t-t'))
    \mqty(\phi^{\text{cl}}(t') \\ \phi^{\text{q}}(t'))\bigg] \fullstop
\end{align}
From this action we can read of the inverse Green functions, 
\begin{align}
    \qty[G^{-1}]^A(t-t') &= \qty(- \frac{i \kappa}{2} +\qty(i \partial_t - \omega_S))\delta(t-t') \comma &
    \qty[G^{-1}]^R(t-t') &= \qty(\frac{i \kappa}{2} + \qty(i \partial_t - \omega_S))\delta(t-t') \comma  \\  
    \qty[G^{-1}]^K(t-t') &= i \kappa(2n_B + 1)\delta(t-t') \fullstop
\end{align}
By Fourier transforming and solving for the inverse, we determine the Green functions in the frequency domain, 
\begin{align}
  G^A[\omega] &=  \frac{1}{\omega - \omega_S - \sfrac{i \kappa}{2}} \comma & 
  G^R[\omega] &= \frac{1}{\omega-\omega_S + \sfrac{i \kappa}{2}} \comma \\ 
  G^K[\omega] &= -G^R[\omega] \cdot \qty[G^{-1}]^K[\omega] \cdot G^A[\omega] \comma & 
     [G^{-1}]^K[\omega] &= i \kappa (2 n_B + 1) \fullstop
\end{align}
The Green functions in the time domain can now be found by a further Fourier transform, see Eq.~\eqref{eq:DHO_greens_functions_time}. 
}

\section{Conventional Input--Output Theory}
\label{sec:conventional_input_output_theory}
Conventional input--output theory describes the system and bath modes of the general setting sketched in Fig.~\ref{fig:cavity_sketch} through the use of Langevin equations for the bath modes. We start here by deriving the time-dependence of a single bath mode,
\begin{align}
    \dv{t} \hat{b}_k &= i \qty[\hat{H},\hat{b}_k] = i \sum_l \qty[\omega_k \hat{b}_l^\dagger \hat{b}_l + \qty(g_l \hat{a}^\dagger \hat{b}_k + g_k^* \hat{b}_k^\dagger \hat{a}), \hat{b}_k]  =- i \sum_l \qty( \delta_{l,k} \hat{b}_l + g_l^* \delta_{l,k} \hat{a}) = -i( \omega_k \hat{b}_k + g_k^* \hat{a}),
\end{align}
this differential equation yields the following solution,
\begin{align}
    \hat{b}_k(t) &= e^{-i \omega_k (t-t_0)} \hat{b}_k(t_0) - i g_k^* \int_{t_0}^t d \tau e^{-i \omega_k (t- \tau)}\hat{a}(\tau) = e^{-i \omega_k (t-t_N)} \hat{b}_k(t_N) - i g_k^* \int_{t_N}^t d \tau e^{-i \omega_k (t- \tau)}\hat{a}(\tau),
\end{align}
where we introduced the times $t_0,t_N$ which are supposed to lie in the far past, far future respectively. 
Now we turn to the time-dependence of the system mode $\hat{a}$,
\begin{align}
    \dv{t} \hat{a} = i \qty[\hat{H}_S + \hat{H}_B + \hat{V},\hat{a}] = i \qty[\hat{H}_S, \hat{a}] + i \qty[\hat{H}_B ,\hat{a}] + i \qty[\hat{V},\hat{a}].
\end{align}
The second term in the equation above vanishes since the bath modes commute with the system mode.
The third term reads, 
\begin{align}
    i \qty[\hat{V},\hat{a}] &=i \sum_k \qty[g_k \hat{a}^\dagger \hat{b}_k + g_k^* \hat{b}_k^\dagger \hat{a},\hat{a}] = -i \sum_k g_k \hat{b}_k =-\sum_k \qty(i g_k e^{-i \omega_k (t-t_0)} \hat{b}_k(t_0) + |g_k|^2 \int_{t_0}^t d \tau e^{i \omega_k (t- \tau)}\hat{a}(\tau))  \label{eq:third_term_eom_a} \\
                            &= -\sqrt{\kappa} \frac{1}{\sqrt{\kappa}} \sum_k i g_k e^{-i \omega_k (t-t_0)} \hat{b}_k(t_0) - \sum_k |g_k|^2 \int_{t_0}^t d \tau e^{i \omega_k (t- \tau)}\hat{a}(\tau),\nonumber
\end{align}
where we introduced the coupling constant $\kappa$ in the last step. We further introduce the input and output operators,
\begin{align}
    \bin(t) &= \frac{1}{\sqrt{\kappa}} \sum_k i g_k e^{-i \omega_k (t-t_0)} \hat{b}_k(t_0), &
    \bout(t) &= \frac{1}{\sqrt{\kappa}} \sum_k i g_k e^{i \omega_k (t_N-t)} \hat{b}_k(t_N), \label{eq:def_input_output}
\end{align}
as well as the bath spectral density, 
\begin{equation}
    \rho(\omega) = \sum_k |g_k|^2 \delta(\omega-\omega_k).
    \label{eq:bath_spectral_density}
\end{equation}
Using the last two quantities, we can express Eq. \eqref{eq:third_term_eom_a} as follows, 
\begin{equation}
    i \qty[\hat{V},\hat{a}] = - \sqrt{\kappa} \bin(t) - \int_{t_0}^{t} d\tau \int_{0}^{\infty} d\omega \, \rho(\omega)e^{-i \omega (t-\tau)} \hat{a}(\tau).
\end{equation}
With these simplifications the equation of motion of the bath mode $\hat{a}$ is,
\begin{equation}
    \dv{t} \hat{a} = i \qty[\hat{H}_S,\hat{a}] - \sqrt{\kappa} \bin(t) - \int_{t_0}^{t} d\tau  \int_0^{\infty} d\omega \, \rho(\omega)e^{-i \omega(t-\tau)} \hat{a}(\tau).
    \label{eq:eom_a_in}
\end{equation}
Analogously we find, 
\begin{equation}
    \dv{t} \hat{a} = i \qty[\hat{H}_S,\hat{a}] - \sqrt{\kappa} \bout(t) + \int_t^{t_N} d\tau  \int_0^{\infty} d\omega \, \rho(\omega)e^{-i \omega(t-\tau)} \hat{a}(\tau).
    \label{eq:eom_a_out}
\end{equation}
Even though at this stage our equations only contain the input and output operators, we can retrieve the bath modes at the times $t_0$ and $t_N$ from the input--output operators, 
\begin{align}
    \hat{b}_k(t_0) &= \frac{\sqrt{\kappa}}{2 \pi i g_k } \int_{\omega_k - \delta\omega/2}^{\omega_k + \delta\omega/2} d\omega \int_{-\infty}^{\infty} dt\, e^{i \omega(t-t_0)} \bin(t) \comma &
    \hat{b}_k(t_N) &= \frac{\sqrt{\kappa}}{2 \pi i g_k} \int_{\omega_k - \delta\omega/2}^{\omega_k + \delta\omega/2} d\omega \int_{-\infty}^{\infty} dt\, e^{-i \omega(t_N-t)} \bout(t) \fullstop \label{eq:bkN_given_out}
\end{align}
From the equations of motion for the system mode $\hat{a}$, Eqs.~\eqref{eq:eom_a_out},~ \eqref{eq:eom_a_in}, we get the following relation between the input and output operators, 
\begin{align}
    \bout(t) = \bin(t) + \frac{1}{\sqrt{\kappa}}\int_{t_0}^{t_N} d\tau \int_0^\infty d\omega\,\rho(\omega) e^{-i \omega (t-\tau)} \hat{a}(\tau),\label{eq:inoutrel_1}
\end{align}
and we can derive their commutation relations, 
\begin{align}
    \qty[\bin(t),\hat{b}_{\text{in}}^\dagger(t')] = \qty[\bout(t),\hat{b}^{\dagger}_{\text{out}}(t')] = \frac{1}{\kappa} \int_0^\infty d\omega\,\rho(\omega) e^{-i \omega(t-t')}.
\end{align}
So far we treated time as a continuous quantity and made no assumptions on the spectrum of the bath. \newtext{In order to connect to the path--integral approach to input--output theory we move to a discretized version of the theory. To that end, we make the following construction: (i) time is discretized such that $t_j = t_0 + j \delta t$ with $j = 0,1,2,\dots, N$, (ii) the bath modes are discrete and spaced according to $\omega_k = m(k) \delta \omega$ with $m(k)$ being some integer--valued function of $k$. We note that this is always possible by choosing a sufficiently small $\delta\omega$. Finally (iii), the timestep $\delta t$ is connected to the discretization of the bath modes such that $\delta t \delta \omega = \frac{2 \pi}{N}$. This last condition is important to preserve the identity 
\begin{align}
\label{eq:ident}
    \int_{\omega_k-\delta \omega/2}^{\omega_k + \delta \omega/2}  d\omega\int dt  e^{-i(\omega-\omega_l)t }  = 2\pi  \delta_{k,l} \fullstop
\end{align}
After discretization, we have
\begin{equation}
    \int_{\omega_k-\delta\omega/2}^{\omega_k+\delta\omega/2} d\omega\sum_{j=0}^N\delta t e^{-i(\omega-\omega_l)t_j} = \delta t\delta\omega\sum_{j=0}^Ne^{-i\delta t\delta\omega j(m(k)-m(l))}=2\pi\delta_{m(k),m(l)} = 2\pi\delta_{k,l}.
\end{equation}
In the first equality, we performed the integral over $\omega$ and took the limit $\delta\omega t_j\rightarrow 0$. In the second equality we employed $\delta t\delta \omega=2\pi/N$ and used the identity
\begin{equation}
    \frac{1}{N}\sum_{j=0}^Ne^{-i\frac{2\pi}{N}j(m-n)}=\delta_{m,n}.
\end{equation}
Finally, we note that $\delta_{m(k),m(l)} = \delta_{k,l}$ by assuming $\omega_k\neq\omega_l$ for all $k\neq l$.
Enforcing $\delta t\delta\omega = 2\pi/N$ thus preserves the identity in Eq.~\eqref{eq:ident} upon discretization.
This relation is crucial to employ Eq.~\eqref{eq:bkN_given_out} in the discretized version of the theory. We note however that the time--step $\delta t$ and the frequency spacing $\delta\omega$ do not have physical relevance. 

}
With these assumptions, the bath modes expressed through the input and output operators from Eq.~\eqref{eq:def_input_output} can be written as follows, 
\begin{align}
    \hat{b}_k(t_0) &= \frac{\sqrt{\kappa}}{i g_k } \sum_{j=0}^{N} \, e^{i \omega_k(t_j-t_0)} \hat{b}_{\text{in}}(t_j) \comma & 
    \hat{b}_k(t_N) &= \frac{\sqrt{\kappa}}{i g_k}  \sum_{j=0}^{N} \, e^{-i \omega_k(t_N-t_j)} \hat{b}_{\text{out}}(t_j) \fullstop \label{eq:bkN_given_out_discrete}
\end{align} 
Additionally to the discretization outlined above, we now make the following two approximations: (i) The Fourier transform of $\hat{a}(\tau)$ is peaked at a specific frequency $\Omega$, where the peak has a fixed width $\gamma$, (ii) the spectral density $\rho(\omega)$ is flat in frequency and around $\Omega$ can be approximated as follows (Markov approximation), 
    \begin{equation}
        \rho(\omega) \approx \rho(\Omega)~~\forall\, \omega \in (\Omega-\gamma, \Omega + \gamma).
    \end{equation}
These assumptions let us simplify the third term in the relation between input and output operators from Eq. \eqref{eq:inoutrel_1} as follows, 
\begin{align}
    \int_{t_0}^{t_N} d\tau \int_0^\infty d\omega\,\rho(\omega) e^{-i \omega (t-\tau)} \hat{a}(\tau) &\approx \rho(\Omega) \int_{t_0}^{t_N} d\tau \int_{-\infty}^{\infty} d\omega e^{-i \omega (t-\tau)} \hat{a}(\tau) & 
    &= \rho(\Omega) \int_{t_0}^{t_N} d\tau  2 \pi \delta(t-\tau) \hat{a}(\tau) = \underbrace{2\pi \rho(\Omega)}_{=:\kappa} \hat{a}(t),
\end{align}
where we extended the range of integration in frequency in the first step. The extension of the integration range is justified since only nearly resonant frequencies are important here and the nonphysical, negative frequencies are negligible. 
Equivalently, using these approximations
we find, 
\begin{align}
    \int_{t_0}^{t} d\tau \int_0^\infty d\omega\,\rho(\omega) e^{-i \omega (t-\tau)} \hat{a}(\tau) &\approx  \frac{\kappa}{2} \hat{a}(t).
\end{align}
With these approximations we can express the equation of motion for the system mode $\hat{a}$ from Eqs. \eqref{eq:eom_a_in} and \eqref{eq:eom_a_out} as follows, see Eq.~\ref{eq:eom_a_out},  
\begin{align}
    \dv{t} \hat{a} &= i \qty[ \hat{H}_S, \hat{a}] - \frac{\kappa}{2} \hat{a} - \sqrt{\kappa} \bin(t) \comma & 
    \dv{t} \hat{a} &= i \qty[ \hat{H}_S, \hat{a}] + \frac{\kappa}{2} \hat{a} - \sqrt{\kappa} \bout(t) \comma 
\end{align}
and get the simplified commutation relations, 
\begin{align}
    \qty[ \bin(t), \hat{b}_{\text{in}}^\dagger(t')] = \qty[\bout(t),\hat{b}_{\text{out}}^\dagger(t')] = \delta(t-t'),
\end{align}
as well as the input--output relation in Eq.~\eqref{eq:in_out_rel}. In this relation we can interpret the outgoing mode as the reflection of the input mode together with an additional signal originating from the cavity. 
We can further simplify our expressions by assuming that all coupling constants are equal, $i g_k := g \in \mathbb{R}$,
and by rewriting the approximated spectral density, $
    \rho(\omega) \approx \rho(\Omega) = g^2 \rho.$
With these simplifications we can write the discretized input-output operators as follows, 
\begin{align}
    \bin(t) &= \frac{1}{\sqrt{2 \pi \rho}} \sum_k e^{-i \omega_k (t-t_0)} \hat{b}_k(t_0) \comma &
    \bout(t) &= \frac{1}{\sqrt{2 \pi \rho}} \sum_k e^{i \omega_k (t_N-t)} \hat{b}_k(t_N) \fullstop
\end{align}
From these expressions we can intuitively understand the input operator as the bath operators freely evolved in time (i.e. in the absence of the coupling $\hat{V}$) from the distant past up until the time $t$. This free evolution reflects the fact that the input cannot be affected by the system, as it describes the bath modes before interacting with the system. Equivalently, the output operator describes the bath operators in the far future, i.e., after interacting with the system, freely evolved back in time up until the time $t$. 
Further details on the conventional input--output formalism can be found in \cite{Gardiner}, \cite{RevModPhys.82.1155} and \cite{walls_book}.
\section{Linear cavity}
\label{sec:harmonic_oscillator}
In this appendix we provide additional information for the treatment of the linear cavity as described in Sec.~\ref{sec:HO_Keldysh}. With the Green functions derived in App.~\ref{sec:Keldysh_Formalism} we can evaluate the moment generating function, 
\begin{align}
    \Lambda_{\text{out}}^0[\chi,\chi'] = e^{i\mathcal{S}_{\rm out}^0[\chi,\chi']} \fullstop
\end{align}
If we let our dynamics start in the far past, $t_0 \to -\infty$, and end in the far future, $t_N \to \infty$, we find,
\begin{align}
    \mathcal{S}_{\rm out}^0[\chi,\chi'] &= -\int dt \int dt' \,  \kappa \qty(\chi(t) G^R(t-t') f(t') - \Bar{f}(t) G^A(t-t')\chi'(t'))  - \int dt \, \qty(i\chi(t) f(t) + i\chi'(t) \Bar{f}(t)+ \chi(t) \chi'(t) n_B). \label{eq:evaluated_action_simple_prob_time_space} 
\end{align}
From the MGF we now have access to the statistics of the output field. We start by calculating the first moment of the output field and its hermitian conjugate, 
\begin{align}
    \langle \hat{b}_{\text{out}}(t) \rangle &= i \left.\frac{\delta \Lambda_{\text{out}}^\text{HO}[\chi,\chi']}{\delta \chi(t)} \right|_{\chi=\chi'=0} = f(t) - i \kappa \int dt'\,f(t') G^R(t-t') \comma \\ 
    \langle \hat{b}_{\text{out}}^\dagger(t) \rangle &= \overline{\langle\hat{b}_{\text{out}}(t)\rangle} = \Bar{f}(t) + i \kappa \int dt'\,\Bar{f}(t') G^A(t'-t) \fullstop
\end{align}
Through these averages we can rewrite the CGF in a concise manner,  
\begin{align}
\label{eq:cumgenapp}
    \mathcal{S}_{\text{out}}^0[\chi,\chi'] = - \int dt \qty(i\chi(t) \langle \hat{b}_{\text{out}}(t) \rangle+ i\chi'(t)\langle \hat{b}_{\text{out}}^\dagger(t) \rangle +  n_B \chi(t) \chi'(t)) \fullstop
\end{align}
From either of these expressions it is now straightforward to compute further moments of the output field. Furthermore, From Eq.~\eqref{eq:cumgenapp} we see that only the first two cumulants are nonvanishing.

\section{Accessing the Output Statistics in Frequency Space}
\label{sec:KIO_in_frequency_space}
We shortly describe how one can access the statistics of the output field in the frequency domain. This is achieved by rewriting the action of the generating functional in the time domain, see Eq.~\eqref{eq:full_general_action}, and a redefinition of the source fields such that functional derivatives of the generating functional yields the moments of the operators $\bin[\omega]$ and $\dbin[-\omega]$.  
We make use of the following identity 
\begin{align}
    \int_{-\infty}^\infty dt f(t) g(t)  = \frac{1}{2 \pi} \int_{-\infty}^\infty d \omega f[\omega] g[-\omega] = \frac{1}{2 \pi} \int_{-\infty}^\infty d\omega f[-\omega] g[\omega] \label{eq:prod_id} \comma 
\end{align}
where we introduced the Fourier transform $f[\omega]$ of a function $f(t)$ and we employ the following convention, 
\begin{align}
    f[\omega] &= \int_{-\infty}^\infty dtf(t) e^{i\omega t} \comma & 
    \overline{f}[\omega] &= \int_{-\infty}^\infty dt\overline{f}(t) e^{i \omega t} \fullstop
\end{align}
Note that this implies the following relations for functions and operators, 
\begin{align}
    \overline{\qty(f[\omega])} &= \overline{f}[-\omega] \comma & \hat{f}^\dagger[\omega] &= \qty(\hat f[-\omega])^\dagger \fullstop \label{eq:fourier_transform_cc_relation}
\end{align}
We extend the integration range, i.e. the initial and final time of our evolution to plus and minus infinity respectively and rewrite the expression of the input--output action through the Fourier transforms of all appearing functions.
\begin{align}
    S[\phi,\chi,\chi'] =&~ S_S[\phi] + \frac{i \kappa}{2\pi} \int_{-\infty}^\infty d\omega \qty[(2n_B + 1) \Bar{\phi}^\text{q}[-\omega] \phi^\text{q}[\omega] - \frac{1}{2} \Bar{\phi}^\text{cl}[-\omega] \phi^\text{q}[\omega] + \frac{1}{2} \Bar{\phi}^\text{q}[-\omega]\phi^\text{cl}[\omega]] \nonumber \\
    &+ \frac{i \sqrt{2 \kappa}}{2\pi} \int_{-\infty}^{\infty} d \omega \qty[f[\omega] \Bar{\phi}^{\text{q}}[-\omega] - \Bar{f}[-\omega] \phi^{\text{q}}[\omega]] \nonumber\\
      &- \int_{-\infty}^{\infty} d \omega  \chi[-\omega] \qty[f[\omega] + \sqrt{\frac{\kappa}{2}}\qty(2 n_B +1)\phi^\text{q}[\omega] + \sqrt{\frac{\kappa}{2}}\phi^{\text{cl}}[\omega]] \\
    &- \int_{-\infty}^{\infty} d \omega  \chi'[\omega] \qty[\Bar{f}[-\omega] - \sqrt{\frac{\kappa}{2}}\qty(2 n_B + 1) \Bar{\phi}^{\text{q}}[-\omega] + \sqrt{\frac{\kappa}{2}} \Bar{\phi}^{\text{cl}}[\omega]] \nonumber \\
    &+ i \int_{-\infty}^{\infty} d \omega  \chi[-\omega] \chi'[\omega] n_B.\nonumber
\end{align}
In order to understand how we can access the statistics of the output field in frequency space, we revisit the introduction of the source fields and re express them through their Fourier transforms, 
\begin{align}
    \Lambda_{\text{out}} [\chi,\chi']   
    &= \int \mathcal{D}[\phi,\varphi_{\text{in}},\varphi_{\text{out}}] e^{i S^{\text{io}[\phi,\varphi_{\text{in}},\varphi_{\text{out}}]}}  \times e^{-i \int_{t_0}^{t_N} dt \qty[\chi(t) \varphi_{\text{out}}^+(t) + \chi'(t) \Bar{\varphi}_{\text{out}}^-(t)]} \nonumber  \\ 
    &= \int \mathcal{D}[\phi,\varphi_{\text{in}},\varphi_{\text{out}}] e^{i S^{\text{io}[\phi,\varphi_{\text{in}},\varphi_{\text{out}}]}}  \times e^{-\frac{i}{2\pi} \int_{-\infty}^{\infty} d\omega \qty[\chi[-\omega] \varphi_{\text{out}}^+[\omega] + \chi'[\omega] \Bar{\varphi}_{\text{out}}^-[-\omega]]} \fullstop
\end{align}
We now relabel the source fields as follows, 
\begin{align}
    \xi[\omega] &= \frac{\chi[-\omega]}{2\pi}\comma &  \xi'[\omega] &= \frac{\chi'[\omega]}{2\pi} \comma
\end{align}
since this will guarantee that we evaluate operator averages of $\bout$ at $\omega$ and operators averages of $\dbout$ at $-\omega$ which in our convention of the Fourier transform, \cite{RevModPhys.82.1155}, results in the averages being complex conjugates of each other, 
\begin{align}
  i \left. \frac{\delta \Lambda_\text{out}[\xi,\xi']}{\delta \xi[\Omega]} \right|_{\xi=\xi'=0} &= i \int \mathcal{D}[\phi,\varphi_{\text{in}},\varphi_{\text{out}}] (-i) \varphi_{\text{out}}^+[\Omega] e^{i S^{\text{io}[\phi,\varphi_{\text{in}},\varphi_{\text{out}}]}} = \langle \bout[\Omega] \rangle \comma \\ 
   i \left. \frac{\delta \Lambda_\text{out}[\xi,\xi']}{\delta \xi'[\Omega]} \right|_{\xi=\xi'=0} &= i \int \mathcal{D}[\phi,\varphi_{\text{in}},\varphi_{\text{out}}] (-i) \Bar{\varphi}_{\text{out}}^-[-\Omega] e^{i S^{\text{io}[\phi,\varphi_{\text{in}},\varphi_{\text{out}}]}} = \langle \dbout[-\Omega] \rangle =\overline{\qty( \langle \bout[\Omega] \rangle)} \fullstop
\end{align}
More generally, 
\begin{align}
   \left. \qty(\prod_{l=1}^N i \frac{\delta }{\delta \xi[\omega_l]}) \qty(\prod_{p=1}^M i \frac{\delta }{\delta \xi'[\omega_p]})\Lambda_{\text{out}}\qty[\xi,\xi'] \right|_{\xi = \xi'=0}  
   = \Biggl \langle \qty[\prod_p \hat{b}^\dagger_{\text{out}}[-\omega_p]] \qty[\prod_l \hat{b}_{\text{out}}[\omega_l]]\Biggr \rangle \fullstop
\end{align}

In the specific case of the linear cavity this reformulation leads to the following MGF and CGF, 
\begin{align}
    \Lambda_\text{out}^0[\xi,\xi'] &= e^{\mathcal{S}_\text{out}^0[\xi,\xi']} \comma  \\ 
    \mathcal{S}^0_\text{out}[\xi,\xi'] &= \begin{aligned}[t]
    -i \int d\omega &(\xi[\omega]f[\omega]\qty(1-i\kappa G^R[\omega])  
    +\xi'[\omega]\Bar{f}[-\omega]\qty(1+i\kappa G^A[\omega])
    -i n_B \xi[\omega]\xi'[-\omega]) \fullstop
    \end{aligned}
\end{align}
We then find the following cumulants for the output field statistics in frequency space, 
\begin{align}
    \langle\!\langle \bout[\omega] \rangle\!\rangle &= f[\omega](1-i \kappa G^R[\omega]) \comma & \langle\!\langle \dbout[-\omega] \rangle\!\rangle &= \Bar{f}[-\omega]\qty(1 + i \kappa G^A[\omega])\comma &  
    \langle\!\langle \dbout[-\omega]\bout[\omega']\rangle\!\rangle &=n_B \delta(\omega+\omega') \fullstop
\end{align}
\section{Interaction action}
\label{sec:Kerr_oscillator_details}
We consider a general interaction term in the Hamiltonian,
\begin{align}
    H_\mathrm{int} = \sum_{\alpha,\beta} \lambda_{\alpha,\beta} (\dahat)^\alpha \ahat^\beta \fullstop
\end{align} 
This leads to the following interaction term in the action, given here in the discrete version to show the time-ordering,
\begin{align}
\label{eq:actiongen}
    S_\mathrm{int}[\phi] &= \sum_{j=1}^{N-1} \delta t \sum_{\alpha,\beta} \lambda_{\alpha,\beta} \qty(\qty(\Bar{\phi}^{-}_{j-1})^\alpha \qty(\phi^-_j)^\beta- \qty({\Bar{\phi}^+}_{j})^\alpha \qty(\phi^+_{j-1})^\beta) \fullstop
\end{align} 
In the continuum limit, the term takes the new form
\begin{align}
    S_\mathrm{int}[\phi] = \sum_{\alpha,\beta} \lambda_{\alpha,\beta} \int dt \frac{1}{2^{\frac{\alpha+\beta}{2}}} \qty[\qty(\Bar{\phi}^\mathrm{cl}(t)-\Bar{\phi}^\mathrm{q}(t))^\alpha \qty(\phi^\mathrm{cl}(t) -\phi^\mathrm{q}(t))^\beta-\qty(\Bar{\phi}^\mathrm{cl}(t) + \Bar{\phi}^\mathrm{q}(t))^\alpha\qty(\phi^\mathrm{cl}+\phi^\mathrm{q})^\beta]  \comma \label{eq:interaction_action_cl_q_fields}
\end{align} 
For the Kerr interaction from Eq.~\eqref{eq:Kerr_Hamiltonian}, the interaction action in Eq.~\eqref{eq:interaction_action_cl_q_fields} can be expressed as follows, 
\begin{align}
    S_\mathrm{int}[\phi] &= -U \int dt \qty[\Bar{\phi}^\mathrm{cl} \Bar{\phi}^\mathrm{q} \phi^\mathrm{cl} \phi^\mathrm{cl} + \Bar{\phi}^\mathrm{cl} \Bar{\phi}^\mathrm{q} \phi^\mathrm{q} \phi^\mathrm{q} + \mathrm{c.c}] 
    = -U \int dt \qty[\qty(\Bar{\phi}^\mathrm{cl} \phi^\mathrm{cl} + \Bar{\phi}^\mathrm{q} \phi^\mathrm{q}) \qty(\Bar{\phi}^\mathrm{q}(t) \phi^\mathrm{cl}(t) + \Bar{\phi}^\mathrm{cl}(t) \phi^\mathrm{q}(t))] \fullstop \label{eq:interaction_action_simple_form}
\end{align}
\end{widetext}

\section{Shifted System Fields}
\label{sec:shift}
In order to achieve the factorization of the MGF in Eq.~\eqref{eq:moment_generating_functional_factorization}, we introduced the shift to the integration variables in Eq.~\eqref{eq:shift}.
\begin{align}
    \delta \phi^{\text{cl}} &= \phi^\text{cl} - \langle \phi^\text{cl} \rangle[\chi,\chi'] \comma& \delta \phi^\text{q} &= \phi^\text{q} - \langle \phi^\text{q} \rangle[\chi,\chi'] \comma \nonumber \\
   \delta  \overline{\phi^{\text{cl}}} &= \overline{\phi^\text{cl}} - \langle \overline{\phi^\text{cl}} \rangle[\chi,\chi'] \comma& \delta \overline{\phi^\text{q}} &= \overline{\phi^\text{q}} - \langle \overline{\phi^\text{q}} \rangle[\chi,\chi'] \fullstop \nonumber
\end{align}
We note that $\phi$ and $\Bar{\phi}$ are shifted independently of each other since taking the expectation value and complex conjugation do not commute here, see Eqs.~\eqref{eq:phiclavgwithsourcefield},\eqref{eq:phibarclavgwithsourcefield}, 
\begin{align}
\overline{\langle \phi^\text{cl} \rangle[\chi,\chi']} \neq \langle \overline{\phi^\text{cl}} \rangle[\chi,\chi'] \comma  
\end{align}
and Eqs.~\eqref{eq:phiqavg},\eqref{eq:phibarqavg},
\begin{align}
    \overline{\langle \phi^\text{q} \rangle[\chi,\chi']} \neq \langle \overline{\phi^\text{q}} \rangle[\chi,\chi'] \fullstop
\end{align}
Nevertheless, as we now show, we may replace integration over $\phi$ by integration over $\delta\phi$, i.e.,
\begin{equation}
\label{eq:shiftint}
    \int \mathcal{D}[\phi] f[\phi,\overline{\phi}] = \int \mathcal{D}[\delta \phi]f[\delta\phi+\langle \phi\rangle,\delta\overline{\phi}+\langle\overline{\phi}\rangle],
\end{equation}
even when $\langle \phi\rangle$ and $\langle \overline{\phi}\rangle$ are not related by complex conjugation.
To this end, we first note that
\begin{align}
\label{eq:pathd}
    \int \mathcal{D}[\phi] = \prod_{j=0}^{N-1}&\frac{1}{\pi}\int_{-\infty}^\infty d(\Re{\phi^{\rm cl}})\int_{-\infty}^\infty d(\Im{\phi^{\rm cl}})\\\nonumber\times&\frac{1}{\pi}\int_{-\infty}^\infty d(\Re{\phi^{\rm q}})\int_{-\infty}^\infty d(\Im{\phi^{\rm q}}).
\end{align}
We may now write 
\begin{equation}
    \Re{\phi^{\rm cl}} = \frac{1}{2}\left(\delta\phi^{\rm cl}+\delta  \overline{\phi^{\text{cl}}}+\langle \phi^\text{cl} \rangle[\chi,\chi']+\langle\overline{\phi^\text{cl}} \rangle[\chi,\chi'] \right).
\end{equation}
Identifying $x = (\delta\phi^{\rm cl}+\delta  \overline{\phi^{\text{cl}}})/2$ , the first integral on the right-hand side in Eq.~\eqref{eq:pathd} may then be written as
\begin{align}
   & \int_{-\infty}^\infty d(\Re{\phi^{\rm cl}}) f(\Re{\phi^{\rm cl}}) \nonumber \\&= \int_{i\delta-\infty}^{i\delta+\infty} dx f\left(x+\frac{\langle \phi^\text{cl} \rangle+\langle\overline{\phi^\text{cl}}\rangle}{2}\right)\\&=\int_{-\infty}^{\infty} dx f\left(x+\frac{\langle \phi^\text{cl} \rangle+\langle\overline{\phi^\text{cl}}\rangle}{2}\right)\nonumber.
\end{align}
Here $\int_{i\delta-\infty}^{i\delta+\infty}dx$ denotes the contour integral along the straight line with constant complex part $\delta = -\Im{\langle \phi^\text{cl} \rangle+\langle\overline{\phi^\text{cl}}\rangle}/2$. In the last equality, we made the assumption that the contour of integration can be deformed back to the real line without crossing any singularities of the integrand.

The other integrals in Eq.~\eqref{eq:pathd} may be treated analogously. This demonstrates that, under the integral, we may shift the real and imaginary parts of fields by complex values, as long as the integrand is analytic. This implies that we can shift fields and their complex conjugates independently, proving Eq.~\eqref{eq:shiftint}

\section{Cumulant Expansion}
\label{sec:cumulant_exp}
We can view the interaction part of the action as a random variable and expand its generating functional to a desired order in the cumulants: 
Following \cite{PhysRevResearch.5.043192} we rewrite the interacting part of the MGF,
    \begin{align}
    &\Lambda_{\mathrm{out}}^{\mathrm{int}}[\chi,\chi'] = \langle e^{i S_{\mathrm{int}}[\delta \phi,\chi,\chi']} \rangle_{0} \nonumber \\ 
    &= 
    \langle 1 + i  S_{\mathrm{int}}[\delta \phi,\chi,\chi']  
    - \frac{1}{2} S_{\mathrm{int}}[\delta \phi,\chi,\chi']^2 + \dots \rangle_{0} \\
   &=\exp  \ln  \langle 1 + i  S_{\mathrm{int}}[\delta \phi,\chi,\chi'] - \frac{1}{2} S_{\mathrm{int}}[\delta \phi,\chi,\chi']^2 + \dots \rangle_{0}\nonumber
    \fullstop
    \end{align}
    Now we employ the expansion of the logarithm
    $$\ln(1+x) = \sum_{n=1}^\infty  (-1)^{n+1}\frac{x^n}{n!}\comma$$
    to get, 
    \begin{align}
       &\Lambda_{\mathrm{out}}^{\mathrm{int}}[\chi,\chi'] = \exp \bigg\{\langle  i S_{\mathrm{int}}[\delta \phi,\chi,\chi'] \rangle_{0}  \\ 
       &-\frac{1}{2}(\langle  S_{\mathrm{int}}^2[\delta \phi,\chi,\chi'] \rangle_{0}-\langle  S_{\mathrm{int}}[\delta \phi,\chi,\chi'] ^2\rangle_{0}) + \mathcal{O}(U^3)\bigg\} \nonumber \fullstop
    \end{align}
    The CGF then becomes, 
    \begin{align}
     &\mathcal{S}_{\mathrm{out}}^\mathrm{int}[\chi,\chi'] = \ln(\Lambda_{\mathrm{out}}^{\mathrm{int}}[\chi,\chi']) 
     \\&= \langle\! \langle  i S_{\mathrm{int}}[\delta \phi,\chi,\chi'] \rangle \!\rangle_{0} -\frac{1}{2}\langle\!\langle  S_{\mathrm{int}}^2[\delta \phi,\chi,\chi'] \rangle\! \rangle_{0}+ \mathcal{O}(U^3) \fullstop\nonumber
    \end{align}
    More generally, one can find the following relation, 
    \begin{align}
        \mathcal{S}_\text{out}^\text{int}[\chi,\chi'] = \sum_{n=1}^\infty \frac{(i\lambda)^n}{n!} \langle\!\langle \qty( S_\text{int}[\delta \phi,\chi,\chi'] / \lambda)^n \rangle\!\rangle \fullstop
    \end{align}
    By rewriting the MGF, 
    \begin{align}
        \Lambda_{\text{out}}^{\text{int}}[\chi,\chi'] = \langle e^{i \lambda X} \rangle_{0} \comma 
    \end{align}
    here we put $X = S_\text{int}[\delta \phi, \chi,\chi']/\lambda$ with $\lambda$ being the parameter we expand in, and recalling the definition of the CGF 
    \begin{align}
        \mathcal{S}_\text{out}^\text{int}[\chi,\chi'] = \ln \langle e^{i \lambda X} \rangle = \sum_{n=1}^\infty \frac{(i\lambda)^n}{n!} \langle\!\langle X^n \rangle\!\rangle \fullstop
    \end{align}

 \section{Diagrammatic Rules for the Kerr Oscillator}
 \label{sec:origin_diag_rules}
In this appendix we explain the origins of the diagram rules outlined in the main text in Sec.~\ref{sec:diagram_rules}.
%%%%%%% Rule 1 %%%%%%%%%%% 
\subsection{}
The rule determining the number of arrows that point to and away from the vertex comes directly from the interaction Hamiltonian $\hat{H}_\text{int}$. The Kerr term consists of two creation and two annihilation operators which, on the level of fields, yields a complex $\phi^4$-theory with two barred and two unbarred fields in every term proportional to the Kerr parameter $U$, see Eq.~\eqref{eq:Kerr_interaction_action_pure}. With barred fields represented by outgoing arrows and unbarred fields by incoming arrows, see Fig.~\ref{fig:operator_diags}, we end up with the rule that every vertex needs to connect to two ingoing and two outgoing arrows. For a more general interaction Hamiltonian, 
\begin{align}
    \hat{H}_\text{int} = (\hat{a}^\dagger)^n \hat{a}^m +H.c. \comma 
\end{align}
we can expect the first rule to change to $n$ outgoing arrows and $m$ incoming arrows or vice versa at every vertex.
%%%%%%% Rule 2 %%%%%%%%%%% 
\subsection{}
The second rule on the number of dashed lines that connect to a vertex is due to the specific structure of the interaction action given in terms of the quantum (dashed lines) and classical fields (full lines) in Eq.~\eqref{eq:Kerr_interaction_action_pure}. We notice that every term in the interaction action contains either one or three quantum fields and since we denote quantum fields by dashed lines, this yields the restriction. For different interaction actions, inspection of the interaction action under the Keldysh rotation is expected to yield a similar rule.  
%%%%%%% Rule 3 %%%%%%%%%%% 
\subsection{}
Neglecting all loops except the ones of the Keldysh equal--time Green function is a convention to simplify calculations. We fix this convention to reduce the number of diagrams to keep track of by considering all possible loops inside the classical--classical connection. See App.~\ref{sec:loops} for an indepth explanation. 
%%%%%%% Rule 4 %%%%%%%%%%% 
\subsection{}
Rule number four reflects the causality structure built into the propagators. Every diagram where two connected vertices attach to a \newtext{dashed} line each contains a product of Green functions that is equal to zero:
\begin{center}
\begin{tikzpicture}
        \pgfsetxvec{\pgfpoint{0.5cm}{0}}
        \pgfsetyvec{\pgfpoint{0}{0.5cm}}
        \coordinate (V1) at (2,4); 
        \coordinate (V2) at (6,4);
        \coordinate (V3) at (2,1.5); 
        \coordinate (V4) at (6,1.5);
        \coordinate (V5) at (2,-1); 
        \coordinate (V6) at (6,-1);
        \coordinate (V7) at (2,-3.5); 
        \coordinate (V8) at (6,-3.5);
        \coordinate (V9) at (-3.5,0.25); 
        \coordinate (V10) at (0.5,0.25);
        \vertexcoordinates{V1}
        \vertexcoordinates{V2}
        \vertexcoordinates{V3}
        \vertexcoordinates{V4}
        \vertexcoordinates{V5}
        \vertexcoordinates{V6}
        \vertexcoordinates{V7}
        \vertexcoordinates{V8}
        \vertexcoordinates{V9}
        \vertexcoordinates{V10}
        \Gadvdown[0.5]{V1}{V2}
        \Gretup[0.5]{V1}{V2}
        \node at ($(V2) + (3.5,0)$) {$\propto G^R(\tau)G^A(\tau) = 0$};
        \Gretup[0.5]{V3}{V4}
        \Gretdown[0.5]{V4}{V3}
        \node at ($(V4) + (3.5,0)$) {$\propto G^R(\tau)G^R(-\tau) = 0$};
        \Gadvup[0.5]{V6}{V5}
        \Gretdown[0.5]{V6}{V5}
        \node at ($(V6) + (3.7,0)$) {$\propto G^R(-\tau)G^A(-\tau) = 0$};
        \Gadvup[0.5]{V8}{V7}
        \Gadvdown[0.5]{V7}{V8}
        \node at ($(V8) + (3.5,0)$) {$\propto G^A(-\tau)G^A(\tau) = 0$};
        \path (V9) -- (V10) coordinate[midway] (mid);
        \coordinate (midd) at ($(mid)  + (90:1)$);
        \draw[thick] (V9) to[out=45, in=180] (midd);
        \draw[dash pattern=on \dashdist\pgflinewidth off \dashdist\pgflinewidth,thick] (midd) to[out=0, in=135] (V10);
        \path (V10) -- (V9) coordinate[midway] (miid);
        \coordinate (miidd) at ($(mid)  + (270:1)$);
        \draw[thick] (V10) to[out=225, in=0] (miidd);
        \draw[dash pattern=on \dashdist\pgflinewidth off \dashdist\pgflinewidth,thick] (miidd) to[out=180, in=-45] (V9);
        \node at (1.25,0.2) {$\leadsto$};
    \end{tikzpicture}
    \end{center}
We see that every possible configuration of arrows in a diagram where two vertices are connected as is depicted in the left side, leads to the corresponding term being zero.
%%%%%%% Rule 5 %%%%%%%%%%% 
\subsection{}
Rule number five relies on the sign difference between the product of the terms $\langle \phi^\text{q}\rangle^{'}_0$ and $F \langle \Bar{\phi}^\text{q}\rangle^{'}_0$ as well as $\langle \Bar{\phi}^\text{q}\rangle^{'}_0$ and $F\langle \phi^\text{q}\rangle^{'}_0$ as they appear in our diagrams, see Fig.~\ref{fig:operator_diags}. For every diagram that fulfills the premise of the rule, there exists an associated diagram that we obtain by changing the position of the \newtext{dashed} line:  

\begin{center}
\begin{tikzpicture}
    \pgfsetxvec{\pgfpoint{0.5cm}{0}}
        \pgfsetyvec{\pgfpoint{0}{0.5cm}}
        \coordinate (V1) at (0,2); 
        \coordinate (V2) at (8.5,2);
        \vertexcoordinates{V1};
        \vertexcoordinates{V2};
        \createoutput[0.5]{V1TL}{V1}
        \destroyoutputK[0.5]{V1BL}{V1}
        \node at (3.2,2) {$\leadsto F \langle \Bar{\phi}^\text{q} \rangle^{'}_0 \langle \phi^\text{q} \rangle^{'}_0 $};
        \createoutputK[0.5]{V2TL}{V2}
        \destroyoutput[0.5]{V2BL}{V2}
        \node at (12,2) {$\leadsto - F \langle \Bar{\phi}^\text{q} \rangle^{'}_0 \langle \phi^\text{q} \rangle^{'}_0 $};
    \end{tikzpicture}
\end{center}    
Should both these diagrams have the same multiplicity, which is always the case if the vertex connects to an even number of detectors, they will cancel out resulting in rule 5\,a). Should the multiplicity of the associated diagram be different, which may happen for three detectors at the same vertex, the vertex with higher multiplicity (which has to be two) survives with multiplicity one, explaining rule 5\,b) as well as the special multiplicities of vertices with three detectors. 
%%%%%%% Rule 6 %%%%%%%%%%% 
\newtext{\subsection{}
Rule number six admits a simple intuitive interpretation: in the absence of external driving ($f=0$), the full system is in thermal equilibrium. In this regime, the statistical properties of the input and output fields must coincide, and no additional contribution originating from the interaction term of the Hamiltonian can arise. Consequently, diagrams violating this condition must cancel identically.

More formally, rule number six follows from a combination of the previously established rules and the causal structure of the Green’s -- function propagators. To derive it systematically, consider an arbitrary diagram containing $n$ vertices. For any given assignment of vertex times 
$\tau_1,\dots,\tau_n$, there necessarily exists a vertex with the smallest time, which we label $\tau_1$. Owing to its earliest time, this vertex has a restricted set of admissible attachments. One may then consider all possible configurations for that vertex and verify that their total contribution vanishes.

To illustrate this procedure explicitly, we consider the case in which the earliest-time vertex is connected to a single filled detector symbol, carries one dotted line, and no loops. Keeping the remainder of the diagram fixed, this vertex can assume four distinct configurations, shown in Fig.~\ref{fig:operator_diags}. We demonstrate below that the sum of these four contributions is identically zero.
\begin{center}
\begin{tikzpicture}
    \pgfsetxvec{\pgfpoint{0.4cm}{0}}
        \pgfsetyvec{\pgfpoint{0}{0.4cm}}
        \coordinate (V1) at (0,2); 
        \coordinate (V2) at (7,2);
        \coordinate (V3) at (5,-4);
        \coordinate (V4) at (12,-4);
        \vertexcoordinates{V1};
        \vertexcoordinates{V2};
        \vertexcoordinates{V3};
        \vertexcoordinates{V4};
        %% math symbols %%
        \begin{scope}
        \pgfmathsetmacro{\s}{3.5}
        \pgfmathsetmacro{\ss}{2.75}    
        \node at ($(V1) + (\s,0)$) {$+$};
        \node at ($(V3) + (-\s,0)$) {$+$};
        \node at ($(V3) + (\s,0)$) {$+$};
        \node at ($(V4) + (\s,0)$) {$= 0$};
        %% labels %% 
        \node at ($(V1) + (0,\ss)$) {$(a)$}; 
        \node at ($(V2) + (0,\ss)$) {$(b)$};
        \node at ($(V3) + (0,\ss)$) {$(c)$};
        \node at ($(V4) + (0,\ss)$) {$(d)$};
        \end{scope}
        %%%%%%%%%%%%%%%%%%%%%%%%%%%%%%%%%%%%%%%%
        \createoutput[0.4]{V1TL}{V1}
        % Calculate the midpoint
        \path (V1TR) -- (V1) coordinate[midway] (mid1);
        \path (V1TR) -- (V1) coordinate[pos = 0.6] (midd1);
        \draw[thick, arrows = {-{Latex[length=10pt,scale = 0.4]}}] (V1TR) -- (midd1);
        % Draw the solid line from midpoint to end
        \draw[thick] (V1) -- (mid1);

        \path (V1BR) -- (V1) coordinate[midway] (mid11);
        \path (V1BR) -- (V1) coordinate[pos = 0.4] (midd11);
        \draw[thick, arrows = {-{Latex[length=10pt,scale = 0.4]}}] (V1) -- (midd11);
        % Draw the solid line from midpoint to end
        \draw[thick] (mid11) -- (V1BR);

        \path (V1BL) -- (V1) coordinate[midway] (mid111);
        \path (V1BL) -- (V1) coordinate[pos = 0.4] (midd111);
        \draw[thick, arrows = {-{Latex[length=10pt,scale = 0.4]}}] (V1) -- (midd111);
        % Draw the solid line from midpoint to end
        \draw[thick] (mid111) -- (V1BL);
        
        %%%%%%%%%%%%%%%%%%%%%%%%%%%%%%%%%%%%%%%%
        \createoutputK[0.4]{V2TL}{V2}
        % Calculate the midpoint
        \path (V2TR) -- (V2) coordinate[midway] (mid2);
        \path (V2TR) -- (V2) coordinate[pos = 0.6] (midd2);
        \draw[thick, arrows = {-{Latex[length=10pt,scale = 0.4]}}] (V2TR) -- (midd2);
        % Draw the solid line from midpoint to end
        \draw[thick,dash pattern=on \dashdist\pgflinewidth off \dashdist\pgflinewidth] (V2) -- (mid2);

        \path (V2BR) -- (V2) coordinate[midway] (mid22);
        \path (V2BR) -- (V2) coordinate[pos = 0.4] (midd22);
        \draw[thick, arrows = {-{Latex[length=10pt,scale = 0.4]}}] (V2) -- (midd22);
        % Draw the solid line from midpoint to end
        \draw[thick] (mid22) -- (V2BR);

        \path (V2BL) -- (V2) coordinate[midway] (mid222);
        \path (V2BL) -- (V2) coordinate[pos = 0.4] (midd222);
        \draw[thick, arrows = {-{Latex[length=10pt,scale = 0.4]}}] (V2) -- (midd222);
        % Draw the solid line from midpoint to end
        \draw[thick] (mid222) -- (V2BL);
       %%%%%%%%%%%%%%%%%%%%%%%%%%%%%%%%%%%%%%%%
        \createoutputK[0.4]{V3TL}{V3}
        % Calculate the midpoint
        \path (V3TR) -- (V3) coordinate[midway] (mid3);
        \path (V3TR) -- (V3) coordinate[pos = 0.6] (midd3);
        \draw[thick, arrows = {-{Latex[length=10pt,scale = 0.4]}}] (V3TR) -- (midd3);
        % Draw the solid line from midpoint to end
        \draw[thick] (V3) -- (mid3);

        \path (V3BR) -- (V3) coordinate[midway] (mid33);
        \path (V3BR) -- (V3) coordinate[pos = 0.4] (midd33);
        \draw[thick,dash pattern=on \dashdist\pgflinewidth off \dashdist\pgflinewidth, arrows = {-{Latex[length=10pt,scale = 0.4]}}] (V3) -- (midd33);
        % Draw the solid line from midpoint to end
        \draw[thick] (mid33) -- (V3BR);

        \path (V3BL) -- (V3) coordinate[midway] (mid333);
        \path (V3BL) -- (V3) coordinate[pos = 0.4] (midd333);
        \draw[thick, arrows = {-{Latex[length=10pt,scale = 0.4]}}] (V3) -- (midd333);
        % Draw the solid line from midpoint to end
        \draw[thick] (mid333) -- (V3BL);
        \createoutputK[0.4]{V4TL}{V4}
        % Calculate the midpoint
        \path (V4TR) -- (V4) coordinate[midway] (mid4);
        \path (V4TR) -- (V4) coordinate[pos = 0.6] (midd4);
        \draw[thick, arrows = {-{Latex[length=10pt,scale = 0.4]}}] (V4TR) -- (midd4);
        % Draw the solid line from midpoint to end
        \draw[thick] (V4) -- (mid4);

        \path (V4BR) -- (V4) coordinate[midway] (mid44);
        \path (V4BR) -- (V4) coordinate[pos = 0.4] (midd44);
        \draw[thick, arrows = {-{Latex[length=10pt,scale = 0.4]}}] (V4) -- (midd44);
        % Draw the solid line from midpoint to end
        \draw[thick] (mid44) -- (V4BR);

        \path (V4BL) -- (V4) coordinate[midway] (mid444);
        \path (V4BL) -- (V4) coordinate[pos = 0.4] (midd444);
        \draw[thick,dash pattern=on \dashdist\pgflinewidth off \dashdist\pgflinewidth ,arrows = {-{Latex[length=10pt,scale = 0.4]}}] (V4) -- (midd444);
        % Draw the solid line from midpoint to end
        \draw[thick] (mid444) -- (V4BL);
    \end{tikzpicture}
\end{center}    

The first diagram contains an incoming Keldysh Green’s function $G^K$
. Since the corresponding vertex occurs at the earliest time $\tau_1$
, this incoming propagator reduces to the advanced Green’s function multiplied by the distribution function, $G^K =  - FG^A$
, in accordance with Eq.~\eqref{eq:Greens}. The resulting contribution is therefore 
    \begin{multline}
        (a) = -\langle \phi^\text{q}_0(\tau_1-\tau)\rangle' F i G^A(\tau_1-t) \\ \cdot i G^K(t'-\tau_1) i G^K(t''-\tau_1) \fullstop
    \end{multline}
The second diagram differs from the first only by replacing the incoming Keldysh propagator with an advanced Green’s function and by the presence of a full line emerging from the detector symbol. As shown in Fig.~\ref{fig:operator_diags}, this full line contributes an additional factor of $-F$, such that the two diagrams are in fact identical, yielding 
$(a)=(b)$. The third diagram again contains an incoming Keldysh Green’s function, which for the same reasons as above reduces to 
$-FG^A$. Moreover, due to the causal structure of the propagators, the outgoing retarded Green’s function is equivalent to a Keldysh Green’s function. This leads to 
\begin{multline}
        (c) =  \langle \phi^\text{q}_0(\tau_1-\tau)\rangle' F i G^A(\tau_1-t) \\ \cdot i G^K(t'-\tau_1) \underbrace{F i G^R(t''-\tau_1)}_{=i G^K(t''-\tau_1)} = -(a) 
    \end{multline}
An entirely analogous argument shows that 
$(d)=-(a)$. Hence, all four admissible configurations of this vertex cancel pairwise, and the total contribution of this class of diagrams vanishes identically. A similar proof can be performed for all other vertex configurations that do not include any input fields (any squares). This shows that all diagrams without input fields cancel, justifying rule number 6.
}
%%%%%%% Rule 7 %%%%%%%%%%% 
\subsection{}
At zero temperature the distribution function $F$ is equal to one and the constituting relation of the Keldysh Green function, see Eq.~\eqref{eq:Greens}, simplifies to the difference between the retarded and the advanced Green function,  
\begin{align}
    G^K = G^R - G^A \fullstop
\end{align}
This leads to the fact that all possible connections of two vertices with three propagators sum to zero. Note that this rule does not hold at finite temperatures.

\section{Multiplicity of diagrams}
\label{sec:mult_of_diags}
\begin{figure}[!htbp]
    \centering
    \vspace{2mm}
    \begin{tikzpicture}
    \pgfsetxvec{\pgfpoint{0.8cm}{0}}
        \pgfsetyvec{\pgfpoint{0}{0.8cm}}
    \coordinate (V1) at (2,2); 
    \coordinate (V2) at (6,2);
    \vertexcoordinates{V1}
    \vertexcoordinates{V2}
    \Gkelup[0.8]{V2}{V1}
    \Gkeldown[0.8]{V2}{V1}
    \destroyinput[0.8]{V1TL}{V1}
    \createoutput[0.8]{V1BL}{V1}
    \createinput[0.8]{V2TR}{V2}
    \destroyoutput[0.8]{V2BR}{V2}
    \end{tikzpicture}
    \caption{Diagram contributing to the second order correction of the cumulant $\langle\!\langle \dbout(\tau_1) \bout(\tau_2) \rangle\! \rangle$. This diagram has multiplicity two since the constituting first order diagrams each have multiplicity one (two identical legs), there is only one detector symbol of each type but the connecting lines between the vertices lead to an additional factor of two due to Wick's theorem, see Eq.~\eqref{eq:wick_example}.}
    \label{fig:bdaggerb_second_order}
\end{figure}
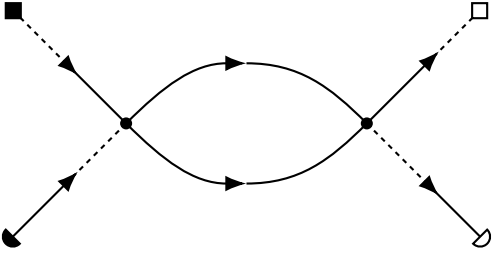
The factors making up the total multiplicity of each diagram have three different origins that are explained in detail below.  
\subsection{}
From the structure of the expression in the interaction action in Eq.~\eqref{eq:Kerr_interaction_action_pure} we find that a term corresponding to a given diagram may appear more than once - a simple algebraic multiplicity. 
The term corresponding to the diagram in Fig.~\ref{fig:diagrams_bout_first_order}\,a)  for example appears twice if one multiplies the terms in Eq.~\eqref{eq:Kerr_interaction_action_pure}, namely as the result of the following two multiplications, 
\begin{align}
 \qty( \langle \Bar{\phi}^\text{q} \rangle^{'}_0(t-\tau) \langle \phi^\text{cl}(t) \rangle_0) &\cdot \qty(\delta \Bar{\phi}^\text{cl} \delta \phi^\text{cl})\comma \\ 
 \qty(\langle \Bar{\phi}^\text{q}\rangle^{'}_0(t-\tau) \delta \phi^\text{cl} ) &\cdot \qty(\delta \Bar{\phi}^\text{cl}\langle \phi^\text{cl}(t) \rangle_0) \fullstop \label{eq:example_multiplicity_two}
\end{align}
The term corresponding to the diagram in Fig.~\ref{fig:diagrams_bout_first_order}\,b) only appears once, resulting in multiplicity one. Each vertex that connects to four distinct leg contributes a factor of two to this algebraic multiplicity. However, as noted above, rule 5 results in a reduced multiplicity for certain vertices that connect to three detectors. 
\subsection{}
As introduced in the main text, correlation functions are derived from the generating functional by taking functional derivatives with respect to the source fields $\chi$ and $\chi'$, see Eq.~\eqref{eq:general_correlation_function}. A correction term to a correlator of the form 
\[ \llangle \dbout(\tau_1) \dots \dbout(\tau_n) \bout(t_1) \dots \bout(t_m) \rrangle\]
therefore arises from functional differentiation of all terms in the CGF that contain $n$ times the source field $\chi'$ and $m$ times the source field $\chi$.  Since the only source-field dependent terms in our final expression for the interaction functional are the field averages of the quantum fields $\langle \phi^\text{q}(t)\rangle_0[\chi']$ and $\langle \Bar{\phi}^\text{q}(t) \rangle_0[\chi]$, we ultimately take functional derivatives of powers of these expressions which are contained in the diagrams through the detector symbols. As soon as the source field appears more than once, the order of differentiation with respect to the source fields at different times matters and we have to consider all possible permutations $\sigma$ of the times leading to the additional multiplicative factor, 
\begin{align}
    &\qty(\prod_{k=1}^m \frac{\delta}{\delta \chi(t_k)}) \qty(\langle \phi^\text{q}(t) \rangle_0[\chi])^m \nonumber \\ 
    &= \sum_{\sigma } \qty(\prod_{k=1}^m \frac{\delta\langle \phi^\text{q}(t) \rangle_0[\chi]}{\delta \chi(t_{\sigma(k)})}) \\ 
    &= m! \qty(\prod_{k=1}^m \frac{\delta\langle \phi^\text{q}(t) \rangle_0[\chi]}{\delta \chi(t_{k})}) \fullstop \nonumber
\end{align}
Due to the exact correspondence between the number of detector symbols of each type and the number of source fields, the number $m$ corresponds to the number of empty detector symbols in this case. The reasoning for the primed source fields is completely analogous. 
\subsection{}
The last multiplicity is due to the evaluation of the shifted field averages using Wicks theorem \cite{Altland2010}. 
We consider as an example the second order diagram from Fig.~\ref{fig:bdaggerb_second_order}. This diagram corresponds to the following mathematical expression in the interaction action, 
\begin{multline}
    \underbrace{\langle \phi^\text{cl}(t_1)\rangle_0 \langle \phi^\text{q}\rangle^{'}_0(t_1-\tau_1) \delta \phi^\mathrm{cl}(t_1) \delta \phi^\mathrm{cl}(t_1)}_{\text{left vertex}} \\ 
    \cdot  \underbrace{\langle \Bar{\phi}^\text{q}\rangle^{'}_0(t_2-\tau_2) \langle \Bar{\phi}^\text{cl}(t_2) \rangle_0 \delta \Bar{\phi}^\mathrm{cl}(t_2) \delta \Bar{\phi}^\mathrm{cl}(t_2)}_{\text{right vertex}} \fullstop
\end{multline}
In order to evaluate the diagram, we have to perform the averaging over the shifted fields with respect to the action of the linear cavity, see Eq.~\eqref{eq:S_int_avgs}. This averaging is done using Wicks theorem since the linear cavity is a Gaussian system. Here we find for the two lines connecting the vertices,  
\begin{multline}
    \langle  \delta \phi^\mathrm{cl}(t_1)\delta \phi^\mathrm{cl}(t_1)  \delta \Bar{\phi}^\mathrm{cl}(t_2)\delta \Bar{\phi}^\mathrm{cl}(t_2)\rangle_0 \\ 
    = 2 \langle\delta \phi^\mathrm{cl}(t_1) \delta \Bar{\phi}^\mathrm{cl}(t_2)\rangle_0^2 
    = 2  [iG^K(t_1-t_2)]^2 \fullstop \label{eq:wick_example}
\end{multline}
The multiplicity factor here corresponds to the number of non--zero two--point correlation functions  that can be made up from the shifted fields present in the diagram, i.e. the number of ways the arrows between vertices can be connected without changing the diagram.

\section{Loops}
\label{sec:loops}
Loops in our diagrams correspond to equal-time Green functions whose interpretation is not entirely unambiguous in the Keldysh formalism. In order to determine their value for the fields that appear in the interaction action, we trace back the correlators of the classical and quantum fields to correlators of the plus and minus fields situated on the Keldysh contour. 
\begin{align}
    \langle \delta \phi^\mathrm{cl}(t) \delta \Bar{\phi}^\mathrm{q}(t') \rangle &= \frac{i}{2} \left( G^{\mathbb{T}} - G^{\tilde{\mathbb{T}}} + G^{>} - G^{<}\right) \comma  \label{eq:clq_correlator_withpmfields} \\ 
    \langle \delta \phi^\mathrm{q}(t) \delta \Bar{\phi}^\mathrm{cl}(t') \rangle &= \frac{i}{2} \left( G^{\mathbb{T}} - G^{\tilde{\mathbb{T}}} - G^{>} + G^{<}\right)  \comma \label{eq:qcl_correlator_withpmfields}\\ 
    \langle \delta \phi^\mathrm{cl}(t) \delta \Bar{\phi}^\mathrm{cl}(t') \rangle &= \frac{i}{2} \left( G^{\mathbb{T}} + G^{\tilde{\mathbb{T}}} + G^{>} + G^{<}\right) \comma \label{eq:clcl_correlator_withpmfields} \\ 
    \langle \delta \phi^\mathrm{q}(t) \delta \Bar{\phi}^\mathrm{q}(t') \rangle &= \frac{i}{2} \left( G^{\mathbb{T}} + G^{\tilde{\mathbb{T}}} - G^{>} - G^{<}\right)   \fullstop \label{eq:qq_correlator_withpmfields}
\end{align}
Here we introduced the following correlators of the plus and minus fields, 
\begin{align}
\label{eq:deltaphiG1}
    \langle \delta \phi^-(t) \delta \Bar{\phi}^+(t') \rangle &= i G^>(t-t') \comma \\
    \langle \delta \phi^+(t) \delta \Bar{\phi}^-(t') \rangle &= i G^<(t-t') \comma \\ 
    \langle \delta \phi^+(t) \delta \Bar{\phi}^+(t') \rangle &= i G^{\mathbb{T}}(t-t') \comma  \\
    \langle \delta \phi^-(t) \delta \Bar{\phi}^-(t') \rangle &= i G^{\tilde{\mathbb{T}}}(t-t') \label{eq:deltaphiG4}\fullstop
\end{align}
Returning to the discrete action, see Eq.~\eqref{eq:action_system_all_modes}, we see that the barred fields appear one time-step after the non-barred field along the Keldysh contour. This implies that what we consider to be equal times in the continuum notation actually corresponds to $\delta\bar{\phi}(t_i+\delta t_i)\delta\phi(t_i)$ \newtext{in the discrete notation. In the continuum limit, $\delta t_i$ approaches zero from above (below) on the forward (backward) contour (because time decreases along the backward contour). The correct equal-time limit in Eqs.~\eqref{eq:deltaphiG1}-\eqref{eq:deltaphiG4} thus reads
\begin{align}
    \label{eq:G01}
    i G^>(0) &= \langle \hat{a}\hat{a}^\dagger\rangle_0 = n_B+1 \comma \\ 
    i G^<(0) &= \langle \hat{a}^\dagger\hat{a}\rangle_0 =n_B  \comma \\ 
    i G^\mathbb{T}(0^+) &= \langle \hat{a}^\dagger\hat{a}\rangle_0 =n_B \comma \\ 
    i G^{\tilde{\mathbb{T}}}(0^-) &= \langle \hat{a}^\dagger\hat{a}\rangle_0 =n_B \comma
     \label{eq:G04}
\end{align}
where $0^+$ ($0^-$) denotes the limit of approaching zero from above (below). We note that this is not identical to the value one would get by setting $t=t'$ in the discrete Green's functions.} In all but the greater Green functions, the creation operator stands to the left of the annihilation operator, because the barred fields appear after the unbarred fields along the Keldysh contour. The greater Green function has the opposite operator ordering, since the barred field is inserted in the forward branch, i.e. before the unbarred field which is inserted along the backward branch.

\newtext{We then find the following values for the equal time correlators of the classical and quantum fields, 
\begin{align}
  \lim_{t\rightarrow t'}  \langle \delta \phi^\mathrm{cl}(t) \delta \Bar{\phi}^\mathrm{q}(t') \rangle 
    &= \frac{1}{2} \comma   \label{eq:equaltime_cl_barq} \\ 
  \lim_{t\rightarrow t'}  \langle \delta \phi^\mathrm{q}(t) \delta \Bar{\phi}^\mathrm{cl}(t') \rangle
    &=  -\frac{1}{2} \comma \label{eq:equaltime_q_barcl}\\ 
    \lim_{t\rightarrow t'}\langle \delta \phi^\mathrm{cl}(t) \delta \Bar{\phi}^\mathrm{cl}(t') \rangle 
    & =  2 n_B + \frac{1}{2} \comma \\ 
   \lim_{t\rightarrow t'} \langle \delta \phi^\mathrm{q}(t) \delta \Bar{\phi}^\mathrm{q}(t') \rangle  
    &= -\frac{1}{2}  \comma
\end{align}
where we used Eqs.~\eqref{eq:clq_correlator_withpmfields}-\eqref{eq:qq_correlator_withpmfields} and took the limits according to Eqs.~\eqref{eq:G01}-\eqref{eq:G04}.}
The first two equations above tell us that for every diagram that contains an equal time advanced Green function, a similar diagram must exist with opposite sign due to an equal time retarded Green function (whenever a loop of an advanced Green function was possible, a loop of a retarded Green function must also fulfill all the diagram rules). These diagrams will invariably sum to zero and hence we fix the rule that no diagrams with loops consisting of a full and a dashed line exist. Interestingly we find that the quantum-quantum correlator does not vanish at equal times but attains a finite value of $-\frac{1}{2}$. This highlights the fact that the continuum notation should always be considered as a notational shorthand for the exact, discrete path integral. Since any diagram that allows for a quantum-quantum loop (a fully dashed circle) also fulfills all the diagram rules with a classical-classical loop in its stead we fix the following convention: We only allow diagrams with classical-classical loops and assign the value $2n_B$ to these loops. We thereby effectively consider the sum of the diagram with the classical-classical loop and the equivalent diagram with the quantum--quantum loop whose contributions will sum to $2n_B$. Apart from the obvious simplification of reducing the number of allowed diagrams, this convention allows us to neglect all diagrams with loops in the zero temperature limit.  

\begin{widetext}
\section{Resummation of loop diagrams}
\label{sec:resummation}
To get rid of all diagrams that contain loops, we write
\begin{equation}
    S_0[\phi]+S_{\rm int}[\phi] = \tilde{S}_0[\phi]+\tilde{S}_{\rm int}[\phi],
\end{equation}
where (we assume a rotating frame)
\begin{equation}
    \tilde{S}_0[\phi]   
    = \int dt \bigg[\Bar{\phi}^{\text{q}}(t) \left(i \partial_t - \Delta-4Un_B+i\frac{\kappa}{2}\right) \phi^{\text{cl}}(t)+ \Bar{\phi}^{\text{cl}}(t) \left(i \partial_t -\Delta-4Un_B-i\frac{\kappa}{2}\right) \phi^{\text{q}}(t) + i \kappa  F \Bar{\phi}^{\text{q}}(t) \phi^{\text{q}}(t)\bigg] \comma 
    \label{eq:s0tilde}
\end{equation}
and
\begin{equation}
\label{eq:actintshift}
     \tilde{S}_\mathrm{int}[\phi]  = -U \int dt \qty(\Bar{\phi}^\mathrm{cl} \phi^\mathrm{cl} + \Bar{\phi}^\mathrm{q} \phi^\mathrm{q}-4n_B) \qty(\Bar{\phi}^\mathrm{q} \phi^\mathrm{cl} + \Bar{\phi}^\mathrm{cl} \phi^\mathrm{q}).
\end{equation}
Following the derivation of the main text with this division of the action results in two differences. First, the detuning is shifted by $4Un_B$, as can be seen by Eq.~\eqref{eq:s0tilde}. Second, for any diagram that involves the equal time correlator $\langle\delta\bar{\phi}^{\rm cl}\delta\phi^{\rm cl}\rangle_0$, there will be an identical term but with the replacement $\langle\delta\bar{\phi}^{\rm cl}\delta\phi^{\rm cl}\rangle_0\rightarrow -2n_B = -\langle\delta\bar{\phi}^{\rm cl}\delta\phi^{\rm cl}\rangle_0$. These terms thus cancel, eliminating any diagram that contains a loop. Note that the factor of $4n_B$ in Eq.~\eqref{eq:actintshift} arises because each of the relevant diagrams appears twice (because of the $\phi^{\rm cl}/\bar{\phi}^{\rm cl}$ appearing in the second bracket), resulting in a multiplicity of two for each vertex including a loop. 

Diagrammatically we can understand the inclusion of this shift as a dressing of the propagators $G^{A/R/K}$. Every propagator, advanced, retarded and Keldysh Green function turns into an infinite series loop diagrams that begin and end with the corresponding fields, see Eqs.~\eqref{eq:resummed_Green_functions_diags_K},~\eqref{eq:resummed_Green_functions_diags_R},~\eqref{eq:resummed_Green_functions_diags_A}. We show this equality for the retarded Green function below, 
\begin{figure}[H]
    \centering
    \begin{tikzpicture}
     %% This rescales the diagrams: 
    \pgfsetxvec{\pgfpoint{0.25cm}{0}}
    \pgfsetyvec{\pgfpoint{0}{0.25cm}}
        %%%%%%%%%%%%%%%%%%%%%%%%%%%%%%%%%%%%%%%%%%%%%%%%%%%%%%%%%%%
        %\draw[step=0.25cm,gray,very thin] (0,-10) grid (60,12);
    %% Dressed retarded Green function 
    % Coordinates
        \coordinate (VV1) at (6,5); 
        \coordinate (VV2) at (10,5); 
        \coordinate (VV3) at (12,5); 
        \coordinate (VV4) at (16,5);
    % Diagrams and math symbols 
        \node at (3.5,5) {$\displaystyle \sum\limits_{n=0}^\infty 2^n$};
        \node at (10.5,5) {$\Biggl[$};
        \Gret[0.4]{VV2}{VV1}
        \clloop[0.4]{VV3}{100}
        \node at (17,5) {$\Biggr]^n$};
        \Gret[0.4]{VV4}{VV3}
        \node at (38.63,6.65) {$ \displaystyle = \sum\limits_{n=0}^\infty (-4i U n_B)^n  \overbrace{\int dt_1 dt_2 \dots dt_n i G^R(t-t_1) i G^R(t_1-t_2) \dots i G^R(t_n-t')}^{\displaystyle := \qty[iG^R]^{\circ (n+1)}}$};
        \node at (38.25,-1.75) {$ \displaystyle = \sum\limits_{n=0}^\infty (-4i U n_B)^n \theta(t-t') e^{-i \Delta (t-t') }e^{\sfrac{\kappa}{2}(t-t')} \underbrace{\int_t^{t'} dt_1 \int_{t_1}^{t'} dt_2 \dots \int_{t_{n-1}}^{t'} dt_n}_{\displaystyle \frac{1}{n!}\qty[\int_{t}^{t'} dt]^n}$};
        \node at (40.3,-8) {$ \displaystyle = \theta(t-t') e^{-i(\Delta +4n_B U) (t-t')} e^{-\sfrac{\kappa}{2}(t-t')} = i G^R(t-t')_{\Delta \to \Delta + 4 n_B U}= i G^R_U(t-t') \fullstop$};
    \end{tikzpicture}
\end{figure}
The derivation for the advanced Green function is analogous, for the Keldysh Green function, we proceed as follows,   
\begin{figure}[H]
    \centering
    \begin{tikzpicture}
     %% This rescales the diagrams: 
    \pgfsetxvec{\pgfpoint{0.25cm}{0}}
    \pgfsetyvec{\pgfpoint{0}{0.25cm}}
        %%%%%%%%%%%%%%%%%%%%%%%%%%%%%%%%%%%%%%%%%%%%%%%%%%%%%%%%%%%
        %\draw[step=0.25cm,gray,very thin] (-5,-10) grid (60,12);
    %% Dressed Keldysh Green function 
    % Coordinates 
        \coordinate (VVV1) at (6,10); 
        \coordinate (VVV2) at (10,10); 
        \coordinate (VVV3) at (12,10); 
        \coordinate (VVV4) at (16,10);
        \coordinate (VVV5) at (18,10); 
        \coordinate (VVV6) at (22,10); 
    % Diagrams and math symbols 
        \node at (2.25,10) {$\displaystyle \sum\limits_{l,p=0}^\infty 2^{l+p}$};
        \node at (5.5,10) {$\displaystyle \Biggl[$};
        \Gadv[0.4]{VVV2}{VVV1}
        \clloop[0.4]{VVV2}{100}
        \node at (11.5,10) {$\displaystyle \Biggr]^l$};
        \Gkel[0.4]{VVV4}{VVV3}
        \node at (16.5,10) {$\displaystyle \Biggl[$};
        \Gret[0.4]{VVV6}{VVV5}; 
        \clloop[0.4]{VVV5}{100}; 
        \node at (22.5,10) {$\displaystyle \Biggr]^p$};
        \node at (37,10) {$\displaystyle = \sum_{n=0}^\infty (-4 i U n_B) \sum_{k=0}^n \qty[i G^A]^{\circ (n-k)} \circ i G^K \circ \qty[i G^R]^{\circ k}$};
        \node at (45.25,5) {$\displaystyle = \sum_{n=0}^\infty (-4 i U n_B) \sum_{k=0}^n F \qty(\qty[i G^A]^{\circ (n-k)} \circ \qty[i G^R]^{\circ (k+1)} - \qty[i G^A]^{\circ(n-k+1)} \circ \qty[i G^R]^{\circ})$};
        \node at (44.05,0) {$\displaystyle = F \sum_{n=0}^\infty (-4 i U n_B) \qty(\qty[i G^R]^{\circ (n+1)}-\qty[i G^A]^{\circ(n+1)}) = F (i G^{R}_U - iG^{A}_U) = i G^K_U \fullstop $};
    \end{tikzpicture}
\end{figure}
In the second to last line we used the fact that the sum over $k$ is a telescopic series. 
\section{Second Order Coherence Function}
\label{sec:g2_through_cumulants}
The moments can be obtained from the cumulants according to~\cite{Peccati2011}
\begin{align}
   \langle \hat{O}_1 \dots \hat{O}_n \rangle = \sum_{\pi} \prod_{B \in \pi} \langle\! \langle \hat{O}_i \colon i \in B \rangle\! \rangle \fullstop \label{eq:moments_from_cumulants}
\end{align}
Where $\pi$ runs through the list of all partitions of $\{1,\dots,n \}$ and $B$ runs through the list of all blocks of the partition $\pi$. 
A straightforward application of this formula to the numerator of the $g^{(2)}$ function in Eq.~\eqref{eq:g2_def} yields, 
\begin{align}
    &\langle \dbout(t) \dbout(t+\tau) \bout(t+\tau) \bout(t) \rangle = \nonumber \\
    &~~\llangle \dbout(t) \dbout(t+\tau) \bout(t+\tau) \bout(t) \rrangle 
    + \llangle \dbout(t) \rrangle \llangle \dbout(t+\tau) \bout(t+\tau) \bout(t) \rrangle \nonumber \\
    &~~+ \llangle \dbout(t+\tau) \rrangle \llangle \dbout(t) \bout(t+\tau) \bout(t) \rrangle 
    + \llangle \bout(t+\tau) \rrangle \llangle \dbout(t+\tau) \dbout(t) \bout(t) \rrangle \nonumber \\
    &~~+ \llangle \bout(t) \rrangle \llangle \dbout(t) \dbout(t+\tau) \bout(t+\tau)  \rrangle 
    +\llangle \dbout(t) \dbout(t+\tau) \rrangle \llangle \bout(t+\tau) \bout(t) \rrangle \nonumber \\
    &~~+\llangle \dbout(t) \bout(t+\tau) \rrangle \llangle \dbout(t+\tau) \bout(t) \rrangle 
    +\llangle \dbout(t) \bout(t) \rrangle \llangle \dbout(t+\tau) \bout(t+\tau)  \rrangle \nonumber \\
    &~~+ \llangle \dbout(t) \rrangle \llangle \dbout(t+\tau) \rrangle \llangle \bout(t+\tau) \bout(t) \rrangle
    + \llangle \dbout(t) \rrangle \llangle \bout(t+\tau) \rrangle \llangle \dbout(t+\tau) \bout(t) \rrangle \nonumber \\
    &~~+ \llangle \dbout(t) \rrangle \llangle \bout(t) \rrangle \llangle \dbout(t+\tau) \bout(t+\tau)  \rrangle 
    + \llangle \dbout(t+\tau) \rrangle \llangle \bout(t+\tau) \rrangle \llangle \dbout(t) \bout(t) \rrangle \nonumber \\
    &~~+ \llangle \dbout(t+\tau) \rrangle \llangle \bout(t) \rrangle \llangle \dbout(t) \bout(t+\tau)  \rrangle 
    + \llangle \bout(t+\tau) \rrangle \llangle \bout(t) \rrangle \llangle \dbout(t) \dbout(t+\tau) \rrangle \nonumber \\
    &~~+  \llangle \dbout(t) \rrangle \llangle \dbout(t+\tau)\rrangle \llangle \bout(t+\tau) \rrangle \llangle \bout(t) \rrangle \comma
\end{align}
where we applied normal-- and time--ordering to all the cumulants. 
\newtext{
\section{Comparison to Existing Approach}
In this appendix, we compare our Schwinger–Keldysh path-integral approach to determining output-field correlation functions with the established method based on (i) perturbation theory, (ii) the standard input–output relation, and (iii) the quantum regression theorem (QRT). This comparison is provided to clarify that our method does not yield new physics that is inaccessible to the conventional framework; however, the conventional approach becomes practically infeasible already at modest orders of perturbation.
As a prototype, we consider the four-time output correlator relevant for evaluating the second-order coherence function
\begin{align}
    \langle \dbout(t_1) \dbout(t_2) \bout(t_3) \bout(t_4) \rangle \comma
\end{align}
with time ordering $t_1 \leq t_2$ and $t_4 \leq t_3$, see Eq.~\eqref{eq:time_ordering2}, and where the intracavity system is the driven Kerr oscillator studied in Sec. \ref{sec:kerr_output_statistics}.

Using the standard input–output identity in Eq.~\eqref{eq:in_out_rel}, the desired correlator expands into 14 distinct terms containing different combinations of input and system operators:
\begin{align}
    &\langle \dbout(t_1) \dbout(t_2) \bout(t_3) \bout(t_4) \rangle  = \nonumber
    \langle \dbin(t_1) \dbin(t_2) \bin(t_3) \bin(t_4) \rangle \nonumber  
    +\kappa^{1/2}
    \big( \langle \dbin(t_1) \dbin(t_2) \ahat(t_3) \bin(t_4) \rangle\\& + \langle \dbin(t_1) \dbin(t_2) \bin(t_3) \ahat(t_4) \rangle  
    +  \langle \dbin(t_1) \dahat(t_2) \bin(t_3) \bin(t_4) \rangle + \langle \dahat(t_1) \dbin(t_2) \bin(t_3) \bin(t_4) \rangle \big)
     \nonumber \\ 
    &+ \kappa 
    \big(\langle \dbin(t_1) \dbin(t_2) \ahat(t_3) \ahat(t_4) \rangle + \langle\dbin(t_1) \dahat(t_2) \ahat(t_3) \bin(t_4) \rangle +  \langle \dbin(t_1) \dahat(t_2) \bin(t_3) \ahat(t_4) \rangle + \langle \dahat(t_1) \dbin(t_2) \ahat(t_3) \bin(t_4) \rangle\nonumber \\ 
    &+  \langle \dahat(t_1) \dbin(t_2) \bin(t_3) \ahat(t_4) \rangle + \langle \dahat(t_1) \dahat(t_2) \bin(t_3) \bin(t_4) \rangle \big) 
    \nonumber  
    + \kappa^{3/2} 
        \big(\langle \dbin(t_1) \dahat(t_2) \ahat(t_3) \ahat(t_4) \rangle + \langle \dahat(t_1) \dbin(t_2) \ahat(t_3) \ahat(t_4) \rangle  \\ 
        & +\langle\dahat(t_1) \dahat(t_2) \bin(t_3) \ahat(t_4) \rangle + \langle \dahat(t_1) \dahat(t_2) \ahat(t_3) \bin(t_4)\rangle \big)
    + \kappa^2 
        \langle \dahat(t_1) \dahat(t_2) \ahat(t_3) \ahat(t_4) \rangle \fullstop
   \label{eq:comparison_in_out_decomposition}
\end{align}
Except for the purely-input and purely-system correlators, all remaining terms are mixed correlators involving both $\ahat$ and $\bin$. Each of the 14 mixed correlators must be decomposed into system correlators only.
To perform this reduction, one introduces shifted input operators
$\delta \hat{b}_\text{in} = \bin - \langle \bin \rangle$ and applies the decomposition rules of Ref.~\cite{PhysRevA.31.3761}.
As an illustration, consider the term  $\langle \dahat(t_1) \dbin(t_2) \bin(t_3) \bin(t_4) \rangle$, which decomposes into eight contributions:
\begin{align}
\langle \dahat(t_1) \dbin(t_2) \bin(t_3) \bin(t_4) \rangle = ~&\langle \dahat(t_1) \delta \hat{b}^\dagger_{\text{in}}(t_2) \delta \hat{b}_{\text{in}}(t_3) \delta \hat{b}_{\text{in}}(t_4) \rangle + \langle \dahat(t_1) \delta \hat{b}^\dagger_{\text{in}}(t_2) \delta \hat{b}_{\text{in}}(t_3)\rangle \langle \hat{b}_{\text{in}}(t_4) \rangle \nonumber \\ 
&+ 
\langle \dahat(t_1) \delta \hat{b}^\dagger_{\text{in}}(t_2)\delta \hat{b}_{\text{in}}(t_4) \rangle \langle \bin(t_3) \rangle + \langle \dahat(t_1) \delta \hat{b}_{\text{in}}(t_3) \delta \hat{b}_{\text{in}}(t_4) \rangle \langle \dbin(t_2) \rangle \nonumber \\ 
&+ \langle \dahat(t_1) \delta \hat{b}^\dagger_{\text{in}}(t_2)\rangle \langle \bin(t_3) \rangle \langle \bin(t_4) \rangle + \langle \dahat(t_1) \delta \hat{b}_{\text{in}}(t_3) \rangle \langle \dbin(t_2) \rangle \langle \bin(t_4) \rangle \nonumber \\ 
&+ \langle \dahat(t_1) \delta \hat{b}_{\text{in}}(t_4) \rangle \langle \dbin(t_2)\rangle \langle \bin(t_3) \rangle + \langle \dahat(t_1) \rangle \langle \dbin(t_2) \rangle \langle \bin(t_3) \rangle \langle \bin(t_4) \rangle \fullstop
\end{align}
This is one of the 14 terms from Eq.~\eqref{eq:comparison_in_out_decomposition}. Each term requires such a decomposition, which rapidly proliferates.
For the remaining mixed correlators one needs identities relating them to system-only correlators. For instance, Ref.~\cite{PhysRevA.31.3761} provides the representative identity
\begin{align}
    \langle \dahat(t_1) \delta \hat{b}_\text{in}(t_2) \delta \hat{b}_\text{in}(t_3) \rangle = \begin{cases}
        \kappa n_B^2 \langle \qty[\qty[\dahat(t_1),\ahat(t_2)],\ahat(t_3)] \rangle\comma & t_1>t_2>t_3 \\ 
        0 \comma & t_2 >t_1 \text{ or } t_3 > t_1 \comma
    \end{cases}
\end{align}
However, to the best of our knowledge no complete catalog of such relations exists, and in order to evaluate Eq.~\eqref{eq:comparison_in_out_decomposition} one must first derive most of the necessary identities. This illustrates the severe practical limitations of the standard method when dealing with finite temperatures. Furthermore, as we now show, computing multi-time correlation functions of the output field using established methods is an extremely challenging task already at zero temperature.

Once all mixed correlators are rewritten in terms of multi-time system correlators such as $\langle \dahat(t_1) \dahat(t_2) \ahat(t_3) \ahat(t_4) \rangle $ one applies the quantum regression theorem, see Ref.~\cite{10.1093/acprof:oso/9780199213900.002.14006}. This requires the equations of motion for operator moments $\langle \qty(\dahat)^n \ahat^m \rangle$, 
\begin{multline}
\partial_t \langle \qty(\dahat)^n \ahat^m \rangle = 
-i \omega_S (n-m) \langle  \qty(\dahat)^n \ahat^m \rangle - i U \qty(2 (m-n) \langle \qty(\dahat)^{n+1} \ahat^{m+1}\rangle + (m (m-1) - n(n-1)) \langle \qty(\dahat)^n \ahat^m \rangle) 
\\ - \sqrt{\kappa} m f(t) \langle \qty( \dahat)^n \ahat^{m-1} \rangle - \sqrt{\kappa} n \Bar{f}(t) \langle \qty(\dahat)^{n-1} \ahat^m \rangle + \kappa \qty(\frac{n+m}{2} \langle \qty(\dahat)^{n} \ahat^m \rangle + n_B n m \langle \qty(\dahat)^{n-1} \ahat^{m-1} \rangle) \comma
\end{multline}
where the right-hand side does not close and generates an infinite hierarchy.
Because we are interested in perturbation theory up to order $U^2$, we truncate the hierarchy to the smallest closed set of operators sufficient to yield $\langle \ahat \rangle$ correctly to second order. One suitable choice is
\begin{align}
\qty{A_{\mu}} = \qty{\ahat,  \ahat \ahat \,,~  \dahat \ahat \,,~   \ahat \ahat \ahat  \,,~ \dahat \ahat \ahat \,,~   \dahat \ahat \ahat \ahat  \,,~  \dahat \dahat \ahat \ahat  \,,~  \dahat \dahat \ahat \ahat \ahat , \text{ h.c. } } \fullstop 
\end{align}
These operators satisfy 
\begin{align}
\partial_t \langle A_{\mu}(t) \rangle = \sum_{\lambda} M_{\mu,\lambda} \langle A_{\lambda}(t) \rangle \fullstop
\end{align}
Applying the QRT to a four-time correlator such as 
$\langle \dahat(t_1)\dahat(t_2)\ahat(t_3)\ahat(t_4) \rangle$ where $t_3> t_2 >t_4,t_1$ yields~\cite{PhysRevA.106.022214}, 
\begin{align}
\partial_{t_3} \langle \dahat(t_1) \dahat(t_2) \ahat(t_3) \ahat(t_4) \rangle =  \sum_{\lambda} M_{1,\lambda}  \langle \dahat(t_1) \dahat(t_2) A_{\lambda}(t_3)\ahat(t_4)   \rangle \comma 
\end{align}
where we use $A_1 = \ahat$ and the solution involves correlators at earlier times:
\begin{align}
\langle \dahat(t_1) \dahat(t_2) \ahat(t_3) \ahat(t_4) \rangle = \sum_{\lambda} C_{\lambda} \langle \dahat(t_1) \dahat(t_2) A_{\lambda}(t_2) \ahat(t_4) \rangle \fullstop
\end{align}
To evaluate these new terms, one must extend the operator basis to
$B_\lambda =\dahat A_\lambda$ apply the QRT again, and iterate until all times coincide. The operator basis grows combinatorially with the number of time arguments and with perturbation order. We iterate this process until we can express the result in equal--time correlation functions of the system mode. These correlation functions can then be determined by standard perturbation theory.

In contrast, our path-integral formalism works directly with cumulants of the output field which contain the same information as the output moments (see App. \ref{sec:g2_through_cumulants}), but can be evaluated diagrammatically using the rules of Sec. \ref{sec:diagram_rules}.
\begin{align}
    \llangle \dbout(t_1) \dbout(t_2) \bout(t_3) \bout(t_4) \rrangle \comma
\end{align}
with the same time--ordering as above. 
The procedure for obtaining the second-order approximation of this quantity is described in detail in Sec.~\ref{sec:diagrams}. In the present case, it reduces to identifying all diagrams that (i) satisfy the rules of Sec.~\ref{sec:diagram_rules}, (ii) contain no more than two vertices, and (iii) feature two empty and two filled external half-circles. Neglecting loop diagrams and accounting for their contributions via the partial summation method of Sec.~\ref{sec:loop_summation}, we are left with a total of twelve diagrams, two of which are shown in Fig.~\ref{fig:sout_plot}(c)–(d). Translating these diagrams and assigning their multiplicities according to Sec.~\ref{sec:mult_of_diags} then directly yields the desired second-order approximation. The procedure is fully systematic, requires no derivation of mixed input–system identities, and does not require closing or truncating an operator hierarchy. We stress that while finite temperatures constitute a major complication for established methods, they can be incorporated easily in our path-integral approach by dressing the Green's functions as discussed in Sec.~\ref{sec:loop_summation}.
}
\end{widetext}
\bibliography{library.bib}

@Book{Kamenev2011,
  title     = {Field Theory of Non-Equilibrium Systems},
  publisher = {Cambridge University Press},
  year      = {2011},
  author    = {Alex Kamenev},
  isbn      = {978-0-521-76082-9},
doi = {10.1017/CBO9781139003667}
}

@article{glauber_1963,
  title = {Coherent and Incoherent States of the Radiation Field},
  author = {Glauber, Roy J.},
  journal = {Phys. Rev.},
  volume = {131},
  issue = {6},
  pages = {2766--2788},
  numpages = {0},
  year = {1963},
  month = {Sep},
  publisher = {American Physical Society},
  doi = {10.1103/PhysRev.131.2766},
  url = {https://link.aps.org/doi/10.1103/PhysRev.131.2766}
}

@article{sudarshan_1963,
  title = {Equivalence of Semiclassical and Quantum Mechanical Descriptions of Statistical Light Beams},
  author = {Sudarshan, E. C. G.},
  journal = {Phys. Rev. Lett.},
  volume = {10},
  issue = {7},
  pages = {277--279},
  numpages = {0},
  year = {1963},
  month = {Apr},
  publisher = {American Physical Society},
  doi = {10.1103/PhysRevLett.10.277},
  url = {https://link.aps.org/doi/10.1103/PhysRevLett.10.277}
}

@article{glauber:1963,
  title = {The Quantum Theory of Optical Coherence},
  author = {Glauber, Roy J.},
  journal = {Phys. Rev.},
  volume = {130},
  issue = {6},
  pages = {2529--2539},
  numpages = {0},
  year = {1963},
  month = {Jun},
  publisher = {American Physical Society},
  doi = {10.1103/PhysRev.130.2529},
  url = {https://link.aps.org/doi/10.1103/PhysRev.130.2529}
}

@Article{RevModPhys.82.1155,
  author    = {Clerk, A. A. and Devoret, M. H. and Girvin, S. M. and Marquardt, Florian and Schoelkopf, R. J.},
  title     = {Introduction to quantum noise, measurement, and amplification},
  journal   = {Rev. Mod. Phys.},
  year      = {2010},
  volume    = {82},
  pages     = {1155--1208},
  month     = {Apr},
  doi       = {10.1103/RevModPhys.82.1155},
  issue     = {2},
  numpages  = {0},
  publisher = {American Physical Society},
  url       = {https://link.aps.org/doi/10.1103/RevModPhys.82.1155},
}

@Book{Gardiner,
  title     = {Quantum Noise},
  publisher = {Springer Berlin, Heidelberg},
  year      = {2004},
  author    = {Crispin Gardiner, Peter Zoller},
  series    = {Springer Series in Synergetics},
  edition   = {3},
  isbn      = {978-3-540-22301-6},
}

@Book{Altland2010,
  title     = {Condensed Matter Field Theory},
  publisher = {Cambridge University Press},
  year      = {2010},
  author    = {Altland, Alexander and Simons, Ben D.},
  edition   = {2},
  doi       = {10.1017/CBO9780511789984},
  place     = {Cambridge},
}

@article{Sieberer_2016,
	doi = {10.1088/0034-4885/79/9/096001},
	url = {https://dx.doi.org/10.1088/0034-4885/79/9/096001},
	year = {2016},
	month = {aug},
	publisher = {IOP Publishing},
	volume = {79},
	number = {9},
	pages = {096001},
	author = {L M Sieberer and M Buchhold and S Diehl},
	title = {Keldysh field theory for driven open quantum systems},
	journal = {Rep. Prog. Phys.}
}

@article{PhysRevResearch.3.023041,
  title = {Wigner negativity in the steady-state output of a {K}err parametric oscillator},
  author = {Strandberg, Ingrid and Johansson, G\"oran and Quijandr\'{\i}a, Fernando},
  journal = {Phys. Rev. Res.},
  volume = {3},
  issue = {2},
  pages = {023041},
  numpages = {13},
  year = {2021},
  month = {Apr},
  publisher = {American Physical Society},
  doi = {10.1103/PhysRevResearch.3.023041},
  url = {https://link.aps.org/doi/10.1103/PhysRevResearch.3.023041}
}

@article{PhysRevA.94.033841,
  title = {Exact steady state of a {K}err resonator with one- and two-photon driving and dissipation: Controllable {W}igner-function multimodality and dissipative phase transitions},
  author = {Bartolo, Nicola and Minganti, Fabrizio and Casteels, Wim and Ciuti, Cristiano},
  journal = {Phys. Rev. A},
  volume = {94},
  issue = {3},
  pages = {033841},
  numpages = {11},
  year = {2016},
  month = {Sep},
  publisher = {American Physical Society},
  doi = {10.1103/PhysRevA.94.033841},
  url = {https://link.aps.org/doi/10.1103/PhysRevA.94.033841}
}

@book{gerry_knight_2004, place={Cambridge}, title={Introductory Quantum Optics}, DOI={10.1017/CBO9780511791239}, publisher={Cambridge University Press}, author={Gerry, Christopher and Knight, Peter}, year={2004}}

@article{PhysRevA.95.063805,
  title = {Time-dependent quantum correlations in phase space},
  author = {Krumm, F. and Vogel, W. and Sperling, J.},
  journal = {Phys. Rev. A},
  volume = {95},
  issue = {6},
  pages = {063805},
  numpages = {12},
  year = {2017},
  month = {Jun},
  publisher = {American Physical Society},
  doi = {10.1103/PhysRevA.95.063805},
  url = {https://link.aps.org/doi/10.1103/PhysRevA.95.063805}
}

@Article{kramer2018quantumoptics,
  author    = {Kr{\"a}mer, Sebastian and Plankensteiner, David and Ostermann, Laurin and Ritsch, Helmut},
  title     = {Quantum{O}ptics. jl: A {J}ulia framework for simulating open quantum systems},
  journal   = {Comput. Phys. Commun.},
  year      = {2018},
  volume    = {227},
  pages     = {109--116},
  publisher = {Elsevier},
  url       = {https://doi.org/10.1016/j.cpc.2018.02.004},
}

@article{PhysRevA.31.3761,
  title = {Input and output in damped quantum systems: Quantum stochastic differential equations and the master equation},
  author = {Gardiner, C. W. and Collett, M. J.},
  journal = {Phys. Rev. A},
  volume = {31},
  issue = {6},
  pages = {3761--3774},
  numpages = {0},
  year = {1985},
  month = {Jun},
  publisher = {American Physical Society},
  doi = {10.1103/PhysRevA.31.3761},
  url = {https://link.aps.org/doi/10.1103/PhysRevA.31.3761}
}

@book{Peccati2011,
  title = {Wiener Chaos: Moments,  Cumulants and Diagrams},
  ISBN = {9788847016798},
  ISSN = {2039-1471},
  url = {http://dx.doi.org/10.1007/978-88-470-1679-8},
  DOI = {10.1007/978-88-470-1679-8},
  journal = {Bocconi &amp; Springer Series},
  publisher = {Springer Milan},
  author = {Peccati,  Giovanni and Taqqu,  Murad S.},
  year = {2011}
}

@article{doi:10.1143/JPSJ.17.1100,
author = {Kubo ,Ryogo},
title = {Generalized Cumulant Expansion Method},
journal = {J. Phys. Soc. Jpn.},
volume = {17},
number = {7},
pages = {1100-1120},
year = {1962},
doi = {10.1143/JPSJ.17.1100}
}

@article{PhysRevResearch.5.043192,
  title = {Modeling particle loss in open systems using {K}eldysh path integral and second order cumulant expansion},
  author = {Huang, Chen-How and Giamarchi, Thierry and Cazalilla, Miguel A.},
  journal = {Phys. Rev. Res.},
  volume = {5},
  issue = {4},
  pages = {043192},
  numpages = {14},
  year = {2023},
  month = {Dec},
  publisher = {American Physical Society},
  doi = {10.1103/PhysRevResearch.5.043192},
  url = {https://link.aps.org/doi/10.1103/PhysRevResearch.5.043192}
}

@article{PhysRevLett.100.013605,
  title = {Nonclassical Correlation Properties of Radiation Fields},
  author = {Vogel, Werner},
  journal = {Phys. Rev. Lett.},
  volume = {100},
  issue = {1},
  pages = {013605},
  numpages = {4},
  year = {2008},
  month = {Jan},
  publisher = {American Physical Society},
  doi = {10.1103/PhysRevLett.100.013605},
  url = {https://link.aps.org/doi/10.1103/PhysRevLett.100.013605}
}

@article{PhysRevA.30.1386,
  title = {Squeezing of intracavity and traveling-wave light fields produced in parametric amplification},
  author = {Collett, M. J. and Gardiner, C. W.},
  journal = {Phys. Rev. A},
  volume = {30},
  issue = {3},
  pages = {1386--1391},
  numpages = {0},
  year = {1984},
  month = {Sep},
  publisher = {American Physical Society},
  doi = {10.1103/PhysRevA.30.1386},
  url = {https://link.aps.org/doi/10.1103/PhysRevA.30.1386}
}

@article{PhysRevA.103.033711,
  title = {Driven-dissipative phase transition in a {K}err oscillator: From semiclassical $\mathcal{PT}$ symmetry to quantum fluctuations},
  author = {Zhang, Xin H. H. and Baranger, Harold U.},
  journal = {Phys. Rev. A},
  volume = {103},
  issue = {3},
  pages = {033711},
  numpages = {12},
  year = {2021},
  month = {Mar},
  publisher = {American Physical Society},
  doi = {10.1103/PhysRevA.103.033711},
  url = {https://link.aps.org/doi/10.1103/PhysRevA.103.033711}
}

@Inbook{Yurke2004,
author="Yurke, B.",
editor="Drummond, Peter D.
and Ficek, Zbigniew",
title="Input-Output Theory",
bookTitle="Quantum Squeezing",
year="2004",
publisher="Springer Berlin Heidelberg",
address="Berlin, Heidelberg",
pages="53--96",
isbn="978-3-662-09645-1",
doi="10.1007/978-3-662-09645-1_3",
url="https://doi.org/10.1007/978-3-662-09645-1_3"
}

@article{PhysRevLett.123.123604,
  title = {Input-Output Theory with Quantum Pulses},
  author = {Kiilerich, Alexander Holm and M\o{}lmer, Klaus},
  journal = {Phys. Rev. Lett.},
  volume = {123},
  issue = {12},
  pages = {123604},
  numpages = {6},
  year = {2019},
  month = {Sep},
  publisher = {American Physical Society},
  doi = {10.1103/PhysRevLett.123.123604},
  url = {https://link.aps.org/doi/10.1103/PhysRevLett.123.123604}
}

@article{RevModPhys.86.1391,
  title = {Cavity optomechanics},
  author = {Aspelmeyer, Markus and Kippenberg, Tobias J. and Marquardt, Florian},
  journal = {Rev. Mod. Phys.},
  volume = {86},
  issue = {4},
  pages = {1391--1452},
  numpages = {62},
  year = {2014},
  month = {Dec},
  publisher = {American Physical Society},
  doi = {10.1103/RevModPhys.86.1391},
  url = {https://link.aps.org/doi/10.1103/RevModPhys.86.1391}
}

@article{RevModPhys.93.025005,
  title = {Circuit quantum electrodynamics},
  author = {Blais, Alexandre and Grimsmo, Arne L. and Girvin, S. M. and Wallraff, Andreas},
  journal = {Rev. Mod. Phys.},
  volume = {93},
  issue = {2},
  pages = {025005},
  numpages = {72},
  year = {2021},
  month = {May},
  publisher = {American Physical Society},
  doi = {10.1103/RevModPhys.93.025005},
  url = {https://link.aps.org/doi/10.1103/RevModPhys.93.025005}
}

@article{RevModPhys.87.1379,
  title = {Cavity-based quantum networks with single atoms and optical photons},
  author = {Reiserer, Andreas and Rempe, Gerhard},
  journal = {Rev. Mod. Phys.},
  volume = {87},
  issue = {4},
  pages = {1379--1418},
  numpages = {40},
  year = {2015},
  month = {Dec},
  publisher = {American Physical Society},
  doi = {10.1103/RevModPhys.87.1379},
  url = {https://link.aps.org/doi/10.1103/RevModPhys.87.1379}
}

@article{THOMPSON2023169385,
title = {Field theory of many-body {L}indbladian dynamics},
journal = {Ann. Phys.},
volume = {455},
pages = {169385},
year = {2023},
issn = {0003-4916},
doi = {https://doi.org/10.1016/j.aop.2023.169385},
url = {https://www.sciencedirect.com/science/article/pii/S0003491623001719},
author = {Foster Thompson and Alex Kamenev}
}

@Article{Huebener2021,
  author        = {Hübener, Hannes and De Giovannini, Umberto and Schäfer, Christian and Andberger, Johan and Ruggenthaler, Michael and Faist, Jerome and Rubio, Angel},
  title         = {Engineering quantum materials with chiral optical cavities},
  journal       = {Nat. Mater.},
  year          = {2021},
  volume        = {20},
  number        = {4},
  pages         = {438--442},
  month         = apr,
  issn          = {1476-4660},
  refid         = {Hübener2021},
  url           = {https://doi.org/10.1038/s41563-020-00801-7},
}

@article{Lu:25,
author = {I-Te Lu and Dongbin Shin and Mark Kamper Svendsen and Simone Latini and Hannes H\"{u}bener and Michael Ruggenthaler and Angel Rubio},
journal = {Adv. Opt. Photon.},
number = {2},
pages = {441--525},
publisher = {Optica Publishing Group},
title = {Cavity engineering of solid-state materials without external driving},
volume = {17},
month = {Jun},
year = {2025},
url = {https://opg.optica.org/aop/abstract.cfm?URI=aop-17-2-441},
doi = {10.1364/AOP.544138},
}

@Article{Mivehvar2021,
  author    = {Farokh Mivehvar and Francesco Piazza and Tobias Donner and Helmut Ritsch and},
  title     = {Cavity QED with quantum gases: new paradigms in many-body physics},
  journal   = {Adv. Phys.},
  year      = {2021},
  volume    = {70},
  number    = {1},
  pages     = {1--153},
  doi       = {10.1080/00018732.2021.1969727},
  publisher = {Taylor \& Francis},
  url       = { 
    
        https://doi.org/10.1080/00018732.2021.1969727
    
    

},
}

@Article{Strathearn2018,
  author        = {Strathearn, A. and Kirton, P. and Kilda, D. and Keeling, J. and Lovett, B. W.},
  title         = {Efficient non-{M}arkovian quantum dynamics using time-evolving matrix product operators},
  journal       = {Nat. Commun.},
  year          = {2018},
  volume        = {9},
  number        = {1},
  pages         = {3322},
  month         = aug,
  issn          = {2041-1723},
  refid         = {Strathearn2018},
  url           = {https://doi.org/10.1038/s41467-018-05617-3},
}

@book{breuer_book,
	title = {The theory of open quantum systems},
	author = {Breuer, Heinz-Peter and Petruccione, Francesco},
	year = {2007},
	month = jan,
	edition = {1st},
	publisher = {Oxford University Press},
	address = {Oxford},
	urldate = {2024-05-13},
	isbn = {978-0-19-921390-0 978-0-19-170634-9},
	langid = {english},
doi = {10.1093/acprof:oso/9780199213900.001.0001}
}

@book{walls_book,
    author = {{D. F. Walls and  G. J. Milburn}},
    title = {Quantum optics},
    publisher = {Springer Nature Switzerland AG},
    year = {2025},
edition = {3rd},
doi ={10.1007/978-3-031-84177-4}
}

@article{PhysRevA.29.408,
  title = {Use of cavities in squeezed-state generation},
  author = {Yurke, Bernard},
  journal = {Phys. Rev. A},
  volume = {29},
  issue = {1},
  pages = {408--410},
  numpages = {0},
  year = {1984},
  month = {Jan},
  publisher = {American Physical Society},
  doi = {10.1103/PhysRevA.29.408},
  url = {https://link.aps.org/doi/10.1103/PhysRevA.29.408}
}

@article{PhysRevA.102.023717,
  title = {Quantum interactions with pulses of radiation},
  author = {Kiilerich, Alexander Holm and M\o{}lmer, Klaus},
  journal = {Phys. Rev. A},
  volume = {102},
  issue = {2},
  pages = {023717},
  numpages = {14},
  year = {2020},
  month = {Aug},
  publisher = {American Physical Society},
  doi = {10.1103/PhysRevA.102.023717}
}

@book{Peskin:1995ev,
    author = "Peskin, Michael E. and Schroeder, Daniel V.",
    title = "{An Introduction to quantum field theory}",
    doi = "10.1201/9780429503559",
    isbn = "978-0-201-50397-5, 978-0-429-50355-9, 978-0-429-49417-8",
    publisher = "Addison-Wesley",
    address = "Reading, USA",
    year = "1995"
}

@article{PhysRevResearch.5.033107,
  title = {Third quantization of open quantum systems: Dissipative symmetries and connections to phase-space and {K}eldysh field-theory formulations},
  author = {McDonald, Alexander and Clerk, Aashish A.},
  journal = {Phys. Rev. Res.},
  volume = {5},
  issue = {3},
  pages = {033107},
  numpages = {20},
  year = {2023},
  month = {Aug},
  publisher = {American Physical Society},
  doi = {10.1103/PhysRevResearch.5.033107},
  url = {https://link.aps.org/doi/10.1103/PhysRevResearch.5.033107}
}

@article{PhysRevA.108.053703,
  title = {Analytical approach to higher-order correlation functions in U(1) symmetric systems},
  author = {Lu, Zhi-Guang and Shang, Cheng and Wu, Ying and L\"u, Xin-You},
  journal = {Phys. Rev. A},
  volume = {108},
  issue = {5},
  pages = {053703},
  numpages = {20},
  year = {2023},
  month = {Nov},
  publisher = {American Physical Society},
  doi = {10.1103/PhysRevA.108.053703},
  url = {https://link.aps.org/doi/10.1103/PhysRevA.108.053703}
}

@book{10.1093/acprof:oso/9780199213900.002.14006,
    author = {Heinz-Peter Breuer and Francesco Petruccione},
    isbn = {9780199213900},
    title = {Bibliography},
    booktitle = {The Theory of Open Quantum Systems},
    publisher = {Oxford University Press},
    year = {2007},
    month = {01},
    doi = {10.1093/acprof:oso/9780199213900.002.14006},
    url = {https://doi.org/10.1093/acprof:oso/9780199213900.002.14006},
}

@article{PhysRevA.106.022214,
  title = {Quantum regression theorem for multi-time correlators: A detailed analysis in the Heisenberg picture},
  author = {Khan, Sakil and Agarwalla, Bijay Kumar and Jain, Sachin},
  journal = {Phys. Rev. A},
  volume = {106},
  issue = {2},
  pages = {022214},
  numpages = {16},
  year = {2022},
  month = {Aug},
  publisher = {American Physical Society},
  doi = {10.1103/PhysRevA.106.022214},
  url = {https://link.aps.org/doi/10.1103/PhysRevA.106.022214}
}

@article{Keldysh1964ud,
    author = "Keldysh, L. V.",
    title = "{Diagram Technique for Nonequilibrium Processes}",
    url = "https://www.jetp.ras.ru/cgi-bin/dn/e_020_04_1018.pdf",
    journal = "Sov. Phys. JETP",
    volume = "20",
    pages = "1018--1026",
    year = "1965",
}
\end{document}